\documentclass[aip,jcp,preprint,longbibliography]{revtex4-2}

\usepackage{amsmath,amssymb,amsfonts}

\usepackage{newtxtext,newtxmath}

\usepackage{mathtools}
\usepackage{booktabs}
\usepackage{graphicx}
\usepackage[version=4]{mhchem}
\usepackage{multirow}
\usepackage[section]{placeins}

\usepackage{algorithm}
\usepackage{algpseudocode}
\usepackage[caption=false]{subfig}


\begin{document}
\newcommand{\matA}{\mathbf{A}}
\newcommand{\matB}{\mathbf{B}}
\newcommand{\matX}{\mathbf{X}}
\newcommand{\matY}{\mathbf{Y}}
\newcommand{\matU}{\mathbf{U}}
\newcommand{\matQ}{\mathbf{Q}}
\newcommand{\matI}{\mathbf{I}}
\newcommand{\matK}{\mathbf{K}}
\newcommand{\matM}{\mathbf{M}}

\title{Energy-Specific Bethe-Salpeter Equation Implementation for Efficient Optical Spectrum Calculations}

\author{Christopher Hillenbrand}
\author{Jiachen Li}
\author{Tianyu Zhu}
\email{tianyu.zhu@yale.edu}
\affiliation{Department of Chemistry, Yale University, New Haven, Connecticut 06520, United States}

\begin{abstract}
We present an energy-specific Bethe-Salpeter equation (BSE) implementation for efficient core and valence optical spectrum calculations. In energy-specific BSE, high-lying excitation energies are obtained by constructing trial vectors and expanding the subspace targeting excitation energies above the predefined energy threshold in the Davidson algorithm. To calculate optical spectra over a wide energy range, energy-specific BSE can be applied to multiple consecutive small energy windows, where trial vectors for each subsequent energy window are made orthogonal to the subspace of preceding windows to accelerate the convergence of the Davidson algorithm. For seven small molecules, energy-specific BSE combined with $G_0W_0$ provides small errors around 0.8 eV for absolute and relative $K$-edge excitation energies when starting from a hybrid PBEh solution with 45\% exact exchange. We further showcase the computational efficiency of this approach by simulating the N $1s$ $K$-edge excitation spectrum of the porphine molecule and the valence optical spectrum of silicon nanoclusters involving 6,000 excited states using $G_0W_0$-BSE. This work expands the applicability of the $GW$-BSE formalism for investigating high-energy excited states of large systems.
\end{abstract}

\maketitle
\section{INTRODUCTION}
Accurate description of excited states plays a crucial role in understanding various phenomena and processes in chemistry, biochemistry, and materials science,
such as charge transfer in photoinduced reactions\cite{rappoportPhotoinducedIntramolecularCharge2004,zhangUltrafastInvestigationPhotoinduced2017,zhuChargeRecombinationPhosphorescent2016,zhuChargeTransferMolecular2018,zhuUnravelingFateHost2019}
and quantum confinement in nanoparticles\cite{frederickControlExcitonConfinement2013,kilinaSurfaceChemistrySemiconducting2016}.
Spectroscopy offers some of the best available means to manipulate and observe excited states.
As investigators propose ever more ambitious experiments, establishing the link between spectroscopic experimental data and the underlying physical phenomena accordingly becomes more difficult.
Therefore,
simulating excited-state properties with first-principle approaches assists in the interpretation of spectroscopic data and the design of new experiments.

Of the many quantum chemistry methods have been developed to compute excited-state properties,
linear response time-dependent density functional theory (LR-TDDFT)\cite{casidaTimeDependentDensityFunctional1995,ullrichTimeDependentDensityFunctionalTheory2011} is the most popular and is widely implemented in modern quantum chemistry packages.
In a typical implementation based on the Davidson algorithm\cite{stratmannEfficientImplementationTimedependent1998},
the computational cost of LR-TDDFT with hybrid functionals is $\mathcal{O} (N^4)$,
where $N$ is the total number of basis functions.
LR-TDDFT offers a good balance between computational cost and accuracy and has achieved great success in simulating excited states to model excited-state properties of molecules, liquids and solid-state materials\cite{furcheAdiabaticTimedependentDensity2002,sottileTDDFTMoleculesSolids2005,casidaTimedependentDensityfunctionalTheory2009,laurentTDDFTBenchmarksReview2013,bremondAccuracyTDDFTGeometries2018,sarkarBenchmarkingTDDFTWave2021,jinExcitedStateProperties2023}. 
However, it suffers from several shortcomings.
Conventional exchange-correlation (XC) functionals have incorrect long-range behavior and consequently fail to describe charge-transfer and Rydberg excited states\cite{dreuwLongrangeChargetransferExcited2003,tozerRelationshipLongrangeChargetransfer2003}. 

In Green's function theory\cite{hedinNewMethodCalculating1965,martinInteractingElectrons2016},
excited state energies are commonly obtained via the so-called Bethe-Salpeter equation\cite{salpeterRelativisticEquationBoundState1951,shamManyParticleDerivationEffectiveMass1966,hankeManyParticleEffectsOptical1979} (BSE) formalism. It is closely related to LR-TDDFT and has been widely used to simulate optical spectroscopy for a broad range of systems\cite{strinatiDynamicalShiftBroadening1982,blaseChargetransferExcitationsMolecular2011,jacqueminAssessmentConvergencePartially2016,jacqueminBetheSalpeterFormalism2017,azariasBetheSalpeterStudyCationic2017,rangelAssessmentLowlyingExcitation2017,moninoSpinConservedSpinFlipOptical2021,choSimplifiedGWBSE2022,mckeonOptimallyTunedRangeseparated2022,liCombiningLocalizedOrbital2022,liCombiningRenormalizedSingles2022,vorwerkDisentanglingPhotoexcitationPhotoluminescence2023,zhangEffectDynamicalScreening2023,haberMaximallyLocalizedExciton2023,rauwolfNonlinearLightMatter2024,wuQuasiparticleExcitonicProperties2024,zhou2024all}.
When quasiparticle (QP) energies computed at the $GW$ level are used as the input,
the resulting method is denoted as $GW$-BSE.
The success of the $GW$-BSE approach can be attributed to several reasons.
First, 
the screened interaction formulated with QP energies from the $GW$ calculation is important for describing the electron-hole interaction in realistic systems\cite{blaseBetheSalpeterEquation2018,blaseBetheSalpeterEquation2020}.
Second,
$GW$ QP energies substantially improve upon the Kohn-Sham orbital energies in DFT\cite{kohnSelfConsistentEquationsIncluding1965,parrDensityFunctionalTheoryAtoms1989} for predicting both core and valence energy levels,
which enter the calculation of optical excitation energies via the density-density response function.
Third,
the scaling of a typical BSE implementation\cite{krauseImplementationBetheSalpeterEquation2017} is $\mathcal{O} (N^4)$. This is usually affordable relative to the prerequisite $GW$ calculation.

Among the linear-response and wave function-based alternative methods to $GW$-BSE and LR-TDDFT are the particle-particle random phase approximation (ppRPA)~\cite{vanaggelenExchangecorrelationEnergyPairing2013,yangDoubleRydbergCharge2013,yangExcitationEnergiesParticleparticle2014,yangSingletTripletEnergy2015,yangChargeTransferExcitations2017,liLinearScalingCalculations2023,liMultireferenceDensityFunctional2022,liAccurateExcitationEnergies2024,liParticleParticleRandom2024}, 
non-orthogonal configuration interaction singles (NOCIS)\cite{oosterbaanNonOrthogonalConfigurationInteraction2019,carter-fenkChoiceReferenceOrbitals2022},
equation-of-motion coupled-cluster (EOM-CC)~\cite{loosReferenceEnergiesDouble2019,loosReferenceEnergiesIntramolecular2021,loosHowAccurateAre2021,loosMountaineeringStrategyExcited2024,michalakRankReducedEquationMotionCoupled2024} theory, and algebraic diagrammatic construction (ADC)~\cite{winterBenchmarks00Transitions2013,loosMountaineeringStrategyExcited2018,mazinCoreExcitedStatesXray2023,maierConsistentThirdorderOneparticle2023,loosMountaineeringStrategyExcited2024,sulznerRoleSinglesAmplitudes2024}. Another method, excited state-specific orbital-optimized DFT, promises excitation energies\cite{haitOrbitalOptimizedDensity2021,haitComputingXrayAbsorption2022} much more accurate than those from LR-TDDFT at a computational cost similar to ground-state DFT. 

Despite the success of the $GW$-BSE formalism,
it still faces several challenges.
First,
predicting high-lying excitations, such as core excitations as measured in X-ray absorption spectroscopy (XAS), can be difficult for $GW$-BSE using the standard Davidson algorithm,
because the required subspace is extremely large compared to those for low-lying excitations.
As shown in Ref.\citenum{yaoAllElectronBSEGWMethod2022},
full diagonalization of the BSE matrix was used to access $K$-edge excitation energies,
which scales as $\mathcal{O}(N^6)$ and is computationally prohibitive for large systems.
Second, 
computing the optical spectrum of a large system (e.g.,~a nanocluster) in a wide energy range requires solving for a massive number of excited states simultaneously\cite{kickSuperresolutionTechniquesSimulate2024}. In this regime, the naive Davidson algorithm uses huge amounts of memory and spends an increasing proportion of its time diagonalizing dense matrices.
Lanczos recurrence- and filter diagonalization-based approaches \cite{shaoStructurePreservingLanczos2018,wallExtractionFilterdiagonalizationGeneral1995,bradburyBetheSalpeterEquation2022,bradburyOptimizedAttenuatedInteraction2023} can often provide good estimates of the optical spectrum at reduced cost. However, if one needs to calculate precise transition densities, these methods may not be applicable, since they do not provide a means to converge individual excited states besides continuing the iteration.

In this work,
we present an efficient ``energy-specific'' scheme for computing high-lying core excitation energies and optical spectra with dense excited-state energy levels within $GW$-BSE.
The energy-specific modification to the Davidson algorithm has been employed previously for calculating $K$-edge excitation energies with LR-TDDFT and EOM-CCSD\cite{liangEnergySpecificLinearResponse2011,pengEnergySpecificEquationMotionCoupledCluster2015,lestrangeCalibrationEnergySpecificTDDFT2015,kasperWellTemperedHybridMethod2018}.
Here we combine it with the $GW$-BSE formalism and further develop an efficient algorithm for solving the BSE in multiple energy windows.
In the energy-specific BSE approach,
initial trial vectors are chosen from occupied-virtual orbital pairs with a $GW$ quasiparticle energy difference exceeding some predefined threshold.
To accelerate convergence, states are expanded until convergence in small batches (``windows''). Optical spectra over a wide energy range are obtained by sliding the window up in energy until the entire energy range has been covered.
We demonstrate that energy-specific $GW$-BSE predicts accurate $K$-edge excitation energies of small molecules and the porphine molecule,
requiring only similar computational cost to low-lying excited-state calculations.
In addition, we show that it is capable of calculating precise excitation energies and transition densities in a wide energy range for silicon nanoclusters,
totaling thousands of excited states.

\section{THEORY}

\subsection{Bethe-Salpeter equation}
In the $GW$-BSE approach,
the BSE kernel in the Dyson equation for the two-particle correlation function is the functional derivative of the $GW$ self-energy with respect to the one-particle Green's function,
which expresses the linear response of the one-particle Green’s function to a general non-local perturbation and leads to the dynamical screened interaction\cite{martinInteractingElectrons2016}.
Like the Casida equation in TDDFT\cite{casidaTimeDependentDensityFunctional1995,ullrichTimeDependentDensityFunctionalTheory2011}, the Bethe-Salpeter equation is a linear response eigenvalue problem\cite{baiMinimizationPrinciplesLinear2012,shaoStructurePreservingParallel2016,ghoshConceptsMethodsModern2016,krauseImplementationBetheSalpeterEquation2017,blaseBetheSalpeterEquation2020}
\begin{equation}\label{eq:bse}
    \begin{bmatrix}
        \matA         & \matB        \\
        \overline{\matB} & \overline{\matA}
    \end{bmatrix}
    \begin{bmatrix}
        X \\
        Y
    \end{bmatrix}
    = \Omega
    \begin{bmatrix}
        \matI & 0      \\
        0     & -\matI
    \end{bmatrix}
    \begin{bmatrix}
        X \\
        Y
    \end{bmatrix} 
\end{equation}
where $\Omega$ is the excitation energy and $X$ and $Y$ are the transition amplitudes. The matrices $A$ and $B$ are defined as
\begin{align}
    A_{ia, \, jb} & = \delta_{ij} \delta_{ab} (\varepsilon_a - \varepsilon_i) + v_{ia, \, bj} - W_{ij, \, ba} \label{eq:a} \\
    B_{ia, \, jb} & = v_{ia,\, jb} - W_{ib, \, aj} \label{eq:b} 
\end{align}
where $\{\varepsilon_p\}$ are quasiparticle energies computed at the $GW$ level.
For clarity, we will also assume that $A$ and $B$ are both real and symmetric. This is the case if the orbitals are real.
In Eq.~\ref{eq:a} and the following, 
we use $i$, $j$, $k$, $l$ for occupied orbitals, 
$a$, $b$, $c$, $d$ for virtual orbitals, 
$p$, $q$, $r$, $s$ for general molecular orbitals, 
and $m$ for the index of excited states.
In Eq.~\ref{eq:bse}, $v$ is the Coulomb interaction defined as
$v_{pq, \, rs} = \int dx_{1} \, dx_{2} \, 
\phi_{p}^{*}(x_{1}) \phi_{q}(x_{1}) ||r_1 - r_2 ||^{-1} \phi_{r}^{*}(x_{2}) \phi_{s}(x_{2})$, and $W$ is the screened interaction taken at the static limit ($\omega=0$). In turn, $W$ is given by
\begin{equation}
    W_{pq, \, rs} = \sum_{tu} (\epsilon^{-1})_{pq, \, tu} \; v_{tu, \, rs}
\end{equation}
where the dielectric function $\epsilon$ is calculated from the static response function $\chi$\cite{ghoshConceptsMethodsModern2016,krauseImplementationBetheSalpeterEquation2017}
\begin{align}
    \epsilon_{pq, \, rs} &= \delta_{pr}\delta_{qs} - v_{pq, \, rs} \; \chi_{rs, \, rs}, \\
    \chi_{ia, \, ia} &= \chi_{ai, \, ai} = (\varepsilon_i - \varepsilon_a)^{-1}
\end{align}
The Tamm-Dancoff approximation (TDA) in BSE is obtained by taking $B \approx 0$ in Eq.~\ref{eq:bse}.
As shown in Refs.\citenum{roccaInitioCalculationsOptical2010,ducheminShortRangeLongRangeChargeTransfer2012,faberManybodyGreensFunction2013},
$GW$-BSE/TDA can lead to blue-shifts in nanosized systems and worse estimations for singlet-triplet gaps in organic molecules. With excitation energies and transition amplitudes obtained from Eq.~\ref{eq:bse},
the oscillator strength of the $m$-th excited state is computed as
\begin{equation}
\label{eq:osc_strength}
f_{m}=\frac{2}{3} \Omega_{m} \sum_{r=x,y,z} \left [ 
\sum_{ia}
\left \langle i |\hat{r}| a \right \rangle 
\left ( X^m_{ia} + Y^m_{ia} \right) \right ] ^2
\end{equation}
The only difference in Eq.~\ref{eq:bse} compared to the Casida equation in TDDFT is that the BSE kernel replaces the XC kernel.
A full diagonalization of Eq.~\ref{eq:bse} scales as $\mathcal{O}(N^6)$.
If only low-lying excited states are desired,
Eq.~\ref{eq:bse} can be solved with the canonical Davidson algorithm\cite{stratmannEfficientImplementationTimedependent1998} with $\mathcal{O}(N^4)$ scaling.
However,
the computational efficiency significantly degrades when the number of desired roots becomes large. 

\subsection{Energy-Specific BSE}

\subsubsection{Computational method}
For molecular systems with real orbitals, Eq.~\ref{eq:bse} is usually rewritten as a generalized symmetric eigenproblem of half the original dimension\cite{stratmannEfficientImplementationTimedependent1998,shaoStructurePreservingParallel2016}
\begin{equation}
\label{eq:geneig}
(\matA-\matB)(\matA+\matB)\;(X+Y)=\Omega^{2}(X+Y)
\end{equation}
where $\matA+\matB$ and $\matA-\matB$ are symmetric
positive definite\cite{stratmannEfficientImplementationTimedependent1998,shaoStructurePreservingLanczos2018}, in order to reduce the computational cost. Unless the entire spectrum
is needed, extracting a small fraction of eigenpairs
with a matrix-free solver is usually faster than full diagonalization of $(\matA-\matB)(\matA+\matB)$. Because such methods avoid explicitly constructing
$\matA$ and $\matB$, they can also take advantage of any available cost-saving
approximations to $v$ and $W$ matrices in Eq.~\ref{eq:a} and Eq.~\ref{eq:b}, such as density fitting (used in this work). Traditionally, the Davidson algorithm\cite{stratmannEfficientImplementationTimedependent1998,vecharynskiEfficientBlockPreconditioned2017} (Algorithm \ref{alg:davidson}) is used for solving TDDFT and BSE problems of the form in Eq.~\ref{eq:bse}.

\begin{algorithm}[H]
\caption{Davidson algorithm for linear response\cite{stratmannEfficientImplementationTimedependent1998,vecharynskiEfficientBlockPreconditioned2017}}
\label{alg:davidson}
\begin{algorithmic}[1]
\Require An orthonormal list of starting vectors $\matU = [u_1 \cdots u_{k}]$, a convergence tolerance $\tau > 0$, and the number of desired roots $N \le k$.
\While{not converged} \label{whileloop}
    \State Calculate $(\matA+\matB)\matU$, $(\matA-\matB)\matU$ for any new vectors in $U$.
    \State Form the matrices $\tilde{\matK} \gets \matU^\dagger (\matA+\matB) \matU$ and $\tilde{\matM} \gets \matU^\dagger (\matA-\matB) \matU$. \label{formreducedmat}
    \State Solve\footnote{Users of LAPACK may find it convenient to use \texttt{dsygvd}} $\tilde{\matM}\tilde{\matK}(\tilde{X}+\tilde{Y})=\Omega^2(\tilde{X}+\tilde{Y})$ for $N$ lowest eigenvalues $\Omega_i$ and right eigenvectors $\tilde{X}_i+\tilde{Y}_i$. \label{subspace_diag}
    \State Calculate the left eigenvectors $\tilde{X}_i-\tilde{Y}_i \gets \tilde{K} (\tilde{X}_i+\tilde{Y}_i)/\Omega_i$
    
    \State Obtain the approximate eigenvectors (Ritz vectors) of $(\matA-\matB)(\matA+\matB)$ in the MO space:
    \[
        X_i-Y_i = \matU(\tilde{X}_i-\tilde{Y}_i), \quad
        X_i+Y_i = \matU(\tilde{X}_i+\tilde{Y}_i) \quad \textrm{for all} \; i
    \]
    \State Calculate the residues $
        L_i \gets (\matA-\matB)\matU(\tilde{X}_i-\tilde{Y}_i) - \Omega_i (X_i+Y_i)$, 
        $R_i \gets (\matA+\matB)\matU(\tilde{X}_i+\tilde{Y}_i)  - \Omega_i (X_i-Y_i)$.  
    \State Initialize $\matQ$ to an empty list $[\;]$.
    \For{$i=1,\ldots,N$}

        \If{either $\|L_i\|$ or $\|R_i\|$ exceeds the convergence tolerance $\tau$}\label{convergencecheck}
            \State Add the preconditioned residues $
            (\Omega_i - (\varepsilon_a - \varepsilon_i))^{-1} L_i$ and $(\Omega_i - (\varepsilon_a - \varepsilon_i))^{-1} R_i$
            to $\matQ$.
        \EndIf
    \EndFor

    \If{$\matQ$ is empty}
            \State Get $\matX'$ and $\matY'$ from $\matX'+\matY'$ and $\matX'-\matY'$ and \Return $\{\Omega, \matX', \matY'\}$. \label{returnsol}
    \EndIf

    \State Orthonormalize the vectors of $\matQ$ against $\matU$ and among themselves\footnote{A rank-revealing QR decomposition is convenient since it allows one to discard negligible components of $\matQ$. As $\matU$ grows large, it may be necessary to orthogonalize against $\matU$ before and after QR to avoid restarts.};  then add them to $\matU$. \label{line:newvec}
\EndWhile
\end{algorithmic}
\end{algorithm}

In Ref.~\citenum{liangEnergySpecificLinearResponse2011}, Liang et.~al.~employed an energy screening and bracketing technique to solve the Casida equations for higher-lying states in TDHF/TDDFT. Because the BSE and Casida equations are both of the form in Eq.~\ref{eq:bse}, the technique can be easily generalized to this context. Aside from the matter of the initial trial vectors $U$, it requires only one modification to Algorithm \ref{alg:davidson}, as shown in Algorithm~\ref{alg:esdavidson}. In this way, only Ritz pairs above a given energy minimum are expanded. Note that the modified algorithm is equivalent to the original Davidson algorithm if $E_\mathrm{min}=0$.

\begin{algorithm}[H]
\caption{Energy-specific Davidson is the same as algorithm \ref{alg:davidson} except}
\label{alg:esdavidson}
\begin{algorithmic}[1]
\Require An energy minimum $E_\mathrm{min}$.
\end{algorithmic}
\begin{itemize}
    \item \textbf{Line \ref{subspace_diag}}: Solve the generalized symmetric eigenproblem $\tilde{\matM}\tilde{\matK}(\tilde{X}+\tilde{Y})=\Omega^2(\tilde{X}+\tilde{Y})$ for the  $N$ lowest eigenvalues $\Omega_i$ and right eigenvectors $\tilde{X}_i+\tilde{Y}_i$ such that $\Omega_i \ge E_\mathrm{min}$.
\end{itemize}
\end{algorithm}

When calculating core excitation energies, the relevant states are sometimes interspersed with unwanted states that lie in the same energy region. The energy bracketing tactic of Algorithm \ref{alg:esdavidson} cannot discriminate between core and non-core excitations, and will be inefficient if the unwanted states are numerous and narrowly spaced. We encountered this issue in our calculation of porphine N $K$-edge energies, and found that a simple modification to Algorithm \ref{alg:esdavidson} reduced the cost of this particular calculation by a large factor, as shown in Algorithm~\ref{alg:coreesdavidson}. This modification has no impact on the accuracy of the excited states returned by the algorithm, since convergence is verified the same way as in Algorithm \ref{alg:davidson}.

\begin{algorithm}[H]
\caption{Core-specific Davidson is the same as algorithm \ref{alg:esdavidson} except}
\label{alg:coreesdavidson}
\begin{algorithmic}[1]
\Require A list of occupied orbitals $O$ for which excitations are sought, and a threshold $\tau_\mathrm{occ}$.
\end{algorithmic}
\begin{itemize}
    \item \textbf{Line \ref{convergencecheck}}: The residues of $X$, $Y$ are added to $\matQ$ if $\|L\| > \tau$ or $\|R\| > \tau$, \textbf{but only if}
    $$\sum_{o\in O} \sum_a X_{oa}^2 > \tau_\mathrm{occ},$$
    i.e.,~the total weight of $O$ in $X$ is large enough.
    \item \textbf{Line \ref{returnsol}}: Only return a Ritz pair $\Omega$, $X$, $Y$ if $\sum_{o\in O} \sum_a X_{oa}^2 > \tau_\mathrm{occ}$.
\end{itemize}
\end{algorithm}

\subsubsection{Initial trial vector generation}
Although it is possible to use random starting vectors $\matU$, our implementation chooses them as follows. Given a list of occupied orbitals $O$, all pairs $(o, a), \; o \in O$  below the energy minimum (plus a shift $\delta E \ge 0$), i.e.,~those with $\varepsilon_a - \varepsilon_o < E_\mathrm{min} + \delta E$, are discarded. Finally, for each of the $k$ lowest energy remaining pairs, the elementary vector representing the $o\to a$ transition is added to $\matU$. This is the same method used by Liang et.~al.~in Ref.~\citenum{liangEnergySpecificLinearResponse2011}, except that the starting vectors can be restricted, if necessary, to a given list of occupied orbitals. For all core $K$-edge calculations in this paper, $O$ is set to those occupied MOs with a minimum of 30\% $1s$ orbital character on the target atoms.

\subsubsection{Restarts, deflation, and sliding windows}
\label{section:windows}

During a calculation, it sometimes happens that $\tilde{\matM} = \matU^\dagger (\matA-\matB) \matU$ is non-positive due to loss of orthogonality among the vectors of $\matU$. When this occurs, Algorithm \ref{alg:davidson} cannot proceed, and restarting is necessary. During a restart, a new orthonormal basis $\matU'$ is constructed from the Ritz vectors $\matX'$ and $\matY'$ found in the previous iteration. Then $\matU'$ replaces $\matU$, and the algorithm restarts at line \ref{whileloop}.

Because recalculating $(\matA+\matB)\matU$ and $(\matA-\matB)\matU$ is very costly, it is important to avoid restarts. To prevent loss of orthogonality, our implementation orthogonalizes the new search vectors $[q_1 \cdots q_N]$ against $\matU$ twice---both before and after orthogonalizing the $q_i$ among themselves---before adding them to $\matU$ in line \ref{line:newvec}. Thanks to this, unintentional restarts rarely occur.

As a calculation proceeds, the search space can grow very large, and the cost of solving the eigenproblems within each iteration can exceed that of the (usually more costly) contraction step. When the search space grows too large, it is deflated via an automatic restart that discards most of the Ritz vectors, as above.

Combined with a sliding window method, energy-specific bracketing is also useful when one wishes to obtain as many excited states as possible in a progressive fashion, without committing to a specific number of states. In this variation of Algorithm \ref{alg:esdavidson}, the first several (20--50) states above $E_\mathrm{min}$ are converged and saved. The energy minimum $E_\mathrm{min}$ is then reset to lie just above the highest converged state, and the solver moves onto the next window. Without a time limit, this algorithm runs indefinitely or until failure. Two kinds of failure have been observed:
\begin{enumerate}
    \item If the solver keeps only Ritz vectors from (or surrounding) the current energy window during deflation, subsequent iterations take longer to converge until, eventually, no progress occurs between deflations.
    \item If the solver keeps all Ritz vectors in the current window and below, steady progress continues until memory runs out or the cost of diagonalizing $\tilde{\matM}\tilde{\matK}$ becomes overwhelming.
\end{enumerate}
The sliding window technique is employed in this work only for calculating the silicon nanocluster valence spectrum.

\section{COMPUTATIONAL DETAILS}
We implemented algorithms 1--3 in the fcDMFT library~\cite{fcdmft,zhuEfficientFormulationInitio2020,cuiEfficientImplementationInitio2020,zhuInitioFullCell2021}, a quantum embedding library based on the PySCF quantum chemistry software package~\cite{sunPySCFPythonbasedSimulations2018,sunRecentDevelopmentsPySCF2020}.
All ground-state DFT calculations were carried out using PySCF without density fitting. $GW$~\cite{zhuAllElectronGaussianBasedG0W02021,leiGaussianbasedQuasiparticleSelfconsistent2022} and BSE calculations were performed with fcDMFT, using 3-center 2-electron repulsion integrals generated by PySCF. For each basis set, the corresponding RIFIT auxiliary basis set was used if available; otherwise an RI auxiliary basis was generated with the AutoAux algorithm\cite{stoychevAutomaticGenerationAuxiliary2017}.
All $GW$ and DFT calculations were non-relativistic.
A post-$GW$-BSE correction scheme derived in Ref.~\citenum{kellerRelativisticCorrectionScheme2020} was employed for relativistic corrections to the $K$-edge excitation energies. 
The magnitude of the corrections increases with the atomic number: 
0.12 eV for C$1s$,
0.24 for N$1s$ and 0.42 eV for O$1s$.

In the calculations for $K$-edge excitation energies of seven small molecules including 
\ce{NH3}, \ce{CH2O}, \ce{CO}, \ce{N2}, \ce{N2O}, \ce{C2H4} and \ce{H2O},
geometries were optimized at the CCSD level with the def2-TZVPD basis set\cite{weigendBalancedBasisSets2005} with GAUSSIAN16 A.03\cite{g16}.
XAS experiment reference values were taken from Ref.~\citenum{pengEnergySpecificEquationMotionCoupledCluster2015}.
For subsequent DFT and $GW$-BSE calculations, aug-cc-pVQZ basis set\cite{dunningGaussianBasisSets1989} was used for H and the aug-cc-pCVQZ basis set\cite{woonGaussianBasisSets1995} was used for remaining atoms.
$G_0W_0$ self-energy elements for the seven small molecules were evaluated analytically\cite{vansettenGWMethodQuantumChemistry2013} to obtain QP energies, where $G_0W_0$ was performed on top of the hybrid PBEh solutions with 45\% exact exchange (denoted as PBEh45)~\cite{golzeCoreLevelBindingEnergies2018,golzeAccurateAbsoluteRelative2020,liBenchmarkGWMethods2022,yaoAllElectronBSEGWMethod2022}.
In energy-specific BSE calculations,
the energy thresholds 280 eV, 390 eV, and 530 eV were used for C, N, and O $K$-edge excitation energies respectively.

In the calculations for $K$-edge excitation energies of porphine,
the ground state geometry was optimized at the B3LYP+D3 level\cite{beckeDensityfunctionalThermochemistryIII1993,leeDevelopmentColleSalvettiCorrelationenergy1988,grimmeConsistentAccurateInitio2010} with the 6-31G(d) basis set\cite{hariharanInfluencePolarizationFunctions1973} in GAUSSIAN16 A.03\cite{g16}.
XAS experiment reference values were taken from Refs.~\citenum{dillerInvestigatingMoleculesubstrateInteraction2013,krasnikovXrayAbsorptionPhotoemission2008}.
For subsequent DFT and $GW$-BSE calculations, 
the aug-cc-pCVTZ basis set\cite{woonGaussianBasisSets1995} was used for N atoms,
and the aug-cc-pVDZ basis set\cite{dunningGaussianBasisSets1989} was used for H and C atoms.
To reduce the computational cost,
$G_0W_0$@PBEh45 self-energy elements of occupied orbitals were evaluated with the contour deformation technique\cite{golzeCoreLevelBindingEnergies2018,zhuAllElectronGaussianBasedG0W02021},
while self-energy elements of virtual orbitals were evaluated with the lower-scaling analytic continuation technique\cite{golzeCoreLevelBindingEnergies2018,zhuAllElectronGaussianBasedG0W02021}.
In energy-specific BSE calculations,
the energy threshold 397 eV was used to get N $K$-edge excitation energies.

In the calculation of valence spectrum of the silicon nanocluster,
the nanocluster geometry was taken from Ref.\citenum{venturellaMachineLearningManyBody2024}.
The DFT ground state was calculated with the PBE0 functional\cite{perdewExactExchange1996} and the cc-pVTZ basis set\cite{dunningGaussianBasisSets1989}; and $G_0W_0$ quasiparticle energies were calculated using Pad{\'e} analytic continuation and the diagonal approximation.
In the BSE calculation, the MO space was truncated to include only orbitals with QP energies in the range $[-81.6, E_\mathrm{LUMO} + 27.2]$ eV, and the Tamm-Dancoff approximation was used. The lowest lying 6000 excited states were obtained with the sliding window technique of Section \ref{section:windows}. After obtaining oscillator strengths via Eq.~\ref{eq:osc_strength}, the optical absorption spectrum was plotted with Gaussian broadening ($\sigma=0.05$ eV).
For comparison, the optical absorption spectrum was also calculated with the Lanczos + generalized averaged Gauss quadrature method described in Ref.~\citenum{shaoStructurePreservingLanczos2018}. To obtain the isotropic absorption spectrum, the three Cartesian components of the MO electric dipole moment were used as starting vectors in three runs of 1500 iterations each; the spectra thus obtained were averaged\cite{brabecEfficientAlgorithmsEstimating2015}.

\section{RESULTS}

\subsection{Validation of Energy-Specific Formalism}
\FloatBarrier

\begin{table}[!ht]
\setlength\tabcolsep{7pt}
\caption{Excitation energies and oscillator strengths of \ce{H2O} and \ce{NH3} obtained from $G_0W_0$-BSE@PBEh45.
aug-cc-pVQZ basis set was used for H and aug-cc-pCVQZ basis set was used for remaining atoms. 
Excitation energies are in eV and oscillator strengths are in A.U.
}\label{tab:validation}
\begin{tabular}{clcccc}
\hline
    &                        & \multicolumn{2}{c}{energy-specific}     & \multicolumn{2}{c}{full diagonalization}   \\
\cmidrule(l{0.5em}r{0.5em}){3-4} \cmidrule(l{0.5em}r{0.5em}){5-6}
    &                        & excitation energy & oscillator strength & excitation energy    & oscillator strength \\
\hline
\ce{H2O} & 1s$\to$4a$_1$/3s  & 532.44            & 0.0168              & 532.43               & 0.0167              \\
         & 1s$\to$2b$_1$/3p  & 534.14            & 0.0363              & 534.14               & 0.0364              \\
         & 1s$\to$3p (b$_2$) & 536.53            & 0.0163              & 536.53               & 0.0163              \\
\ce{NH3} & 1s$\to$3s         & 399.65            & 0.0088              & 399.65               & 0.0088              \\
         & 1s$\to$3p (E)     & 401.28            & 0.0337              & 401.28               & 0.0337              \\
         & 1s$\to$3p (A$_1$) & 401.28            & 0.0337              & 401.28               & 0.0337              \\
         & 1s$\to$4s (A$_1$) & 402.70            & 0.0132              & 402.70               & 0.0132              \\
\hline
\end{tabular}
\end{table}

We first verify the accuracy of the core-specific Davidson algorithm (Algorithm \ref{alg:coreesdavidson}).
The calculated $K$-edge excitation energies and oscillator strengths obtained from Algorithm \ref{alg:coreesdavidson} and full diagonalization $GW$-BSE are shown in Table~\ref{tab:validation}.
The excitation energies and oscillator strengths for high-energy $K$-edge excitations from Algorithm \ref{alg:coreesdavidson} are nearly identical to those from full diagonalization; the largest difference between the two is $<0.01$ eV for excitation energies and $<0.0001$ A.U. for oscillator strengths.

\FloatBarrier

\subsection{$K$-edge Excitation Energies}
\FloatBarrier

\begin{table}[!ht]
\setlength\tabcolsep{7pt}
\caption{Mean absolute errors of absolute and relative $K$-edge excitation energies for 31 states of 7 small molecules obtained from different approaches.
$G_0W_0$-BSE/TDA@PBEh45 (literature) calculated with aug-cc-pCV6Z basis set were taken from Ref.\citenum{yaoAllElectronBSEGWMethod2022}.
Results of TDDFT and EOM approaches calculated with d-aug-cc-pCVDZ basis set were taken from Ref.\citenum{pengEnergySpecificEquationMotionCoupledCluster2015}.
Relativistic effects were considered using a post-$GW$-BSE correction scheme in Ref.\citenum{kellerRelativisticCorrectionScheme2020}.
All values are in eV.
}\label{tab:k-edge mol}
\begin{tabular}{c|ccc}
\hline
                                     & absolute (non-relativistic)    & absolute (relativistic) & relative \\
\hline
$G_0W_0$-BSE@PBEh45                  & 0.94                           & 0.74                    & 0.84     \\
$G_0W_0$-BSE/TDA@PBEh45              & 0.93                           & 0.73                    & 0.84     \\
$G_0W_0$-BSE/TDA@PBEh45~\cite{yaoAllElectronBSEGWMethod2022} & 0.81                 & 0.62                        & 0.67     \\
TDDFT@PBE0~\cite{pengEnergySpecificEquationMotionCoupledCluster2015}                       & 11.68                          & 11.54                   & 0.91     \\
TDDFT@BHandHLYP~\cite{pengEnergySpecificEquationMotionCoupledCluster2015}                      & 1.99                           & 1.74                    & 0.59     \\
EOM-MBPT2~\cite{pengEnergySpecificEquationMotionCoupledCluster2015}                            & 3.42                           & 3.70                    & 0.28     \\
EOM-CCSD~\cite{pengEnergySpecificEquationMotionCoupledCluster2015}                             & 3.11                           & 3.40                    & 0.26     \\
\hline
\end{tabular}
\end{table}

We next examine the ability of energy-specific $GW$-BSE to predict  $K$-edge core excitation energies.
Mean absolute errors (MAEs) of 31 excited states from 7 small molecules obtained from different computational methods are tabulated in Table~\ref{tab:k-edge mol}.
As shown in Refs.\citenum{golzeCoreLevelBindingEnergies2018,golzeAccurateAbsoluteRelative2020,liBenchmarkGWMethods2022,yaoAllElectronBSEGWMethod2022},
$G_0W_0$@PBEh45 is a good choice for calculating core-level binding energies, with errors of $\sim 0.3$ eV on the published datasets.
Our
$G_0W_0$-BSE/TDA@PBEh45 results for $K$-edge excitation energies exhibit errors around 0.7 eV\cite{yaoAllElectronBSEGWMethod2022} and generally agree well with literature values in Ref.~\citenum{yaoAllElectronBSEGWMethod2022}.
The small differences from Ref~\citenum{yaoAllElectronBSEGWMethod2022} (around 0.1 eV) may originate from different treatments of relativistic effects
and core-level orbitals described by numerical atomic orbitals (NAOs) in FHI-aims compared to Gaussian-type orbitals (GTOs) in PySCF.
For $K$-edge excitation energies of these molecules obtained from $G_0W_0$-BSE@PBEh45, the Tamm-Dancoff approximation incurs minimal additional error---TDA and full BSE exhibit similar errors for predicting absolute and relative $K$-edge excitation energies.
With relativistic corrections, we notice that
the errors of BSE for absolute $K$-edge excitation energies are reduced by around 0.2 eV.

In contrast, TDDFT@PBE0 exhibited extremely high absolute error for $K$-edge excitation energies due to delocalization error and lack of screening\cite{yu2024accurate}, while BHandHLYP reduces the absolute error to $\sim 2$ eV. The good performance of BHandHLYP, as noted in Refs.~\citenum{lestrangeCalibrationEnergySpecificTDDFT2015,pengEnergySpecificEquationMotionCoupledCluster2015} can be attributed to fortuitous error cancellation caused by its high percentage (50\%) of exact exchange.
EOM-MBPT2 and EOM-CCSD had quite good relative $K$-edge excitation energies with errors around 0.2 eV, but the absolute $K$-edge excitation energies had errors larger than 3 eV.

\begin{table}[!ht]
\setlength\tabcolsep{10pt}
\caption{$K$-edge excitation energies of porphine obtained from BSE and BSE/TDA based on $G_0W_0$@PBEh45 compared to experiments.
The aug-cc-pCVTZ basis set was used for N atoms,
the aug-cc-pVDZ basis set was used for H and C atoms.
Relativistic effects were considered using a post-$GW$-BSE correction scheme in Ref.\citenum{kellerRelativisticCorrectionScheme2020}.
All values are in eV.
}\label{tab:porphine}
\begin{tabular}{c|cccc}
\hline
      & exp\cite{polzonettiElectronicStructurePlatinum2004} & exp\cite{krasnikovXrayAbsorptionPhotoemission2008}  & BSE & BSE/TDA \\
\hline
A$_0$ & 398.2 & 397.90 & 397.86 & 397.86 \\
A$_4$ & 400.3 & 400.00 & 400.01 & 400.02 \\
\hline
\end{tabular}
\end{table}

\begin{figure}
\includegraphics[width=0.6\textwidth]{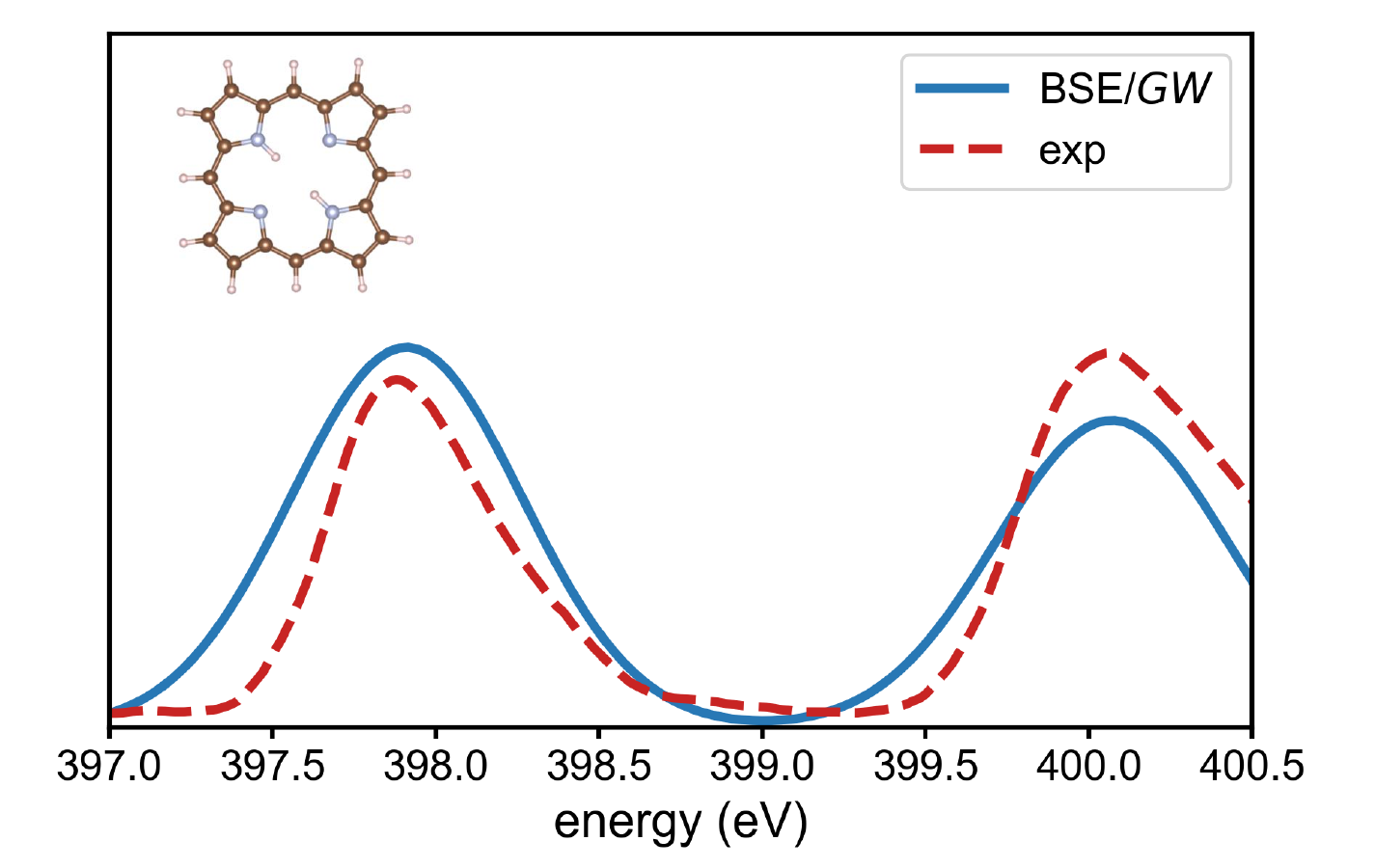}
\caption{XAS spectrum of N1s in porphine calculated with $G_0W_0$-BSE@PBEh45 compared to experiment reference\cite{krasnikovXrayAbsorptionPhotoemission2008}.
The height of the calculated spectrum is scaled to that of experiment spectrum.
}
\label{fig:porphine}
\end{figure}

We also applied the energy-specific $GW$-BSE approach to calculate N $1s$ $K$-edge excitation energies of porphine.
The energies thus obtained ($G_0W_0$-BSE@PBEh45) are shown in Table \ref{tab:porphine}, and a simulated X-ray absorption spectrum is shown in Fig.~\ref{fig:porphine} alongside the experimental spectrum from Ref.~\citenum{krasnikovXrayAbsorptionPhotoemission2008}.
Remarkably, the errors for both A$_0$ and A$_4$ $K$-edge excitation energies compared to the experiments in Refs.~\citenum{krasnikovXrayAbsorptionPhotoemission2008,dillerInvestigatingMoleculesubstrateInteraction2013} are only $0.1\sim0.3$ eV.
In addition to its excellent, though perhaps fortuitous, accuracy in this case,
the actual energy-specific BSE calculation was inexpensive.
With a 48-CPU\footnote{2 x Intel\textsuperscript{\textregistered} Xeon\textsuperscript{\textregistered} Platinum 8268 CPU @ 2.90GHz} node,
the time required to converge 20 roots was 273 minutes (no TDA) and $<5$ minutes (with TDA).
The computational bottleneck in the entire process was by a large margin the $GW$ step,
which uses the contour deformation technique to obtain accurate core-level QP energies but suffers from $O(N^5)$ scaling.
Hopefully, the development of an appropriate Green's function embedding scheme~\cite{zhuEfficientFormulationInitio2020,zhuInitioFullCell2021,li2024restoring,li2024interacting} will reduce the cost of the $GW$ step in the future.
Alternatively, one could try to avoid $GW$ entirely by approximating the $GW$ QP energies with other techniques (e.g. Ref.~\citenum{liCombiningLocalizedOrbital2022}) or predicting them with a machine learning model\cite{caylakMachineLearningQuasiparticle2021,venturellaMachineLearningManyBody2024,venturella2024unified,biswas2024incorporating}.
\FloatBarrier

\subsection{Valence Spectrum of Silicon Nanoclusters}
\FloatBarrier

\begin{figure}

\includegraphics[width=0.9\textwidth]{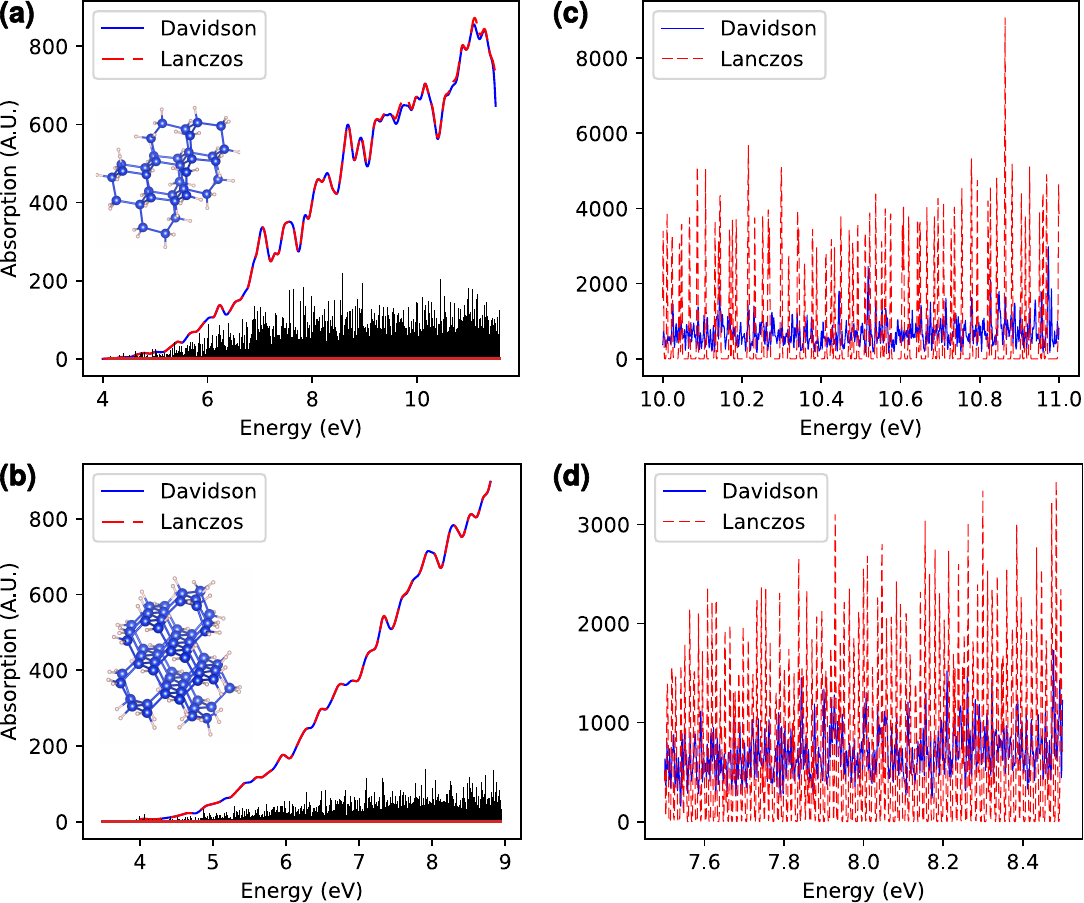}
\caption{Valence optical spectrum of the silicon nanoclusters \ce{Si35H44} and \ce{Si60H64} obtained from $G_0W_0$-BSE@PBE0 with energy-specific Davidson and Lanczos algorithms. (a--b) Plot of both spectra with a broadening of 0.05 eV. Scaled oscillator strengths are shown in black. (c--d) Zoomed-in view with a much narrower broadening of 0.001 eV.}

\label{fig:cluster}
\end{figure}

We also applied the the sliding window technique of Section \ref{section:windows}  to calculate the optical absorption spectrum of the silicon nanoclusters \ce{Si35H44} and \ce{Si60H64} with $GW$-BSE. For the larger cluster, the truncated MO space contained 152 occupied and 816 virtual orbitals, and the problem dimension was $152\times 816 = 124,032$.
The valence spectra of both systems contain a very large number of densely spaced, nearly degenerate states; the energy of the 6000th lowest state of \ce{Si60H54} is only $\sim$8.9 eV.
The optical absorption spectra thus obtained are shown in Fig.~\ref{fig:cluster}.

The sliding window technique described in Section \ref{section:windows} was chosen due to the need to converge a very large number of excited states without advance knowledge of the convergence behavior for the problem.
Aside from this convenience, the technique appears to be rather efficient in terms of computational cost.  When the solver moves up to a higher energy window, the Ritz vectors in that window are already orthogonal to transition amplitudes previously converged in lower energy windows. Because each energy window contains only 40 Ritz values, 
updates to the search space are smaller but more frequent than in Algorithm \ref{alg:davidson}. In our experiments, this sometimes significantly reduces the number of trial vectors necessary to converge a given number of states.
The whole calculation for \ce{Si60H64} took about 60 hours on the same 48-CPU node. The maximum memory usage was 195 GB, most of which comprised the density-fitted ERIs required to compute the action of $v$ and $W$ on transition vectors. With the maximum search space size set to 15,000, deflation occurred three times; the total number of matrix-vector products was about 60,000.

Remarkably, the optical absorption spectra from the Lanczos and Davidson algorithms agree very well for both systems, even though the Lanczos \ce{Si60H64} calculation took only 205 minutes (5.7\% of the time for Davidson) on the same 48-CPU node. Since the excited states are so densely spaced, precise transition densities are apparently irrelevant if one is interested only in the absorption spectrum. However, a closeup of the two optical spectra in Fig.~\ref{fig:cluster}(c--d), this time with a broadening of 0.001 eV, shows how the Lanczos algorithm has not converged individual Ritz values. Thus, the state-wise convergence guarantee provided by the Davidson algorithm is still valuable if one needs to analyze the character of individual excited states.

\section{CONCLUSIONS}
In summary, 
we present an energy-specific BSE approach for efficient optical spectrum calculations of high-lying excited states and wide energy ranges. 
In the energy-specific $GW$-BSE approach, 
trial vectors targeting excitation energies above the predefined energy threshold are constructed for the Davidson algorithm,
then the subspace is expanded with new trial vectors associated with transitions above the target energy threshold.
To calculate optical spectra over a wide energy range,
energy-specific $GW$-BSE can be applied to multiple small energy windows,
where trial vectors orthogonal to the true eigenvectors solved preceding energy windows are constructed for the subsequent energy window to accelerate the convergence of the Davidson algorithm.
For $K$-edge excitation energies of small molecules and porphine,
the energy-specific $G_0W_0$-BSE@PBEh45 approach provides good accuracy with errors around 0.8 eV and high efficiency.
For the silicon nanoclusters,
the energy-specific $G_0W_0$-BSE@PBE0 approach is capable of computing the optical spectra in a wide energy range,
totaling 6000 excited states.
This work expands the applicability of the $GW$-BSE formalism for computing excited states of large systems.

\section*{SUPPORTING INFORMATION}
See the Supporting Information for $K$-edge excitation energies of small molecules obtained from $G_0W_0$-BSE@PBEh45 with different basis sets, 
$K$-edge excitation energies and oscillator strengths of porphine calculated with $G_0W_0$-BSE@PBEh45.

\begin{acknowledgments}
This work was supported by the Air Force Office of Scientific Research under award number FA9550-24-1-0096 (J.L. and T.Z.).
C.H. acknowledges support from the National Science Foundation Engines Development Award: Advancing Quantum Technologies (CT) under award number 2302908. We thank Troy Van Voorhis for inspiring discussion.
We also thank the Yale Center for Research Computing for guidance and use of the research computing infrastructure, specifically Tom Langford.
\end{acknowledgments}

\section*{Data Availability Statement}
The data that support the findings of this study are available from the corresponding author upon reasonable request.

\bibliography{ref}

\begin{thebibliography}{123}%
\makeatletter
\providecommand \@ifxundefined [1]{%
 \@ifx{#1\undefined}
}%
\providecommand \@ifnum [1]{%
 \ifnum #1\expandafter \@firstoftwo
 \else \expandafter \@secondoftwo
 \fi
}%
\providecommand \@ifx [1]{%
 \ifx #1\expandafter \@firstoftwo
 \else \expandafter \@secondoftwo
 \fi
}%
\providecommand \natexlab [1]{#1}%
\providecommand \enquote  [1]{``#1''}%
\providecommand \bibnamefont  [1]{#1}%
\providecommand \bibfnamefont [1]{#1}%
\providecommand \citenamefont [1]{#1}%
\providecommand \href@noop [0]{\@secondoftwo}%
\providecommand \href [0]{\begingroup \@sanitize@url \@href}%
\providecommand \@href[1]{\@@startlink{#1}\@@href}%
\providecommand \@@href[1]{\endgroup#1\@@endlink}%
\providecommand \@sanitize@url [0]{\catcode `\\12\catcode `\$12\catcode
  `\&12\catcode `\#12\catcode `\^12\catcode `\_12\catcode `\%12\relax}%
\providecommand \@@startlink[1]{}%
\providecommand \@@endlink[0]{}%
\providecommand \url  [0]{\begingroup\@sanitize@url \@url }%
\providecommand \@url [1]{\endgroup\@href {#1}{\urlprefix }}%
\providecommand \urlprefix  [0]{URL }%
\providecommand \Eprint [0]{\href }%
\providecommand \doibase [0]{https://doi.org/}%
\providecommand \selectlanguage [0]{\@gobble}%
\providecommand \bibinfo  [0]{\@secondoftwo}%
\providecommand \bibfield  [0]{\@secondoftwo}%
\providecommand \translation [1]{[#1]}%
\providecommand \BibitemOpen [0]{}%
\providecommand \bibitemStop [0]{}%
\providecommand \bibitemNoStop [0]{.\EOS\space}%
\providecommand \EOS [0]{\spacefactor3000\relax}%
\providecommand \BibitemShut  [1]{\csname bibitem#1\endcsname}%
\let\auto@bib@innerbib\@empty
\bibitem [{\citenamefont {Rappoport}\ and\ \citenamefont
  {Furche}(2004)}]{rappoportPhotoinducedIntramolecularCharge2004}%
  \BibitemOpen
  \bibfield  {author} {\bibinfo {author} {\bibfnamefont {D.}~\bibnamefont
  {Rappoport}}\ and\ \bibinfo {author} {\bibfnamefont {F.}~\bibnamefont
  {Furche}},\ }\bibfield  {title} {\enquote {\bibinfo {title} {Photoinduced
  {{Intramolecular Charge Transfer}} in 4-({{Dimethyl}})aminobenzonitrile - {{A
  Theoretical Perspective}}},}\ }\href {https://doi.org/10.1021/ja037806u}
  {\bibfield  {journal} {\bibinfo  {journal} {J. Am. Chem. Soc.}\ }\textbf
  {\bibinfo {volume} {126}},\ \bibinfo {pages} {1277--1284} (\bibinfo {year}
  {2004})}\BibitemShut {NoStop}%
\bibitem [{\citenamefont {Zhang}\ \emph {et~al.}(2017)\citenamefont {Zhang},
  \citenamefont {Sun}, \citenamefont {Zhou}, \citenamefont {Wang},\ and\
  \citenamefont {Zhang}}]{zhangUltrafastInvestigationPhotoinduced2017}%
  \BibitemOpen
  \bibfield  {author} {\bibinfo {author} {\bibfnamefont {S.}~\bibnamefont
  {Zhang}}, \bibinfo {author} {\bibfnamefont {S.}~\bibnamefont {Sun}}, \bibinfo
  {author} {\bibfnamefont {M.}~\bibnamefont {Zhou}}, \bibinfo {author}
  {\bibfnamefont {L.}~\bibnamefont {Wang}},\ and\ \bibinfo {author}
  {\bibfnamefont {B.}~\bibnamefont {Zhang}},\ }\bibfield  {title} {\enquote
  {\bibinfo {title} {Ultrafast investigation of photoinduced charge transfer in
  aminoanthraquinone pharmaceutical product},}\ }\href
  {https://doi.org/10.1038/srep43419} {\bibfield  {journal} {\bibinfo
  {journal} {Sci Rep}\ }\textbf {\bibinfo {volume} {7}},\ \bibinfo {pages}
  {43419} (\bibinfo {year} {2017})}\BibitemShut {NoStop}%
\bibitem [{\citenamefont {Zhu}\ and\ \citenamefont
  {Van~Voorhis}(2016)}]{zhuChargeRecombinationPhosphorescent2016}%
  \BibitemOpen
  \bibfield  {author} {\bibinfo {author} {\bibfnamefont {T.}~\bibnamefont
  {Zhu}}\ and\ \bibinfo {author} {\bibfnamefont {T.}~\bibnamefont
  {Van~Voorhis}},\ }\bibfield  {title} {\enquote {\bibinfo {title} {Charge
  {{Recombination}} in {{Phosphorescent Organic Light-Emitting Diode
  Host}}--{{Guest Systems}} through {{QM}}/{{MM Simulations}}},}\ }\href
  {https://doi.org/10.1021/acs.jpcc.6b05559} {\bibfield  {journal} {\bibinfo
  {journal} {J. Phys. Chem. C}\ }\textbf {\bibinfo {volume} {120}},\ \bibinfo
  {pages} {19987--19994} (\bibinfo {year} {2016})}\BibitemShut {NoStop}%
\bibitem [{\citenamefont {Zhu}, \citenamefont {Van~Voorhis},\ and\
  \citenamefont {{de Silva}}(2018)}]{zhuChargeTransferMolecular2018}%
  \BibitemOpen
  \bibfield  {author} {\bibinfo {author} {\bibfnamefont {T.}~\bibnamefont
  {Zhu}}, \bibinfo {author} {\bibfnamefont {T.}~\bibnamefont {Van~Voorhis}},\
  and\ \bibinfo {author} {\bibfnamefont {P.}~\bibnamefont {{de Silva}}},\
  }\bibfield  {title} {\enquote {\bibinfo {title} {Charge {{Transfer}} in
  {{Molecular Materials}}},}\ }in\ \href
  {https://doi.org/10.1007/978-3-319-42913-7_7-1} {\emph {\bibinfo {booktitle}
  {Handbook of {{Materials Modeling}} : {{Methods}}: {{Theory}} and
  {{Modeling}}}}},\ \bibinfo {editor} {edited by\ \bibinfo {editor}
  {\bibfnamefont {W.}~\bibnamefont {Andreoni}}\ and\ \bibinfo {editor}
  {\bibfnamefont {S.}~\bibnamefont {Yip}}}\ (\bibinfo  {publisher} {Springer
  International Publishing},\ \bibinfo {address} {Cham},\ \bibinfo {year}
  {2018})\ pp.\ \bibinfo {pages} {1--31}\BibitemShut {NoStop}%
\bibitem [{\citenamefont {Zhu}\ and\ \citenamefont
  {Van~Voorhis}(2019)}]{zhuUnravelingFateHost2019}%
  \BibitemOpen
  \bibfield  {author} {\bibinfo {author} {\bibfnamefont {T.}~\bibnamefont
  {Zhu}}\ and\ \bibinfo {author} {\bibfnamefont {T.}~\bibnamefont
  {Van~Voorhis}},\ }\bibfield  {title} {\enquote {\bibinfo {title} {Unraveling
  the {{Fate}} of {{Host Excitons}} in {{Host}}--{{Guest Phosphorescent Organic
  Light-Emitting Diodes}}},}\ }\href {https://doi.org/10.1021/acs.jpcc.9b02820}
  {\bibfield  {journal} {\bibinfo  {journal} {J. Phys. Chem. C}\ }\textbf
  {\bibinfo {volume} {123}},\ \bibinfo {pages} {10311--10318} (\bibinfo {year}
  {2019})}\BibitemShut {NoStop}%
\bibitem [{\citenamefont {Frederick}\ \emph {et~al.}(2013)\citenamefont
  {Frederick}, \citenamefont {Amin}, \citenamefont {Swenson}, \citenamefont
  {Ho},\ and\ \citenamefont {Weiss}}]{frederickControlExcitonConfinement2013}%
  \BibitemOpen
  \bibfield  {author} {\bibinfo {author} {\bibfnamefont {M.~T.}\ \bibnamefont
  {Frederick}}, \bibinfo {author} {\bibfnamefont {V.~A.}\ \bibnamefont {Amin}},
  \bibinfo {author} {\bibfnamefont {N.~K.}\ \bibnamefont {Swenson}}, \bibinfo
  {author} {\bibfnamefont {A.~Y.}\ \bibnamefont {Ho}},\ and\ \bibinfo {author}
  {\bibfnamefont {E.~A.}\ \bibnamefont {Weiss}},\ }\bibfield  {title} {\enquote
  {\bibinfo {title} {Control of {{Exciton Confinement}} in {{Quantum
  Dot}}--{{Organic Complexes}} through {{Energetic Alignment}} of {{Interfacial
  Orbitals}}},}\ }\href {https://doi.org/10.1021/nl304098e} {\bibfield
  {journal} {\bibinfo  {journal} {Nano Lett.}\ }\textbf {\bibinfo {volume}
  {13}},\ \bibinfo {pages} {287--292} (\bibinfo {year} {2013})}\BibitemShut
  {NoStop}%
\bibitem [{\citenamefont {Kilina}, \citenamefont {Tamukong},\ and\
  \citenamefont {Kilin}(2016)}]{kilinaSurfaceChemistrySemiconducting2016}%
  \BibitemOpen
  \bibfield  {author} {\bibinfo {author} {\bibfnamefont {S.~V.}\ \bibnamefont
  {Kilina}}, \bibinfo {author} {\bibfnamefont {P.~K.}\ \bibnamefont
  {Tamukong}},\ and\ \bibinfo {author} {\bibfnamefont {D.~S.}\ \bibnamefont
  {Kilin}},\ }\bibfield  {title} {\enquote {\bibinfo {title} {Surface
  {{Chemistry}} of {{Semiconducting Quantum Dots}}: {{Theoretical
  Perspectives}}},}\ }\href {https://doi.org/10.1021/acs.accounts.6b00196}
  {\bibfield  {journal} {\bibinfo  {journal} {Acc. Chem. Res.}\ }\textbf
  {\bibinfo {volume} {49}},\ \bibinfo {pages} {2127--2135} (\bibinfo {year}
  {2016})}\BibitemShut {NoStop}%
\bibitem [{\citenamefont
  {Casida}(1995)}]{casidaTimeDependentDensityFunctional1995}%
  \BibitemOpen
  \bibfield  {author} {\bibinfo {author} {\bibfnamefont {M.~E.}\ \bibnamefont
  {Casida}},\ }\bibfield  {title} {\enquote {\bibinfo {title} {Time-{{Dependent
  Density Functional Response Theory}} for {{Molecules}}},}\ }in\ \href
  {https://doi.org/10.1142/9789812830586_0005} {\emph {\bibinfo {booktitle}
  {Recent {{Advances}} in {{Density Functional Methods}}}}},\ \bibinfo {series}
  {Recent {{Advances}} in {{Computational Chemistry}}}, Vol.\ \bibinfo {volume}
  {Volume 1}\ (\bibinfo  {publisher} {WORLD SCIENTIFIC},\ \bibinfo {year}
  {1995})\ pp.\ \bibinfo {pages} {155--192}\BibitemShut {NoStop}%
\bibitem [{\citenamefont
  {Ullrich}(2011)}]{ullrichTimeDependentDensityFunctionalTheory2011}%
  \BibitemOpen
  \bibfield  {author} {\bibinfo {author} {\bibfnamefont {C.~A.}\ \bibnamefont
  {Ullrich}},\ }\href@noop {} {\emph {\bibinfo {title} {Time-{{Dependent
  Density-Functional Theory}}: {{Concepts}} and {{Applications}}}}}\ (\bibinfo
  {publisher} {OUP Oxford},\ \bibinfo {year} {2011})\BibitemShut {NoStop}%
\bibitem [{\citenamefont {Stratmann}, \citenamefont {Scuseria},\ and\
  \citenamefont
  {Frisch}(1998)}]{stratmannEfficientImplementationTimedependent1998}%
  \BibitemOpen
  \bibfield  {author} {\bibinfo {author} {\bibfnamefont {R.~E.}\ \bibnamefont
  {Stratmann}}, \bibinfo {author} {\bibfnamefont {G.~E.}\ \bibnamefont
  {Scuseria}},\ and\ \bibinfo {author} {\bibfnamefont {M.~J.}\ \bibnamefont
  {Frisch}},\ }\bibfield  {title} {\enquote {\bibinfo {title} {An efficient
  implementation of time-dependent density-functional theory for the
  calculation of excitation energies of large molecules},}\ }\href
  {https://doi.org/10.1063/1.477483} {\bibfield  {journal} {\bibinfo  {journal}
  {J. Chem. Phys.}\ }\textbf {\bibinfo {volume} {109}},\ \bibinfo {pages}
  {8218--8224} (\bibinfo {year} {1998})}\BibitemShut {NoStop}%
\bibitem [{\citenamefont {Furche}\ and\ \citenamefont
  {Ahlrichs}(2002)}]{furcheAdiabaticTimedependentDensity2002}%
  \BibitemOpen
  \bibfield  {author} {\bibinfo {author} {\bibfnamefont {F.}~\bibnamefont
  {Furche}}\ and\ \bibinfo {author} {\bibfnamefont {R.}~\bibnamefont
  {Ahlrichs}},\ }\bibfield  {title} {\enquote {\bibinfo {title} {Adiabatic
  time-dependent density functional methods for excited state properties},}\
  }\href {https://doi.org/10.1063/1.1508368} {\bibfield  {journal} {\bibinfo
  {journal} {J. Chem. Phys.}\ }\textbf {\bibinfo {volume} {117}},\ \bibinfo
  {pages} {7433--7447} (\bibinfo {year} {2002})}\BibitemShut {NoStop}%
\bibitem [{\citenamefont {Sottile}\ \emph {et~al.}(2005)\citenamefont
  {Sottile}, \citenamefont {Bruneval}, \citenamefont {Marinopoulos},
  \citenamefont {Dash}, \citenamefont {Botti}, \citenamefont {Olevano},
  \citenamefont {Vast}, \citenamefont {Rubio},\ and\ \citenamefont
  {Reining}}]{sottileTDDFTMoleculesSolids2005}%
  \BibitemOpen
  \bibfield  {author} {\bibinfo {author} {\bibfnamefont {F.}~\bibnamefont
  {Sottile}}, \bibinfo {author} {\bibfnamefont {F.}~\bibnamefont {Bruneval}},
  \bibinfo {author} {\bibfnamefont {A.~G.}\ \bibnamefont {Marinopoulos}},
  \bibinfo {author} {\bibfnamefont {L.~K.}\ \bibnamefont {Dash}}, \bibinfo
  {author} {\bibfnamefont {S.}~\bibnamefont {Botti}}, \bibinfo {author}
  {\bibfnamefont {V.}~\bibnamefont {Olevano}}, \bibinfo {author} {\bibfnamefont
  {N.}~\bibnamefont {Vast}}, \bibinfo {author} {\bibfnamefont {A.}~\bibnamefont
  {Rubio}},\ and\ \bibinfo {author} {\bibfnamefont {L.}~\bibnamefont
  {Reining}},\ }\bibfield  {title} {\enquote {\bibinfo {title} {{{TDDFT}} from
  molecules to solids: {{The}} role of long-range interactions},}\ }\href
  {https://doi.org/10.1002/qua.20486} {\bibfield  {journal} {\bibinfo
  {journal} {Int. J. Quantum Chem.}\ }\textbf {\bibinfo {volume} {102}},\
  \bibinfo {pages} {684--701} (\bibinfo {year} {2005})}\BibitemShut {NoStop}%
\bibitem [{\citenamefont
  {Casida}(2009)}]{casidaTimedependentDensityfunctionalTheory2009}%
  \BibitemOpen
  \bibfield  {author} {\bibinfo {author} {\bibfnamefont {M.~E.}\ \bibnamefont
  {Casida}},\ }\bibfield  {title} {\enquote {\bibinfo {title} {Time-dependent
  density-functional theory for molecules and molecular solids},}\ }\href
  {https://doi.org/10.1016/j.theochem.2009.08.018} {\bibfield  {journal}
  {\bibinfo  {journal} {Journal of Molecular Structure: THEOCHEM}\ }\bibinfo
  {series} {Time-Dependent Density-Functional Theory for Molecules and
  Molecular Solids},\ \textbf {\bibinfo {volume} {914}},\ \bibinfo {pages}
  {3--18} (\bibinfo {year} {2009})}\BibitemShut {NoStop}%
\bibitem [{\citenamefont {Laurent}\ and\ \citenamefont
  {Jacquemin}(2013)}]{laurentTDDFTBenchmarksReview2013}%
  \BibitemOpen
  \bibfield  {author} {\bibinfo {author} {\bibfnamefont {A.~D.}\ \bibnamefont
  {Laurent}}\ and\ \bibinfo {author} {\bibfnamefont {D.}~\bibnamefont
  {Jacquemin}},\ }\bibfield  {title} {\enquote {\bibinfo {title} {{{TD-DFT}}
  benchmarks: {{A}} review},}\ }\href {https://doi.org/10.1002/qua.24438}
  {\bibfield  {journal} {\bibinfo  {journal} {Int. J. Quantum Chem.}\ }\textbf
  {\bibinfo {volume} {113}},\ \bibinfo {pages} {2019--2039} (\bibinfo {year}
  {2013})}\BibitemShut {NoStop}%
\bibitem [{\citenamefont {Br{\'e}mond}\ \emph {et~al.}(2018)\citenamefont
  {Br{\'e}mond}, \citenamefont {Savarese}, \citenamefont {Adamo},\ and\
  \citenamefont {Jacquemin}}]{bremondAccuracyTDDFTGeometries2018}%
  \BibitemOpen
  \bibfield  {author} {\bibinfo {author} {\bibfnamefont {E.}~\bibnamefont
  {Br{\'e}mond}}, \bibinfo {author} {\bibfnamefont {M.}~\bibnamefont
  {Savarese}}, \bibinfo {author} {\bibfnamefont {C.}~\bibnamefont {Adamo}},\
  and\ \bibinfo {author} {\bibfnamefont {D.}~\bibnamefont {Jacquemin}},\
  }\bibfield  {title} {\enquote {\bibinfo {title} {Accuracy of {{TD-DFT
  Geometries}}: {{A Fresh Look}}},}\ }\href
  {https://doi.org/10.1021/acs.jctc.8b00311} {\bibfield  {journal} {\bibinfo
  {journal} {J. Chem. Theory Comput.}\ }\textbf {\bibinfo {volume} {14}},\
  \bibinfo {pages} {3715--3727} (\bibinfo {year} {2018})}\BibitemShut {NoStop}%
\bibitem [{\citenamefont {Sarkar}\ \emph {et~al.}(2021)\citenamefont {Sarkar},
  \citenamefont {{Boggio-Pasqua}}, \citenamefont {Loos},\ and\ \citenamefont
  {Jacquemin}}]{sarkarBenchmarkingTDDFTWave2021}%
  \BibitemOpen
  \bibfield  {author} {\bibinfo {author} {\bibfnamefont {R.}~\bibnamefont
  {Sarkar}}, \bibinfo {author} {\bibfnamefont {M.}~\bibnamefont
  {{Boggio-Pasqua}}}, \bibinfo {author} {\bibfnamefont {P.-F.}\ \bibnamefont
  {Loos}},\ and\ \bibinfo {author} {\bibfnamefont {D.}~\bibnamefont
  {Jacquemin}},\ }\bibfield  {title} {\enquote {\bibinfo {title} {Benchmarking
  {{TD-DFT}} and {{Wave Function Methods}} for {{Oscillator Strengths}} and
  {{Excited-State Dipole Moments}}},}\ }\href
  {https://doi.org/10.1021/acs.jctc.0c01228} {\bibfield  {journal} {\bibinfo
  {journal} {J. Chem. Theory Comput.}\ }\textbf {\bibinfo {volume} {17}},\
  \bibinfo {pages} {1117--1132} (\bibinfo {year} {2021})}\BibitemShut {NoStop}%
\bibitem [{\citenamefont {Jin}\ \emph {et~al.}(2023)\citenamefont {Jin},
  \citenamefont {Yu}, \citenamefont {Govoni}, \citenamefont {Xu},\ and\
  \citenamefont {Galli}}]{jinExcitedStateProperties2023}%
  \BibitemOpen
  \bibfield  {author} {\bibinfo {author} {\bibfnamefont {Y.}~\bibnamefont
  {Jin}}, \bibinfo {author} {\bibfnamefont {V.~W.-z.}\ \bibnamefont {Yu}},
  \bibinfo {author} {\bibfnamefont {M.}~\bibnamefont {Govoni}}, \bibinfo
  {author} {\bibfnamefont {A.~C.}\ \bibnamefont {Xu}},\ and\ \bibinfo {author}
  {\bibfnamefont {G.}~\bibnamefont {Galli}},\ }\bibfield  {title} {\enquote
  {\bibinfo {title} {Excited {{State Properties}} of {{Point Defects}} in
  {{Semiconductors}} and {{Insulators Investigated}} with {{Time-Dependent
  Density Functional Theory}}},}\ }\href
  {https://doi.org/10.1021/acs.jctc.3c00986} {\bibfield  {journal} {\bibinfo
  {journal} {J. Chem. Theory Comput.}\ }\textbf {\bibinfo {volume} {19}},\
  \bibinfo {pages} {8689--8705} (\bibinfo {year} {2023})}\BibitemShut {NoStop}%
\bibitem [{\citenamefont {Dreuw}, \citenamefont {Weisman},\ and\ \citenamefont
  {{Head-Gordon}}(2003)}]{dreuwLongrangeChargetransferExcited2003}%
  \BibitemOpen
  \bibfield  {author} {\bibinfo {author} {\bibfnamefont {A.}~\bibnamefont
  {Dreuw}}, \bibinfo {author} {\bibfnamefont {J.~L.}\ \bibnamefont {Weisman}},\
  and\ \bibinfo {author} {\bibfnamefont {M.}~\bibnamefont {{Head-Gordon}}},\
  }\bibfield  {title} {\enquote {\bibinfo {title} {Long-range charge-transfer
  excited states in time-dependent density functional theory require non-local
  exchange},}\ }\href {https://doi.org/10.1063/1.1590951} {\bibfield  {journal}
  {\bibinfo  {journal} {J. Chem. Phys.}\ }\textbf {\bibinfo {volume} {119}},\
  \bibinfo {pages} {2943--2946} (\bibinfo {year} {2003})}\BibitemShut {NoStop}%
\bibitem [{\citenamefont
  {Tozer}(2003)}]{tozerRelationshipLongrangeChargetransfer2003}%
  \BibitemOpen
  \bibfield  {author} {\bibinfo {author} {\bibfnamefont {D.~J.}\ \bibnamefont
  {Tozer}},\ }\bibfield  {title} {\enquote {\bibinfo {title} {Relationship
  between long-range charge-transfer excitation energy error and integer
  discontinuity in {{Kohn}}--{{Sham}} theory},}\ }\href
  {https://doi.org/10.1063/1.1633756} {\bibfield  {journal} {\bibinfo
  {journal} {J. Chem. Phys.}\ }\textbf {\bibinfo {volume} {119}},\ \bibinfo
  {pages} {12697--12699} (\bibinfo {year} {2003})}\BibitemShut {NoStop}%
\bibitem [{\citenamefont {Hedin}(1965)}]{hedinNewMethodCalculating1965}%
  \BibitemOpen
  \bibfield  {author} {\bibinfo {author} {\bibfnamefont {L.}~\bibnamefont
  {Hedin}},\ }\bibfield  {title} {\enquote {\bibinfo {title} {New {{Method}}
  for {{Calculating}} the {{One-Particle Green}}'s {{Function}} with
  {{Application}} to the {{Electron-Gas Problem}}},}\ }\href
  {https://doi.org/10.1103/PhysRev.139.A796} {\bibfield  {journal} {\bibinfo
  {journal} {Phys. Rev.}\ }\textbf {\bibinfo {volume} {139}},\ \bibinfo {pages}
  {A796--A823} (\bibinfo {year} {1965})}\BibitemShut {NoStop}%
\bibitem [{\citenamefont {Martin}, \citenamefont {Reining},\ and\ \citenamefont
  {Ceperley}(2016)}]{martinInteractingElectrons2016}%
  \BibitemOpen
  \bibfield  {author} {\bibinfo {author} {\bibfnamefont {R.~M.}\ \bibnamefont
  {Martin}}, \bibinfo {author} {\bibfnamefont {L.}~\bibnamefont {Reining}},\
  and\ \bibinfo {author} {\bibfnamefont {D.~M.}\ \bibnamefont {Ceperley}},\
  }\href@noop {} {\emph {\bibinfo {title} {Interacting {{Electrons}}}}}\
  (\bibinfo  {publisher} {Cambridge University Press},\ \bibinfo {year}
  {2016})\BibitemShut {NoStop}%
\bibitem [{\citenamefont {Salpeter}\ and\ \citenamefont
  {Bethe}(1951)}]{salpeterRelativisticEquationBoundState1951}%
  \BibitemOpen
  \bibfield  {author} {\bibinfo {author} {\bibfnamefont {E.~E.}\ \bibnamefont
  {Salpeter}}\ and\ \bibinfo {author} {\bibfnamefont {H.~A.}\ \bibnamefont
  {Bethe}},\ }\bibfield  {title} {\enquote {\bibinfo {title} {A {{Relativistic
  Equation}} for {{Bound-State Problems}}},}\ }\href
  {https://doi.org/10.1103/PhysRev.84.1232} {\bibfield  {journal} {\bibinfo
  {journal} {Phys. Rev.}\ }\textbf {\bibinfo {volume} {84}},\ \bibinfo {pages}
  {1232--1242} (\bibinfo {year} {1951})}\BibitemShut {NoStop}%
\bibitem [{\citenamefont {Sham}\ and\ \citenamefont
  {Rice}(1966)}]{shamManyParticleDerivationEffectiveMass1966}%
  \BibitemOpen
  \bibfield  {author} {\bibinfo {author} {\bibfnamefont {L.~J.}\ \bibnamefont
  {Sham}}\ and\ \bibinfo {author} {\bibfnamefont {T.~M.}\ \bibnamefont
  {Rice}},\ }\bibfield  {title} {\enquote {\bibinfo {title} {Many-{{Particle
  Derivation}} of the {{Effective-Mass Equation}} for the {{Wannier
  Exciton}}},}\ }\href {https://doi.org/10.1103/PhysRev.144.708} {\bibfield
  {journal} {\bibinfo  {journal} {Phys. Rev.}\ }\textbf {\bibinfo {volume}
  {144}},\ \bibinfo {pages} {708--714} (\bibinfo {year} {1966})}\BibitemShut
  {NoStop}%
\bibitem [{\citenamefont {Hanke}\ and\ \citenamefont
  {Sham}(1979)}]{hankeManyParticleEffectsOptical1979}%
  \BibitemOpen
  \bibfield  {author} {\bibinfo {author} {\bibfnamefont {W.}~\bibnamefont
  {Hanke}}\ and\ \bibinfo {author} {\bibfnamefont {L.~J.}\ \bibnamefont
  {Sham}},\ }\bibfield  {title} {\enquote {\bibinfo {title} {Many-{{Particle
  Effects}} in the {{Optical Excitations}} of a {{Semiconductor}}},}\ }\href
  {https://doi.org/10.1103/PhysRevLett.43.387} {\bibfield  {journal} {\bibinfo
  {journal} {Phys. Rev. Lett.}\ }\textbf {\bibinfo {volume} {43}},\ \bibinfo
  {pages} {387--390} (\bibinfo {year} {1979})}\BibitemShut {NoStop}%
\bibitem [{\citenamefont
  {Strinati}(1982)}]{strinatiDynamicalShiftBroadening1982}%
  \BibitemOpen
  \bibfield  {author} {\bibinfo {author} {\bibfnamefont {G.}~\bibnamefont
  {Strinati}},\ }\bibfield  {title} {\enquote {\bibinfo {title} {Dynamical
  {{Shift}} and {{Broadening}} of {{Core Excitons}} in {{Semiconductors}}},}\
  }\href {https://doi.org/10.1103/PhysRevLett.49.1519} {\bibfield  {journal}
  {\bibinfo  {journal} {Phys. Rev. Lett.}\ }\textbf {\bibinfo {volume} {49}},\
  \bibinfo {pages} {1519--1522} (\bibinfo {year} {1982})}\BibitemShut {NoStop}%
\bibitem [{\citenamefont {Blase}\ and\ \citenamefont
  {Attaccalite}(2011)}]{blaseChargetransferExcitationsMolecular2011}%
  \BibitemOpen
  \bibfield  {author} {\bibinfo {author} {\bibfnamefont {X.}~\bibnamefont
  {Blase}}\ and\ \bibinfo {author} {\bibfnamefont {C.}~\bibnamefont
  {Attaccalite}},\ }\bibfield  {title} {\enquote {\bibinfo {title}
  {Charge-transfer excitations in molecular donor-acceptor complexes within the
  many-body {{Bethe-Salpeter}} approach},}\ }\href
  {https://doi.org/10.1063/1.3655352} {\bibfield  {journal} {\bibinfo
  {journal} {Appl. Phys. Lett.}\ }\textbf {\bibinfo {volume} {99}},\ \bibinfo
  {pages} {171909} (\bibinfo {year} {2011})}\BibitemShut {NoStop}%
\bibitem [{\citenamefont {Jacquemin}, \citenamefont {Duchemin},\ and\
  \citenamefont {Blase}(2016)}]{jacqueminAssessmentConvergencePartially2016}%
  \BibitemOpen
  \bibfield  {author} {\bibinfo {author} {\bibfnamefont {D.}~\bibnamefont
  {Jacquemin}}, \bibinfo {author} {\bibfnamefont {I.}~\bibnamefont
  {Duchemin}},\ and\ \bibinfo {author} {\bibfnamefont {X.}~\bibnamefont
  {Blase}},\ }\bibfield  {title} {\enquote {\bibinfo {title} {Assessment of the
  convergence of partially self-consistent {{BSE}}/{{GW}} calculations},}\
  }\href {https://doi.org/10.1080/00268976.2015.1119901} {\bibfield  {journal}
  {\bibinfo  {journal} {Mol. Phys.}\ }\textbf {\bibinfo {volume} {114}},\
  \bibinfo {pages} {957--967} (\bibinfo {year} {2016})}\BibitemShut {NoStop}%
\bibitem [{\citenamefont {Jacquemin}, \citenamefont {Duchemin},\ and\
  \citenamefont {Blase}(2017)}]{jacqueminBetheSalpeterFormalism2017}%
  \BibitemOpen
  \bibfield  {author} {\bibinfo {author} {\bibfnamefont {D.}~\bibnamefont
  {Jacquemin}}, \bibinfo {author} {\bibfnamefont {I.}~\bibnamefont
  {Duchemin}},\ and\ \bibinfo {author} {\bibfnamefont {X.}~\bibnamefont
  {Blase}},\ }\bibfield  {title} {\enquote {\bibinfo {title} {Is the
  {{Bethe}}--{{Salpeter Formalism Accurate}} for {{Excitation Energies}}?
  {{Comparisons}} with {{TD-DFT}}, {{CASPT2}}, and {{EOM-CCSD}}},}\ }\href
  {https://doi.org/10.1021/acs.jpclett.7b00381} {\bibfield  {journal} {\bibinfo
   {journal} {J. Phys. Chem. Lett.}\ }\textbf {\bibinfo {volume} {8}},\
  \bibinfo {pages} {1524--1529} (\bibinfo {year} {2017})}\BibitemShut {NoStop}%
\bibitem [{\citenamefont {Azarias}\ \emph {et~al.}(2017)\citenamefont
  {Azarias}, \citenamefont {Duchemin}, \citenamefont {Blase},\ and\
  \citenamefont {Jacquemin}}]{azariasBetheSalpeterStudyCationic2017}%
  \BibitemOpen
  \bibfield  {author} {\bibinfo {author} {\bibfnamefont {C.}~\bibnamefont
  {Azarias}}, \bibinfo {author} {\bibfnamefont {I.}~\bibnamefont {Duchemin}},
  \bibinfo {author} {\bibfnamefont {X.}~\bibnamefont {Blase}},\ and\ \bibinfo
  {author} {\bibfnamefont {D.}~\bibnamefont {Jacquemin}},\ }\bibfield  {title}
  {\enquote {\bibinfo {title} {Bethe-{{Salpeter}} study of cationic dyes:
  {{Comparisons}} with {{ADC}}(2) and {{TD-DFT}}},}\ }\href
  {https://doi.org/10.1063/1.4974097} {\bibfield  {journal} {\bibinfo
  {journal} {J. Chem. Phys.}\ }\textbf {\bibinfo {volume} {146}},\ \bibinfo
  {pages} {034301} (\bibinfo {year} {2017})}\BibitemShut {NoStop}%
\bibitem [{\citenamefont {Rangel}\ \emph {et~al.}(2017)\citenamefont {Rangel},
  \citenamefont {Hamed}, \citenamefont {Bruneval},\ and\ \citenamefont
  {Neaton}}]{rangelAssessmentLowlyingExcitation2017}%
  \BibitemOpen
  \bibfield  {author} {\bibinfo {author} {\bibfnamefont {T.}~\bibnamefont
  {Rangel}}, \bibinfo {author} {\bibfnamefont {S.~M.}\ \bibnamefont {Hamed}},
  \bibinfo {author} {\bibfnamefont {F.}~\bibnamefont {Bruneval}},\ and\
  \bibinfo {author} {\bibfnamefont {J.~B.}\ \bibnamefont {Neaton}},\ }\bibfield
   {title} {\enquote {\bibinfo {title} {An assessment of low-lying excitation
  energies and triplet instabilities of organic molecules with an ab initio
  {{Bethe-Salpeter}} equation approach and the {{Tamm-Dancoff}}
  approximation},}\ }\href {https://doi.org/10.1063/1.4983126} {\bibfield
  {journal} {\bibinfo  {journal} {J. Chem. Phys.}\ }\textbf {\bibinfo {volume}
  {146}},\ \bibinfo {pages} {194108} (\bibinfo {year} {2017})}\BibitemShut
  {NoStop}%
\bibitem [{\citenamefont {Monino}\ and\ \citenamefont
  {Loos}(2021)}]{moninoSpinConservedSpinFlipOptical2021}%
  \BibitemOpen
  \bibfield  {author} {\bibinfo {author} {\bibfnamefont {E.}~\bibnamefont
  {Monino}}\ and\ \bibinfo {author} {\bibfnamefont {P.-F.}\ \bibnamefont
  {Loos}},\ }\bibfield  {title} {\enquote {\bibinfo {title} {Spin-{{Conserved}}
  and {{Spin-Flip Optical Excitations}} from the {{Bethe}}--{{Salpeter Equation
  Formalism}}},}\ }\href {https://doi.org/10.1021/acs.jctc.1c00074} {\bibfield
  {journal} {\bibinfo  {journal} {J. Chem. Theory Comput.}\ }\textbf {\bibinfo
  {volume} {17}},\ \bibinfo {pages} {2852--2867} (\bibinfo {year}
  {2021})}\BibitemShut {NoStop}%
\bibitem [{\citenamefont {Cho}, \citenamefont {Bintrim},\ and\ \citenamefont
  {Berkelbach}(2022)}]{choSimplifiedGWBSE2022}%
  \BibitemOpen
  \bibfield  {author} {\bibinfo {author} {\bibfnamefont {Y.}~\bibnamefont
  {Cho}}, \bibinfo {author} {\bibfnamefont {S.~J.}\ \bibnamefont {Bintrim}},\
  and\ \bibinfo {author} {\bibfnamefont {T.~C.}\ \bibnamefont {Berkelbach}},\
  }\bibfield  {title} {\enquote {\bibinfo {title} {Simplified {{GW}}/{{BSE
  Approach}} for {{Charged}} and {{Neutral Excitation Energies}} of {{Large
  Molecules}} and {{Nanomaterials}}},}\ }\href
  {https://doi.org/10.1021/acs.jctc.2c00087} {\bibfield  {journal} {\bibinfo
  {journal} {J. Chem. Theory Comput.}\ }\textbf {\bibinfo {volume} {18}},\
  \bibinfo {pages} {3438--3446} (\bibinfo {year} {2022})}\BibitemShut {NoStop}%
\bibitem [{\citenamefont {McKeon}\ \emph {et~al.}(2022)\citenamefont {McKeon},
  \citenamefont {Hamed}, \citenamefont {Bruneval},\ and\ \citenamefont
  {Neaton}}]{mckeonOptimallyTunedRangeseparated2022}%
  \BibitemOpen
  \bibfield  {author} {\bibinfo {author} {\bibfnamefont {C.~A.}\ \bibnamefont
  {McKeon}}, \bibinfo {author} {\bibfnamefont {S.~M.}\ \bibnamefont {Hamed}},
  \bibinfo {author} {\bibfnamefont {F.}~\bibnamefont {Bruneval}},\ and\
  \bibinfo {author} {\bibfnamefont {J.~B.}\ \bibnamefont {Neaton}},\ }\bibfield
   {title} {\enquote {\bibinfo {title} {An optimally tuned range-separated
  hybrid starting point for ab initio {{GW}} plus {{Bethe}}--{{Salpeter}}
  equation calculations of molecules},}\ }\href
  {https://doi.org/10.1063/5.0097582} {\bibfield  {journal} {\bibinfo
  {journal} {J. Chem. Phys.}\ }\textbf {\bibinfo {volume} {157}},\ \bibinfo
  {pages} {074103} (\bibinfo {year} {2022})}\BibitemShut {NoStop}%
\bibitem [{\citenamefont {Li}\ \emph {et~al.}(2022{\natexlab{a}})\citenamefont
  {Li}, \citenamefont {Jin}, \citenamefont {Su},\ and\ \citenamefont
  {Yang}}]{liCombiningLocalizedOrbital2022}%
  \BibitemOpen
  \bibfield  {author} {\bibinfo {author} {\bibfnamefont {J.}~\bibnamefont
  {Li}}, \bibinfo {author} {\bibfnamefont {Y.}~\bibnamefont {Jin}}, \bibinfo
  {author} {\bibfnamefont {N.~Q.}\ \bibnamefont {Su}},\ and\ \bibinfo {author}
  {\bibfnamefont {W.}~\bibnamefont {Yang}},\ }\bibfield  {title} {\enquote
  {\bibinfo {title} {Combining localized orbital scaling correction and
  {{Bethe}}--{{Salpeter}} equation for accurate excitation energies},}\ }\href
  {https://doi.org/10.1063/5.0087498} {\bibfield  {journal} {\bibinfo
  {journal} {J. Chem. Phys.}\ }\textbf {\bibinfo {volume} {156}},\ \bibinfo
  {pages} {154101} (\bibinfo {year} {2022}{\natexlab{a}})}\BibitemShut
  {NoStop}%
\bibitem [{\citenamefont {Li}, \citenamefont {Golze},\ and\ \citenamefont
  {Yang}(2022)}]{liCombiningRenormalizedSingles2022}%
  \BibitemOpen
  \bibfield  {author} {\bibinfo {author} {\bibfnamefont {J.}~\bibnamefont
  {Li}}, \bibinfo {author} {\bibfnamefont {D.}~\bibnamefont {Golze}},\ and\
  \bibinfo {author} {\bibfnamefont {W.}~\bibnamefont {Yang}},\ }\bibfield
  {title} {\enquote {\bibinfo {title} {Combining {{Renormalized Singles GW
  Methods}} with the {{Bethe}}--{{Salpeter Equation}} for {{Accurate Neutral
  Excitation Energies}}},}\ }\href {https://doi.org/10.1021/acs.jctc.2c00686}
  {\bibfield  {journal} {\bibinfo  {journal} {J. Chem. Theory Comput.}\
  }\textbf {\bibinfo {volume} {18}},\ \bibinfo {pages} {6637--6645} (\bibinfo
  {year} {2022})}\BibitemShut {NoStop}%
\bibitem [{\citenamefont {Vorwerk}\ and\ \citenamefont
  {Galli}(2023)}]{vorwerkDisentanglingPhotoexcitationPhotoluminescence2023}%
  \BibitemOpen
  \bibfield  {author} {\bibinfo {author} {\bibfnamefont {C.}~\bibnamefont
  {Vorwerk}}\ and\ \bibinfo {author} {\bibfnamefont {G.}~\bibnamefont
  {Galli}},\ }\bibfield  {title} {\enquote {\bibinfo {title} {Disentangling
  photoexcitation and photoluminescence processes in defective {{MgO}}},}\
  }\href {https://doi.org/10.1103/PhysRevMaterials.7.033801} {\bibfield
  {journal} {\bibinfo  {journal} {Phys. Rev. Mater.}\ }\textbf {\bibinfo
  {volume} {7}},\ \bibinfo {pages} {033801} (\bibinfo {year}
  {2023})}\BibitemShut {NoStop}%
\bibitem [{\citenamefont {Zhang}, \citenamefont {Leveillee},\ and\
  \citenamefont {Schleife}(2023)}]{zhangEffectDynamicalScreening2023}%
  \BibitemOpen
  \bibfield  {author} {\bibinfo {author} {\bibfnamefont {X.}~\bibnamefont
  {Zhang}}, \bibinfo {author} {\bibfnamefont {J.~A.}\ \bibnamefont
  {Leveillee}},\ and\ \bibinfo {author} {\bibfnamefont {A.}~\bibnamefont
  {Schleife}},\ }\bibfield  {title} {\enquote {\bibinfo {title} {Effect of
  dynamical screening in the {{Bethe-Salpeter}} framework: {{Excitons}} in
  crystalline naphthalene},}\ }\href
  {https://doi.org/10.1103/PhysRevB.107.235205} {\bibfield  {journal} {\bibinfo
   {journal} {Phys. Rev. B}\ }\textbf {\bibinfo {volume} {107}},\ \bibinfo
  {pages} {235205} (\bibinfo {year} {2023})}\BibitemShut {NoStop}%
\bibitem [{\citenamefont {Haber}\ \emph {et~al.}(2023)\citenamefont {Haber},
  \citenamefont {Qiu}, \citenamefont {{da Jornada}},\ and\ \citenamefont
  {Neaton}}]{haberMaximallyLocalizedExciton2023}%
  \BibitemOpen
  \bibfield  {author} {\bibinfo {author} {\bibfnamefont {J.~B.}\ \bibnamefont
  {Haber}}, \bibinfo {author} {\bibfnamefont {D.~Y.}\ \bibnamefont {Qiu}},
  \bibinfo {author} {\bibfnamefont {F.~H.}\ \bibnamefont {{da Jornada}}},\ and\
  \bibinfo {author} {\bibfnamefont {J.~B.}\ \bibnamefont {Neaton}},\ }\bibfield
   {title} {\enquote {\bibinfo {title} {Maximally localized exciton {{Wannier}}
  functions for solids},}\ }\href {https://doi.org/10.1103/PhysRevB.108.125118}
  {\bibfield  {journal} {\bibinfo  {journal} {Phys. Rev. B}\ }\textbf {\bibinfo
  {volume} {108}},\ \bibinfo {pages} {125118} (\bibinfo {year}
  {2023})}\BibitemShut {NoStop}%
\bibitem [{\citenamefont {Rauwolf}, \citenamefont {Klopper},\ and\
  \citenamefont {Holzer}(2024)}]{rauwolfNonlinearLightMatter2024}%
  \BibitemOpen
  \bibfield  {author} {\bibinfo {author} {\bibfnamefont {N.}~\bibnamefont
  {Rauwolf}}, \bibinfo {author} {\bibfnamefont {W.}~\bibnamefont {Klopper}},\
  and\ \bibinfo {author} {\bibfnamefont {C.}~\bibnamefont {Holzer}},\
  }\bibfield  {title} {\enquote {\bibinfo {title} {Non-linear light--matter
  interactions from the {{Bethe}}--{{Salpeter}} equation},}\ }\href
  {https://doi.org/10.1063/5.0191499} {\bibfield  {journal} {\bibinfo
  {journal} {J. Chem. Phys.}\ }\textbf {\bibinfo {volume} {160}},\ \bibinfo
  {pages} {061101} (\bibinfo {year} {2024})}\BibitemShut {NoStop}%
\bibitem [{\citenamefont {Wu}\ \emph {et~al.}(2024)\citenamefont {Wu},
  \citenamefont {Hou}, \citenamefont {Li}, \citenamefont {He},\ and\
  \citenamefont {Qiu}}]{wuQuasiparticleExcitonicProperties2024}%
  \BibitemOpen
  \bibfield  {author} {\bibinfo {author} {\bibfnamefont {J.}~\bibnamefont
  {Wu}}, \bibinfo {author} {\bibfnamefont {B.}~\bibnamefont {Hou}}, \bibinfo
  {author} {\bibfnamefont {W.}~\bibnamefont {Li}}, \bibinfo {author}
  {\bibfnamefont {Y.}~\bibnamefont {He}},\ and\ \bibinfo {author}
  {\bibfnamefont {D.~Y.}\ \bibnamefont {Qiu}},\ }\bibfield  {title} {\enquote
  {\bibinfo {title} {Quasiparticle and excitonic properties of monolayer $1t'
  \text{WT}_{\text{e2}}$ within many-body perturbation theory},}\ }\href
  {https://doi.org/10.1103/PhysRevB.110.075133} {\bibfield  {journal} {\bibinfo
   {journal} {Phys. Rev. B}\ }\textbf {\bibinfo {volume} {110}},\ \bibinfo
  {pages} {075133} (\bibinfo {year} {2024})}\BibitemShut {NoStop}%
\bibitem [{\citenamefont {Zhou}\ \emph {et~al.}(2024)\citenamefont {Zhou},
  \citenamefont {Yao}, \citenamefont {Blum}, \citenamefont {Ren},\ and\
  \citenamefont {Kanai}}]{zhou2024all}%
  \BibitemOpen
  \bibfield  {author} {\bibinfo {author} {\bibfnamefont {R.}~\bibnamefont
  {Zhou}}, \bibinfo {author} {\bibfnamefont {Y.}~\bibnamefont {Yao}}, \bibinfo
  {author} {\bibfnamefont {V.}~\bibnamefont {Blum}}, \bibinfo {author}
  {\bibfnamefont {X.}~\bibnamefont {Ren}},\ and\ \bibinfo {author}
  {\bibfnamefont {Y.}~\bibnamefont {Kanai}},\ }\bibfield  {title} {\enquote
  {\bibinfo {title} {All-electron {BSE}@{$GW$} method with numeric
  atom-centered orbitals for extended systems},}\ }\href@noop {} {\bibfield
  {journal} {\bibinfo  {journal} {arXiv preprint arXiv:2406.11122}\ } (\bibinfo
  {year} {2024})}\BibitemShut {NoStop}%
\bibitem [{\citenamefont {Blase}, \citenamefont {Duchemin},\ and\ \citenamefont
  {Jacquemin}(2018)}]{blaseBetheSalpeterEquation2018}%
  \BibitemOpen
  \bibfield  {author} {\bibinfo {author} {\bibfnamefont {X.}~\bibnamefont
  {Blase}}, \bibinfo {author} {\bibfnamefont {I.}~\bibnamefont {Duchemin}},\
  and\ \bibinfo {author} {\bibfnamefont {D.}~\bibnamefont {Jacquemin}},\
  }\bibfield  {title} {\enquote {\bibinfo {title} {The {{Bethe}}--{{Salpeter}}
  equation in chemistry: Relations with {{TD-DFT}}, applications and
  challenges},}\ }\href {https://doi.org/10.1039/C7CS00049A} {\bibfield
  {journal} {\bibinfo  {journal} {Chem. Soc. Rev.}\ }\textbf {\bibinfo {volume}
  {47}},\ \bibinfo {pages} {1022--1043} (\bibinfo {year} {2018})}\BibitemShut
  {NoStop}%
\bibitem [{\citenamefont {Blase}\ \emph {et~al.}(2020)\citenamefont {Blase},
  \citenamefont {Duchemin}, \citenamefont {Jacquemin},\ and\ \citenamefont
  {Loos}}]{blaseBetheSalpeterEquation2020}%
  \BibitemOpen
  \bibfield  {author} {\bibinfo {author} {\bibfnamefont {X.}~\bibnamefont
  {Blase}}, \bibinfo {author} {\bibfnamefont {I.}~\bibnamefont {Duchemin}},
  \bibinfo {author} {\bibfnamefont {D.}~\bibnamefont {Jacquemin}},\ and\
  \bibinfo {author} {\bibfnamefont {P.-F.}\ \bibnamefont {Loos}},\ }\bibfield
  {title} {\enquote {\bibinfo {title} {The {{Bethe}}--{{Salpeter Equation
  Formalism}}: {{From Physics}} to {{Chemistry}}},}\ }\href
  {https://doi.org/10.1021/acs.jpclett.0c01875} {\bibfield  {journal} {\bibinfo
   {journal} {J. Phys. Chem. Lett.}\ }\textbf {\bibinfo {volume} {11}},\
  \bibinfo {pages} {7371--7382} (\bibinfo {year} {2020})}\BibitemShut {NoStop}%
\bibitem [{\citenamefont {Kohn}\ and\ \citenamefont
  {Sham}(1965)}]{kohnSelfConsistentEquationsIncluding1965}%
  \BibitemOpen
  \bibfield  {author} {\bibinfo {author} {\bibfnamefont {W.}~\bibnamefont
  {Kohn}}\ and\ \bibinfo {author} {\bibfnamefont {L.~J.}\ \bibnamefont
  {Sham}},\ }\bibfield  {title} {\enquote {\bibinfo {title} {Self-{{Consistent
  Equations Including Exchange}} and {{Correlation Effects}}},}\ }\href
  {https://doi.org/10.1103/PhysRev.140.A1133} {\bibfield  {journal} {\bibinfo
  {journal} {Phys. Rev.}\ }\textbf {\bibinfo {volume} {140}},\ \bibinfo {pages}
  {A1133--A1138} (\bibinfo {year} {1965})}\BibitemShut {NoStop}%
\bibitem [{\citenamefont {Parr}\ and\ \citenamefont
  {Weitao}(1989)}]{parrDensityFunctionalTheoryAtoms1989}%
  \BibitemOpen
  \bibfield  {author} {\bibinfo {author} {\bibfnamefont {R.~G.}\ \bibnamefont
  {Parr}}\ and\ \bibinfo {author} {\bibfnamefont {Y.}~\bibnamefont {Weitao}},\
  }\href@noop {} {\emph {\bibinfo {title} {Density-{{Functional Theory}} of
  {{Atoms}} and {{Molecules}}}}}\ (\bibinfo  {publisher} {Oxford University
  Press},\ \bibinfo {year} {1989})\BibitemShut {NoStop}%
\bibitem [{\citenamefont {Krause}\ and\ \citenamefont
  {Klopper}(2017)}]{krauseImplementationBetheSalpeterEquation2017}%
  \BibitemOpen
  \bibfield  {author} {\bibinfo {author} {\bibfnamefont {K.}~\bibnamefont
  {Krause}}\ and\ \bibinfo {author} {\bibfnamefont {W.}~\bibnamefont
  {Klopper}},\ }\bibfield  {title} {\enquote {\bibinfo {title} {Implementation
  of the {{Bethe}}-{{Salpeter}} equation in the {{TURBOMOLE}} program},}\
  }\href {https://doi.org/10.1002/jcc.24688} {\bibfield  {journal} {\bibinfo
  {journal} {J. Comput. Chem.}\ }\textbf {\bibinfo {volume} {38}},\ \bibinfo
  {pages} {383--388} (\bibinfo {year} {2017})}\BibitemShut {NoStop}%
\bibitem [{\citenamefont {{van Aggelen}}, \citenamefont {Yang},\ and\
  \citenamefont {Yang}(2013)}]{vanaggelenExchangecorrelationEnergyPairing2013}%
  \BibitemOpen
  \bibfield  {author} {\bibinfo {author} {\bibfnamefont {H.}~\bibnamefont {{van
  Aggelen}}}, \bibinfo {author} {\bibfnamefont {Y.}~\bibnamefont {Yang}},\ and\
  \bibinfo {author} {\bibfnamefont {W.}~\bibnamefont {Yang}},\ }\bibfield
  {title} {\enquote {\bibinfo {title} {Exchange-correlation energy from pairing
  matrix fluctuation and the particle-particle random-phase approximation},}\
  }\href {https://doi.org/10.1103/PhysRevA.88.030501} {\bibfield  {journal}
  {\bibinfo  {journal} {Phys. Rev. A}\ }\textbf {\bibinfo {volume} {88}},\
  \bibinfo {pages} {030501} (\bibinfo {year} {2013})}\BibitemShut {NoStop}%
\bibitem [{\citenamefont {Yang}, \citenamefont {{van Aggelen}},\ and\
  \citenamefont {Yang}(2013)}]{yangDoubleRydbergCharge2013}%
  \BibitemOpen
  \bibfield  {author} {\bibinfo {author} {\bibfnamefont {Y.}~\bibnamefont
  {Yang}}, \bibinfo {author} {\bibfnamefont {H.}~\bibnamefont {{van
  Aggelen}}},\ and\ \bibinfo {author} {\bibfnamefont {W.}~\bibnamefont
  {Yang}},\ }\bibfield  {title} {\enquote {\bibinfo {title} {Double,
  {{Rydberg}} and charge transfer excitations from pairing matrix fluctuation
  and particle-particle random phase approximation},}\ }\href
  {https://doi.org/10.1063/1.4834875} {\bibfield  {journal} {\bibinfo
  {journal} {J. Chem. Phys.}\ }\textbf {\bibinfo {volume} {139}},\ \bibinfo
  {pages} {224105} (\bibinfo {year} {2013})}\BibitemShut {NoStop}%
\bibitem [{\citenamefont {Yang}\ \emph {et~al.}(2014)\citenamefont {Yang},
  \citenamefont {Peng}, \citenamefont {Lu},\ and\ \citenamefont
  {Yang}}]{yangExcitationEnergiesParticleparticle2014}%
  \BibitemOpen
  \bibfield  {author} {\bibinfo {author} {\bibfnamefont {Y.}~\bibnamefont
  {Yang}}, \bibinfo {author} {\bibfnamefont {D.}~\bibnamefont {Peng}}, \bibinfo
  {author} {\bibfnamefont {J.}~\bibnamefont {Lu}},\ and\ \bibinfo {author}
  {\bibfnamefont {W.}~\bibnamefont {Yang}},\ }\bibfield  {title} {\enquote
  {\bibinfo {title} {Excitation energies from particle-particle random phase
  approximation: {{Davidson}} algorithm and benchmark studies},}\ }\href
  {https://doi.org/10.1063/1.4895792} {\bibfield  {journal} {\bibinfo
  {journal} {J. Chem. Phys.}\ }\textbf {\bibinfo {volume} {141}},\ \bibinfo
  {pages} {124104} (\bibinfo {year} {2014})}\BibitemShut {NoStop}%
\bibitem [{\citenamefont {Yang}\ \emph {et~al.}(2015)\citenamefont {Yang},
  \citenamefont {Peng}, \citenamefont {Davidson},\ and\ \citenamefont
  {Yang}}]{yangSingletTripletEnergy2015}%
  \BibitemOpen
  \bibfield  {author} {\bibinfo {author} {\bibfnamefont {Y.}~\bibnamefont
  {Yang}}, \bibinfo {author} {\bibfnamefont {D.}~\bibnamefont {Peng}}, \bibinfo
  {author} {\bibfnamefont {E.~R.}\ \bibnamefont {Davidson}},\ and\ \bibinfo
  {author} {\bibfnamefont {W.}~\bibnamefont {Yang}},\ }\bibfield  {title}
  {\enquote {\bibinfo {title} {Singlet--{{Triplet Energy Gaps}} for
  {{Diradicals}} from {{Particle}}--{{Particle Random Phase Approximation}}},}\
  }\href {https://doi.org/10.1021/jp512727a} {\bibfield  {journal} {\bibinfo
  {journal} {J. Phys. Chem. A}\ }\textbf {\bibinfo {volume} {119}},\ \bibinfo
  {pages} {4923--4932} (\bibinfo {year} {2015})}\BibitemShut {NoStop}%
\bibitem [{\citenamefont {Yang}\ \emph {et~al.}(2017)\citenamefont {Yang},
  \citenamefont {Dominguez}, \citenamefont {Zhang}, \citenamefont {Lutsker},
  \citenamefont {Niehaus}, \citenamefont {Frauenheim},\ and\ \citenamefont
  {Yang}}]{yangChargeTransferExcitations2017}%
  \BibitemOpen
  \bibfield  {author} {\bibinfo {author} {\bibfnamefont {Y.}~\bibnamefont
  {Yang}}, \bibinfo {author} {\bibfnamefont {A.}~\bibnamefont {Dominguez}},
  \bibinfo {author} {\bibfnamefont {D.}~\bibnamefont {Zhang}}, \bibinfo
  {author} {\bibfnamefont {V.}~\bibnamefont {Lutsker}}, \bibinfo {author}
  {\bibfnamefont {T.~A.}\ \bibnamefont {Niehaus}}, \bibinfo {author}
  {\bibfnamefont {T.}~\bibnamefont {Frauenheim}},\ and\ \bibinfo {author}
  {\bibfnamefont {W.}~\bibnamefont {Yang}},\ }\bibfield  {title} {\enquote
  {\bibinfo {title} {Charge transfer excitations from particle-particle random
  phase approximation---{{Opportunities}} and challenges arising from
  two-electron deficient systems},}\ }\href {https://doi.org/10.1063/1.4977928}
  {\bibfield  {journal} {\bibinfo  {journal} {J. Chem. Phys.}\ }\textbf
  {\bibinfo {volume} {146}},\ \bibinfo {pages} {124104} (\bibinfo {year}
  {2017})}\BibitemShut {NoStop}%
\bibitem [{\citenamefont {Li}\ \emph {et~al.}(2023)\citenamefont {Li},
  \citenamefont {Yu}, \citenamefont {Chen},\ and\ \citenamefont
  {Yang}}]{liLinearScalingCalculations2023}%
  \BibitemOpen
  \bibfield  {author} {\bibinfo {author} {\bibfnamefont {J.}~\bibnamefont
  {Li}}, \bibinfo {author} {\bibfnamefont {J.}~\bibnamefont {Yu}}, \bibinfo
  {author} {\bibfnamefont {Z.}~\bibnamefont {Chen}},\ and\ \bibinfo {author}
  {\bibfnamefont {W.}~\bibnamefont {Yang}},\ }\bibfield  {title} {\enquote
  {\bibinfo {title} {Linear {{Scaling Calculations}} of {{Excitation Energies}}
  with {{Active-Space Particle}}--{{Particle Random-Phase Approximation}}},}\
  }\href {https://doi.org/10.1021/acs.jpca.3c02834} {\bibfield  {journal}
  {\bibinfo  {journal} {J. Phys. Chem. A}\ }\textbf {\bibinfo {volume} {127}},\
  \bibinfo {pages} {7811--7822} (\bibinfo {year} {2023})}\BibitemShut {NoStop}%
\bibitem [{\citenamefont {Li}, \citenamefont {Chen},\ and\ \citenamefont
  {Yang}(2022)}]{liMultireferenceDensityFunctional2022}%
  \BibitemOpen
  \bibfield  {author} {\bibinfo {author} {\bibfnamefont {J.}~\bibnamefont
  {Li}}, \bibinfo {author} {\bibfnamefont {Z.}~\bibnamefont {Chen}},\ and\
  \bibinfo {author} {\bibfnamefont {W.}~\bibnamefont {Yang}},\ }\bibfield
  {title} {\enquote {\bibinfo {title} {Multireference {{Density Functional
  Theory}} for {{Describing Ground}} and {{Excited States}} with {{Renormalized
  Singles}}},}\ }\href {https://doi.org/10.1021/acs.jpclett.1c03913} {\bibfield
   {journal} {\bibinfo  {journal} {J. Phys. Chem. Lett.}\ }\textbf {\bibinfo
  {volume} {13}},\ \bibinfo {pages} {894--903} (\bibinfo {year}
  {2022})}\BibitemShut {NoStop}%
\bibitem [{\citenamefont {Li}\ \emph {et~al.}(2024{\natexlab{a}})\citenamefont
  {Li}, \citenamefont {Jin}, \citenamefont {Yu}, \citenamefont {Yang},\ and\
  \citenamefont {Zhu}}]{liAccurateExcitationEnergies2024}%
  \BibitemOpen
  \bibfield  {author} {\bibinfo {author} {\bibfnamefont {J.}~\bibnamefont
  {Li}}, \bibinfo {author} {\bibfnamefont {Y.}~\bibnamefont {Jin}}, \bibinfo
  {author} {\bibfnamefont {J.}~\bibnamefont {Yu}}, \bibinfo {author}
  {\bibfnamefont {W.}~\bibnamefont {Yang}},\ and\ \bibinfo {author}
  {\bibfnamefont {T.}~\bibnamefont {Zhu}},\ }\bibfield  {title} {\enquote
  {\bibinfo {title} {Accurate {{Excitation Energies}} of {{Point Defects}} from
  {{Fast Particle}}--{{Particle Random Phase Approximation Calculations}}},}\
  }\href {https://doi.org/10.1021/acs.jpclett.4c00184} {\bibfield  {journal}
  {\bibinfo  {journal} {J. Phys. Chem. Lett.}\ }\textbf {\bibinfo {volume}
  {15}},\ \bibinfo {pages} {2757--2764} (\bibinfo {year}
  {2024}{\natexlab{a}})}\BibitemShut {NoStop}%
\bibitem [{\citenamefont {Li}\ \emph {et~al.}(2024{\natexlab{b}})\citenamefont
  {Li}, \citenamefont {Jin}, \citenamefont {Yu}, \citenamefont {Yang},\ and\
  \citenamefont {Zhu}}]{liParticleParticleRandom2024}%
  \BibitemOpen
  \bibfield  {author} {\bibinfo {author} {\bibfnamefont {J.}~\bibnamefont
  {Li}}, \bibinfo {author} {\bibfnamefont {Y.}~\bibnamefont {Jin}}, \bibinfo
  {author} {\bibfnamefont {J.}~\bibnamefont {Yu}}, \bibinfo {author}
  {\bibfnamefont {W.}~\bibnamefont {Yang}},\ and\ \bibinfo {author}
  {\bibfnamefont {T.}~\bibnamefont {Zhu}},\ }\bibfield  {title} {\enquote
  {\bibinfo {title} {Particle--{{Particle Random Phase Approximation}} for
  {{Predicting Correlated Excited States}} of {{Point Defects}}},}\ }\href
  {https://doi.org/10.1021/acs.jctc.4c00829} {\bibfield  {journal} {\bibinfo
  {journal} {J. Chem. Theory Comput.}\ }\textbf {\bibinfo {volume} {20}},\
  \bibinfo {pages} {7979--7989} (\bibinfo {year}
  {2024}{\natexlab{b}})}\BibitemShut {NoStop}%
\bibitem [{\citenamefont {Oosterbaan}, \citenamefont {White},\ and\
  \citenamefont
  {{Head-Gordon}}(2019)}]{oosterbaanNonOrthogonalConfigurationInteraction2019}%
  \BibitemOpen
  \bibfield  {author} {\bibinfo {author} {\bibfnamefont {K.~J.}\ \bibnamefont
  {Oosterbaan}}, \bibinfo {author} {\bibfnamefont {A.~F.}\ \bibnamefont
  {White}},\ and\ \bibinfo {author} {\bibfnamefont {M.}~\bibnamefont
  {{Head-Gordon}}},\ }\bibfield  {title} {\enquote {\bibinfo {title}
  {Non-{{Orthogonal Configuration Interaction}} with {{Single Substitutions}}
  for {{Core-Excited States}}: {{An Extension}} to {{Doublet Radicals}}},}\
  }\href {https://doi.org/10.1021/acs.jctc.8b01259} {\bibfield  {journal}
  {\bibinfo  {journal} {J. Chem. Theory Comput.}\ }\textbf {\bibinfo {volume}
  {15}},\ \bibinfo {pages} {2966--2973} (\bibinfo {year} {2019})}\BibitemShut
  {NoStop}%
\bibitem [{\citenamefont {{Carter-Fenk}}\ and\ \citenamefont
  {{Head-Gordon}}(2022)}]{carter-fenkChoiceReferenceOrbitals2022}%
  \BibitemOpen
  \bibfield  {author} {\bibinfo {author} {\bibfnamefont {K.}~\bibnamefont
  {{Carter-Fenk}}}\ and\ \bibinfo {author} {\bibfnamefont {M.}~\bibnamefont
  {{Head-Gordon}}},\ }\bibfield  {title} {\enquote {\bibinfo {title} {On the
  choice of reference orbitals for linear-response calculations of
  solution-phase {{K-edge X-ray}} absorption spectra},}\ }\href
  {https://doi.org/10.1039/D2CP04077H} {\bibfield  {journal} {\bibinfo
  {journal} {Phys. Chem. Chem. Phys.}\ }\textbf {\bibinfo {volume} {24}},\
  \bibinfo {pages} {26170--26179} (\bibinfo {year} {2022})}\BibitemShut
  {NoStop}%
\bibitem [{\citenamefont {Loos}\ \emph {et~al.}(2019)\citenamefont {Loos},
  \citenamefont {{Boggio-Pasqua}}, \citenamefont {Scemama}, \citenamefont
  {Caffarel},\ and\ \citenamefont
  {Jacquemin}}]{loosReferenceEnergiesDouble2019}%
  \BibitemOpen
  \bibfield  {author} {\bibinfo {author} {\bibfnamefont {P.-F.}\ \bibnamefont
  {Loos}}, \bibinfo {author} {\bibfnamefont {M.}~\bibnamefont
  {{Boggio-Pasqua}}}, \bibinfo {author} {\bibfnamefont {A.}~\bibnamefont
  {Scemama}}, \bibinfo {author} {\bibfnamefont {M.}~\bibnamefont {Caffarel}},\
  and\ \bibinfo {author} {\bibfnamefont {D.}~\bibnamefont {Jacquemin}},\
  }\bibfield  {title} {\enquote {\bibinfo {title} {Reference {{Energies}} for
  {{Double Excitations}}},}\ }\href {https://doi.org/10.1021/acs.jctc.8b01205}
  {\bibfield  {journal} {\bibinfo  {journal} {J. Chem. Theory Comput.}\
  }\textbf {\bibinfo {volume} {15}},\ \bibinfo {pages} {1939--1956} (\bibinfo
  {year} {2019})}\BibitemShut {NoStop}%
\bibitem [{\citenamefont {Loos}\ \emph
  {et~al.}(2021{\natexlab{a}})\citenamefont {Loos}, \citenamefont {Comin},
  \citenamefont {Blase},\ and\ \citenamefont
  {Jacquemin}}]{loosReferenceEnergiesIntramolecular2021}%
  \BibitemOpen
  \bibfield  {author} {\bibinfo {author} {\bibfnamefont {P.-F.}\ \bibnamefont
  {Loos}}, \bibinfo {author} {\bibfnamefont {M.}~\bibnamefont {Comin}},
  \bibinfo {author} {\bibfnamefont {X.}~\bibnamefont {Blase}},\ and\ \bibinfo
  {author} {\bibfnamefont {D.}~\bibnamefont {Jacquemin}},\ }\bibfield  {title}
  {\enquote {\bibinfo {title} {Reference {{Energies}} for {{Intramolecular
  Charge-Transfer Excitations}}},}\ }\href
  {https://doi.org/10.1021/acs.jctc.1c00226} {\bibfield  {journal} {\bibinfo
  {journal} {J. Chem. Theory Comput.}\ }\textbf {\bibinfo {volume} {17}},\
  \bibinfo {pages} {3666--3686} (\bibinfo {year}
  {2021}{\natexlab{a}})}\BibitemShut {NoStop}%
\bibitem [{\citenamefont {Loos}\ \emph
  {et~al.}(2021{\natexlab{b}})\citenamefont {Loos}, \citenamefont {Matthews},
  \citenamefont {Lipparini},\ and\ \citenamefont
  {Jacquemin}}]{loosHowAccurateAre2021}%
  \BibitemOpen
  \bibfield  {author} {\bibinfo {author} {\bibfnamefont {P.-F.}\ \bibnamefont
  {Loos}}, \bibinfo {author} {\bibfnamefont {D.~A.}\ \bibnamefont {Matthews}},
  \bibinfo {author} {\bibfnamefont {F.}~\bibnamefont {Lipparini}},\ and\
  \bibinfo {author} {\bibfnamefont {D.}~\bibnamefont {Jacquemin}},\ }\bibfield
  {title} {\enquote {\bibinfo {title} {How accurate are {{EOM-CC4}} vertical
  excitation energies?}}\ }\href {https://doi.org/10.1063/5.0055994} {\bibfield
   {journal} {\bibinfo  {journal} {J. Chem. Phys.}\ }\textbf {\bibinfo {volume}
  {154}},\ \bibinfo {pages} {221103} (\bibinfo {year}
  {2021}{\natexlab{b}})}\BibitemShut {NoStop}%
\bibitem [{\citenamefont {Loos}\ and\ \citenamefont
  {Jacquemin}(2024)}]{loosMountaineeringStrategyExcited2024}%
  \BibitemOpen
  \bibfield  {author} {\bibinfo {author} {\bibfnamefont {P.-F.}\ \bibnamefont
  {Loos}}\ and\ \bibinfo {author} {\bibfnamefont {D.}~\bibnamefont
  {Jacquemin}},\ }\bibfield  {title} {\enquote {\bibinfo {title} {A
  mountaineering strategy to excited states: {{Accurate}} vertical transition
  energies and benchmarks for substituted benzenes},}\ }\href
  {https://doi.org/10.1002/jcc.27358} {\bibfield  {journal} {\bibinfo
  {journal} {J. Comput. Chem.}\ }\textbf {\bibinfo {volume} {45}},\ \bibinfo
  {pages} {1791--1805} (\bibinfo {year} {2024})}\BibitemShut {NoStop}%
\bibitem [{\citenamefont {Michalak}\ and\ \citenamefont
  {Lesiuk}(2024)}]{michalakRankReducedEquationMotionCoupled2024}%
  \BibitemOpen
  \bibfield  {author} {\bibinfo {author} {\bibfnamefont {P.}~\bibnamefont
  {Michalak}}\ and\ \bibinfo {author} {\bibfnamefont {M.}~\bibnamefont
  {Lesiuk}},\ }\bibfield  {title} {\enquote {\bibinfo {title} {Rank-{{Reduced
  Equation-of-Motion Coupled Cluster Triples}}: An {{Accurate}} and
  {{Affordable Way}} of {{Calculating Electronic Excitation Energies}}},}\
  }\href {https://doi.org/10.1021/acs.jctc.4c00959} {\bibfield  {journal}
  {\bibinfo  {journal} {J. Chem. Theory Comput.}\ }\textbf {\bibinfo {volume}
  {20}},\ \bibinfo {pages} {8970--8983} (\bibinfo {year} {2024})}\BibitemShut
  {NoStop}%
\bibitem [{\citenamefont {Winter}\ \emph {et~al.}(2013)\citenamefont {Winter},
  \citenamefont {Graf}, \citenamefont {Leutwyler},\ and\ \citenamefont
  {H{\"a}ttig}}]{winterBenchmarks00Transitions2013}%
  \BibitemOpen
  \bibfield  {author} {\bibinfo {author} {\bibfnamefont {N.~O.~C.}\
  \bibnamefont {Winter}}, \bibinfo {author} {\bibfnamefont {N.~K.}\
  \bibnamefont {Graf}}, \bibinfo {author} {\bibfnamefont {S.}~\bibnamefont
  {Leutwyler}},\ and\ \bibinfo {author} {\bibfnamefont {C.}~\bibnamefont
  {H{\"a}ttig}},\ }\bibfield  {title} {\enquote {\bibinfo {title} {Benchmarks
  for 0--0 transitions of aromatic organic molecules: {{DFT}}/{{B3LYP}},
  {{ADC}}(2), {{CC2}}, {{SOS-CC2}} and {{SCS-CC2}} compared to high-resolution
  gas-phase data},}\ }\href {https://doi.org/10.1039/C2CP42694C} {\bibfield
  {journal} {\bibinfo  {journal} {Phys. Chem. Chem. Phys.}\ }\textbf {\bibinfo
  {volume} {15}},\ \bibinfo {pages} {6623--6630} (\bibinfo {year}
  {2013})}\BibitemShut {NoStop}%
\bibitem [{\citenamefont {Loos}\ \emph {et~al.}(2018)\citenamefont {Loos},
  \citenamefont {Scemama}, \citenamefont {Blondel}, \citenamefont {Garniron},
  \citenamefont {Caffarel},\ and\ \citenamefont
  {Jacquemin}}]{loosMountaineeringStrategyExcited2018}%
  \BibitemOpen
  \bibfield  {author} {\bibinfo {author} {\bibfnamefont {P.-F.}\ \bibnamefont
  {Loos}}, \bibinfo {author} {\bibfnamefont {A.}~\bibnamefont {Scemama}},
  \bibinfo {author} {\bibfnamefont {A.}~\bibnamefont {Blondel}}, \bibinfo
  {author} {\bibfnamefont {Y.}~\bibnamefont {Garniron}}, \bibinfo {author}
  {\bibfnamefont {M.}~\bibnamefont {Caffarel}},\ and\ \bibinfo {author}
  {\bibfnamefont {D.}~\bibnamefont {Jacquemin}},\ }\bibfield  {title} {\enquote
  {\bibinfo {title} {A {{Mountaineering Strategy}} to {{Excited States}}:
  {{Highly Accurate Reference Energies}} and {{Benchmarks}}},}\ }\href
  {https://doi.org/10.1021/acs.jctc.8b00406} {\bibfield  {journal} {\bibinfo
  {journal} {J. Chem. Theory Comput.}\ }\textbf {\bibinfo {volume} {14}},\
  \bibinfo {pages} {4360--4379} (\bibinfo {year} {2018})}\BibitemShut {NoStop}%
\bibitem [{\citenamefont {Mazin}\ and\ \citenamefont
  {Sokolov}(2023)}]{mazinCoreExcitedStatesXray2023}%
  \BibitemOpen
  \bibfield  {author} {\bibinfo {author} {\bibfnamefont {I.~M.}\ \bibnamefont
  {Mazin}}\ and\ \bibinfo {author} {\bibfnamefont {A.~Y.}\ \bibnamefont
  {Sokolov}},\ }\bibfield  {title} {\enquote {\bibinfo {title} {Core-{{Excited
  States}} and {{X-ray Absorption Spectra}} from {{Multireference Algebraic
  Diagrammatic Construction Theory}}},}\ }\href
  {https://doi.org/10.1021/acs.jctc.3c00477} {\bibfield  {journal} {\bibinfo
  {journal} {J. Chem. Theory Comput.}\ }\textbf {\bibinfo {volume} {19}},\
  \bibinfo {pages} {4991--5006} (\bibinfo {year} {2023})}\BibitemShut {NoStop}%
\bibitem [{\citenamefont {Maier}, \citenamefont {Bauer},\ and\ \citenamefont
  {Dreuw}(2023)}]{maierConsistentThirdorderOneparticle2023}%
  \BibitemOpen
  \bibfield  {author} {\bibinfo {author} {\bibfnamefont {R.}~\bibnamefont
  {Maier}}, \bibinfo {author} {\bibfnamefont {M.}~\bibnamefont {Bauer}},\ and\
  \bibinfo {author} {\bibfnamefont {A.}~\bibnamefont {Dreuw}},\ }\bibfield
  {title} {\enquote {\bibinfo {title} {Consistent third-order one-particle
  transition and excited-state properties within the algebraic-diagrammatic
  construction scheme for the polarization propagator},}\ }\href
  {https://doi.org/10.1063/5.0151765} {\bibfield  {journal} {\bibinfo
  {journal} {J. Chem. Phys.}\ }\textbf {\bibinfo {volume} {159}},\ \bibinfo
  {pages} {014104} (\bibinfo {year} {2023})}\BibitemShut {NoStop}%
\bibitem [{\citenamefont {S{\"u}lzner}\ and\ \citenamefont
  {H{\"a}ttig}(2024)}]{sulznerRoleSinglesAmplitudes2024}%
  \BibitemOpen
  \bibfield  {author} {\bibinfo {author} {\bibfnamefont {N.}~\bibnamefont
  {S{\"u}lzner}}\ and\ \bibinfo {author} {\bibfnamefont {C.}~\bibnamefont
  {H{\"a}ttig}},\ }\bibfield  {title} {\enquote {\bibinfo {title} {Role of
  {{Singles Amplitudes}} in {{ADC}}(2) and {{CC2}} for {{Low-Lying
  Electronically Excited States}}},}\ }\href
  {https://doi.org/10.1021/acs.jctc.3c01355} {\bibfield  {journal} {\bibinfo
  {journal} {J. Chem. Theory Comput.}\ }\textbf {\bibinfo {volume} {20}},\
  \bibinfo {pages} {2462--2474} (\bibinfo {year} {2024})}\BibitemShut {NoStop}%
\bibitem [{\citenamefont {Hait}\ and\ \citenamefont
  {{Head-Gordon}}(2021)}]{haitOrbitalOptimizedDensity2021}%
  \BibitemOpen
  \bibfield  {author} {\bibinfo {author} {\bibfnamefont {D.}~\bibnamefont
  {Hait}}\ and\ \bibinfo {author} {\bibfnamefont {M.}~\bibnamefont
  {{Head-Gordon}}},\ }\bibfield  {title} {\enquote {\bibinfo {title} {Orbital
  {{Optimized Density Functional Theory}} for {{Electronic Excited States}}},}\
  }\href {https://doi.org/10.1021/acs.jpclett.1c00744} {\bibfield  {journal}
  {\bibinfo  {journal} {J. Phys. Chem. Lett.}\ }\textbf {\bibinfo {volume}
  {12}},\ \bibinfo {pages} {4517--4529} (\bibinfo {year} {2021})}\BibitemShut
  {NoStop}%
\bibitem [{\citenamefont {Hait}\ \emph {et~al.}(2022)\citenamefont {Hait},
  \citenamefont {Oosterbaan}, \citenamefont {{Carter-Fenk}},\ and\
  \citenamefont {{Head-Gordon}}}]{haitComputingXrayAbsorption2022}%
  \BibitemOpen
  \bibfield  {author} {\bibinfo {author} {\bibfnamefont {D.}~\bibnamefont
  {Hait}}, \bibinfo {author} {\bibfnamefont {K.~J.}\ \bibnamefont
  {Oosterbaan}}, \bibinfo {author} {\bibfnamefont {K.}~\bibnamefont
  {{Carter-Fenk}}},\ and\ \bibinfo {author} {\bibfnamefont {M.}~\bibnamefont
  {{Head-Gordon}}},\ }\bibfield  {title} {\enquote {\bibinfo {title} {Computing
  x-ray absorption spectra from linear-response particles atop optimized
  holes},}\ }\href {https://doi.org/10.1063/5.0092987} {\bibfield  {journal}
  {\bibinfo  {journal} {J. Chem. Phys.}\ }\textbf {\bibinfo {volume} {156}},\
  \bibinfo {pages} {201104} (\bibinfo {year} {2022})}\BibitemShut {NoStop}%
\bibitem [{\citenamefont {Yao}\ \emph {et~al.}(2022)\citenamefont {Yao},
  \citenamefont {Golze}, \citenamefont {Rinke}, \citenamefont {Blum},\ and\
  \citenamefont {Kanai}}]{yaoAllElectronBSEGWMethod2022}%
  \BibitemOpen
  \bibfield  {author} {\bibinfo {author} {\bibfnamefont {Y.}~\bibnamefont
  {Yao}}, \bibinfo {author} {\bibfnamefont {D.}~\bibnamefont {Golze}}, \bibinfo
  {author} {\bibfnamefont {P.}~\bibnamefont {Rinke}}, \bibinfo {author}
  {\bibfnamefont {V.}~\bibnamefont {Blum}},\ and\ \bibinfo {author}
  {\bibfnamefont {Y.}~\bibnamefont {Kanai}},\ }\bibfield  {title} {\enquote
  {\bibinfo {title} {All-{{Electron BSE}}@{{GW Method}} for {{K-Edge Core
  Electron Excitation Energies}}},}\ }\href
  {https://doi.org/10.1021/acs.jctc.1c01180} {\bibfield  {journal} {\bibinfo
  {journal} {J. Chem. Theory Comput.}\ } (\bibinfo {year} {2022}),\
  10.1021/acs.jctc.1c01180}\BibitemShut {NoStop}%
\bibitem [{\citenamefont {Kick}\ \emph {et~al.}(2024)\citenamefont {Kick},
  \citenamefont {Alexander}, \citenamefont {Beiersdorfer},\ and\ \citenamefont
  {Van~Voorhis}}]{kickSuperresolutionTechniquesSimulate2024}%
  \BibitemOpen
  \bibfield  {author} {\bibinfo {author} {\bibfnamefont {M.}~\bibnamefont
  {Kick}}, \bibinfo {author} {\bibfnamefont {E.}~\bibnamefont {Alexander}},
  \bibinfo {author} {\bibfnamefont {A.}~\bibnamefont {Beiersdorfer}},\ and\
  \bibinfo {author} {\bibfnamefont {T.}~\bibnamefont {Van~Voorhis}},\
  }\bibfield  {title} {\enquote {\bibinfo {title} {Super-resolution techniques
  to simulate electronic spectra of large molecular systems},}\ }\href
  {https://doi.org/10.1038/s41467-024-52368-5} {\bibfield  {journal} {\bibinfo
  {journal} {Nat Commun}\ }\textbf {\bibinfo {volume} {15}},\ \bibinfo {pages}
  {8001} (\bibinfo {year} {2024})}\BibitemShut {NoStop}%
\bibitem [{\citenamefont {Shao}\ \emph {et~al.}(2018)\citenamefont {Shao},
  \citenamefont {{da Jornada}}, \citenamefont {Lin}, \citenamefont {Yang},
  \citenamefont {Deslippe},\ and\ \citenamefont
  {Louie}}]{shaoStructurePreservingLanczos2018}%
  \BibitemOpen
  \bibfield  {author} {\bibinfo {author} {\bibfnamefont {M.}~\bibnamefont
  {Shao}}, \bibinfo {author} {\bibfnamefont {F.~H.}\ \bibnamefont {{da
  Jornada}}}, \bibinfo {author} {\bibfnamefont {L.}~\bibnamefont {Lin}},
  \bibinfo {author} {\bibfnamefont {C.}~\bibnamefont {Yang}}, \bibinfo {author}
  {\bibfnamefont {J.}~\bibnamefont {Deslippe}},\ and\ \bibinfo {author}
  {\bibfnamefont {S.~G.}\ \bibnamefont {Louie}},\ }\bibfield  {title} {\enquote
  {\bibinfo {title} {A {{Structure Preserving Lanczos Algorithm}} for
  {{Computing}} the {{Optical Absorption Spectrum}}},}\ }\href
  {https://doi.org/10.1137/16M1102641} {\bibfield  {journal} {\bibinfo
  {journal} {SIAM J. Matrix Anal. Appl.}\ }\textbf {\bibinfo {volume} {39}},\
  \bibinfo {pages} {683--711} (\bibinfo {year} {2018})}\BibitemShut {NoStop}%
\bibitem [{\citenamefont {Wall}\ and\ \citenamefont
  {Neuhauser}(1995)}]{wallExtractionFilterdiagonalizationGeneral1995}%
  \BibitemOpen
  \bibfield  {author} {\bibinfo {author} {\bibfnamefont {M.~R.}\ \bibnamefont
  {Wall}}\ and\ \bibinfo {author} {\bibfnamefont {D.}~\bibnamefont
  {Neuhauser}},\ }\bibfield  {title} {\enquote {\bibinfo {title} {Extraction,
  through filter-diagonalization, of general quantum eigenvalues or classical
  normal mode frequencies\,from\,a\,small\,number\,of\,residues or a short-time
  segment of a signal. {{I}}. {{Theory}} and application to a quantum-dynamics
  model},}\ }\href {https://doi.org/10.1063/1.468999} {\bibfield  {journal}
  {\bibinfo  {journal} {J. Chem. Phys.}\ }\textbf {\bibinfo {volume} {102}},\
  \bibinfo {pages} {8011--8022} (\bibinfo {year} {1995})}\BibitemShut {NoStop}%
\bibitem [{\citenamefont {Bradbury}\ \emph {et~al.}(2022)\citenamefont
  {Bradbury}, \citenamefont {Nguyen}, \citenamefont {Caram},\ and\
  \citenamefont {Neuhauser}}]{bradburyBetheSalpeterEquation2022}%
  \BibitemOpen
  \bibfield  {author} {\bibinfo {author} {\bibfnamefont {N.~C.}\ \bibnamefont
  {Bradbury}}, \bibinfo {author} {\bibfnamefont {M.}~\bibnamefont {Nguyen}},
  \bibinfo {author} {\bibfnamefont {J.~R.}\ \bibnamefont {Caram}},\ and\
  \bibinfo {author} {\bibfnamefont {D.}~\bibnamefont {Neuhauser}},\ }\bibfield
  {title} {\enquote {\bibinfo {title} {Bethe--{{Salpeter}} equation spectra for
  very large systems},}\ }\href {https://doi.org/10.1063/5.0100213} {\bibfield
  {journal} {\bibinfo  {journal} {J. Chem. Phys.}\ }\textbf {\bibinfo {volume}
  {157}} (\bibinfo {year} {2022}),\ 10.1063/5.0100213}\BibitemShut {NoStop}%
\bibitem [{\citenamefont {Bradbury}\ \emph {et~al.}(2023)\citenamefont
  {Bradbury}, \citenamefont {Allen}, \citenamefont {Nguyen}, \citenamefont
  {Ibrahim},\ and\ \citenamefont
  {Neuhauser}}]{bradburyOptimizedAttenuatedInteraction2023}%
  \BibitemOpen
  \bibfield  {author} {\bibinfo {author} {\bibfnamefont {N.~C.}\ \bibnamefont
  {Bradbury}}, \bibinfo {author} {\bibfnamefont {T.}~\bibnamefont {Allen}},
  \bibinfo {author} {\bibfnamefont {M.}~\bibnamefont {Nguyen}}, \bibinfo
  {author} {\bibfnamefont {K.~Z.}\ \bibnamefont {Ibrahim}},\ and\ \bibinfo
  {author} {\bibfnamefont {D.}~\bibnamefont {Neuhauser}},\ }\bibfield  {title}
  {\enquote {\bibinfo {title} {Optimized attenuated interaction: {{Enabling}}
  stochastic {{Bethe}}--{{Salpeter}} spectra for large systems},}\ }\href
  {https://doi.org/10.1063/5.0146555} {\bibfield  {journal} {\bibinfo
  {journal} {J. Chem. Phys.}\ }\textbf {\bibinfo {volume} {158}} (\bibinfo
  {year} {2023}),\ 10.1063/5.0146555}\BibitemShut {NoStop}%
\bibitem [{\citenamefont {Liang}\ \emph {et~al.}(2011)\citenamefont {Liang},
  \citenamefont {Fischer}, \citenamefont {Frisch},\ and\ \citenamefont
  {Li}}]{liangEnergySpecificLinearResponse2011}%
  \BibitemOpen
  \bibfield  {author} {\bibinfo {author} {\bibfnamefont {W.}~\bibnamefont
  {Liang}}, \bibinfo {author} {\bibfnamefont {S.~A.}\ \bibnamefont {Fischer}},
  \bibinfo {author} {\bibfnamefont {M.~J.}\ \bibnamefont {Frisch}},\ and\
  \bibinfo {author} {\bibfnamefont {X.}~\bibnamefont {Li}},\ }\bibfield
  {title} {\enquote {\bibinfo {title} {Energy-{{Specific Linear Response
  TDHF}}/{{TDDFT}} for {{Calculating High-Energy Excited States}}},}\ }\href
  {https://doi.org/10.1021/ct200485x} {\bibfield  {journal} {\bibinfo
  {journal} {J. Chem. Theory Comput.}\ }\textbf {\bibinfo {volume} {7}},\
  \bibinfo {pages} {3540--3547} (\bibinfo {year} {2011})}\BibitemShut {NoStop}%
\bibitem [{\citenamefont {Peng}\ \emph {et~al.}(2015)\citenamefont {Peng},
  \citenamefont {Lestrange}, \citenamefont {Goings}, \citenamefont {Caricato},\
  and\ \citenamefont
  {Li}}]{pengEnergySpecificEquationMotionCoupledCluster2015}%
  \BibitemOpen
  \bibfield  {author} {\bibinfo {author} {\bibfnamefont {B.}~\bibnamefont
  {Peng}}, \bibinfo {author} {\bibfnamefont {P.~J.}\ \bibnamefont {Lestrange}},
  \bibinfo {author} {\bibfnamefont {J.~J.}\ \bibnamefont {Goings}}, \bibinfo
  {author} {\bibfnamefont {M.}~\bibnamefont {Caricato}},\ and\ \bibinfo
  {author} {\bibfnamefont {X.}~\bibnamefont {Li}},\ }\bibfield  {title}
  {\enquote {\bibinfo {title} {Energy-{{Specific Equation-of-Motion
  Coupled-Cluster Methods}} for {{High-Energy Excited States}}: {{Application}}
  to {{K-edge X-ray Absorption Spectroscopy}}},}\ }\href
  {https://doi.org/10.1021/acs.jctc.5b00459} {\bibfield  {journal} {\bibinfo
  {journal} {J. Chem. Theory Comput.}\ }\textbf {\bibinfo {volume} {11}},\
  \bibinfo {pages} {4146--4153} (\bibinfo {year} {2015})}\BibitemShut {NoStop}%
\bibitem [{\citenamefont {Lestrange}, \citenamefont {Nguyen},\ and\
  \citenamefont {Li}(2015)}]{lestrangeCalibrationEnergySpecificTDDFT2015}%
  \BibitemOpen
  \bibfield  {author} {\bibinfo {author} {\bibfnamefont {P.~J.}\ \bibnamefont
  {Lestrange}}, \bibinfo {author} {\bibfnamefont {P.~D.}\ \bibnamefont
  {Nguyen}},\ and\ \bibinfo {author} {\bibfnamefont {X.}~\bibnamefont {Li}},\
  }\bibfield  {title} {\enquote {\bibinfo {title} {Calibration of
  {{Energy-Specific TDDFT}} for {{Modeling K-edge XAS Spectra}} of {{Light
  Elements}}},}\ }\href {https://doi.org/10.1021/acs.jctc.5b00169} {\bibfield
  {journal} {\bibinfo  {journal} {J. Chem. Theory Comput.}\ }\textbf {\bibinfo
  {volume} {11}},\ \bibinfo {pages} {2994--2999} (\bibinfo {year}
  {2015})}\BibitemShut {NoStop}%
\bibitem [{\citenamefont {Kasper}\ \emph {et~al.}(2018)\citenamefont {Kasper},
  \citenamefont {{Williams-Young}}, \citenamefont {Vecharynski}, \citenamefont
  {Yang},\ and\ \citenamefont {Li}}]{kasperWellTemperedHybridMethod2018}%
  \BibitemOpen
  \bibfield  {author} {\bibinfo {author} {\bibfnamefont {J.~M.}\ \bibnamefont
  {Kasper}}, \bibinfo {author} {\bibfnamefont {D.~B.}\ \bibnamefont
  {{Williams-Young}}}, \bibinfo {author} {\bibfnamefont {E.}~\bibnamefont
  {Vecharynski}}, \bibinfo {author} {\bibfnamefont {C.}~\bibnamefont {Yang}},\
  and\ \bibinfo {author} {\bibfnamefont {X.}~\bibnamefont {Li}},\ }\bibfield
  {title} {\enquote {\bibinfo {title} {A {{Well-Tempered Hybrid Method}} for
  {{Solving Challenging Time-Dependent Density Functional Theory}} ({{TDDFT}})
  {{Systems}}},}\ }\href {https://doi.org/10.1021/acs.jctc.8b00141} {\bibfield
  {journal} {\bibinfo  {journal} {J. Chem. Theory Comput.}\ }\textbf {\bibinfo
  {volume} {14}},\ \bibinfo {pages} {2034--2041} (\bibinfo {year}
  {2018})}\BibitemShut {NoStop}%
\bibitem [{\citenamefont {Bai}\ and\ \citenamefont
  {Li}(2012)}]{baiMinimizationPrinciplesLinear2012}%
  \BibitemOpen
  \bibfield  {author} {\bibinfo {author} {\bibfnamefont {Z.}~\bibnamefont
  {Bai}}\ and\ \bibinfo {author} {\bibfnamefont {R.-C.}\ \bibnamefont {Li}},\
  }\bibfield  {title} {\enquote {\bibinfo {title} {Minimization {{Principles}}
  for the {{Linear Response Eigenvalue Problem I}}: {{Theory}}},}\ }\href
  {https://doi.org/10.1137/110838960} {\bibfield  {journal} {\bibinfo
  {journal} {SIAM J. Matrix Anal. Appl.}\ }\textbf {\bibinfo {volume} {33}},\
  \bibinfo {pages} {1075--1100} (\bibinfo {year} {2012})}\BibitemShut {NoStop}%
\bibitem [{\citenamefont {Shao}\ \emph {et~al.}(2016)\citenamefont {Shao},
  \citenamefont {{da Jornada}}, \citenamefont {Yang}, \citenamefont
  {Deslippe},\ and\ \citenamefont
  {Louie}}]{shaoStructurePreservingParallel2016}%
  \BibitemOpen
  \bibfield  {author} {\bibinfo {author} {\bibfnamefont {M.}~\bibnamefont
  {Shao}}, \bibinfo {author} {\bibfnamefont {F.~H.}\ \bibnamefont {{da
  Jornada}}}, \bibinfo {author} {\bibfnamefont {C.}~\bibnamefont {Yang}},
  \bibinfo {author} {\bibfnamefont {J.}~\bibnamefont {Deslippe}},\ and\
  \bibinfo {author} {\bibfnamefont {S.~G.}\ \bibnamefont {Louie}},\ }\bibfield
  {title} {\enquote {\bibinfo {title} {Structure preserving parallel algorithms
  for solving the {{Bethe}}--{{Salpeter}} eigenvalue problem},}\ }\href
  {https://doi.org/10.1016/j.laa.2015.09.036} {\bibfield  {journal} {\bibinfo
  {journal} {Linear Algebra Its Appl.}\ }\textbf {\bibinfo {volume} {488}},\
  \bibinfo {pages} {148--167} (\bibinfo {year} {2016})}\BibitemShut {NoStop}%
\bibitem [{\citenamefont {Ghosh}\ and\ \citenamefont
  {Chattaraj}(2016)}]{ghoshConceptsMethodsModern2016}%
  \BibitemOpen
  \bibfield  {author} {\bibinfo {author} {\bibfnamefont {S.~K.}\ \bibnamefont
  {Ghosh}}\ and\ \bibinfo {author} {\bibfnamefont {P.~K.}\ \bibnamefont
  {Chattaraj}},\ }\href@noop {} {\emph {\bibinfo {title} {Concepts and
  {{Methods}} in {{Modern Theoretical Chemistry}}: {{Electronic Structure}} and
  {{Reactivity}}}}}\ (\bibinfo  {publisher} {CRC Press},\ \bibinfo {year}
  {2016})\BibitemShut {NoStop}%
\bibitem [{\citenamefont {Rocca}, \citenamefont {Lu},\ and\ \citenamefont
  {Galli}(2010)}]{roccaInitioCalculationsOptical2010}%
  \BibitemOpen
  \bibfield  {author} {\bibinfo {author} {\bibfnamefont {D.}~\bibnamefont
  {Rocca}}, \bibinfo {author} {\bibfnamefont {D.}~\bibnamefont {Lu}},\ and\
  \bibinfo {author} {\bibfnamefont {G.}~\bibnamefont {Galli}},\ }\bibfield
  {title} {\enquote {\bibinfo {title} {Ab initio calculations of optical
  absorption spectra: {{Solution}} of the {{Bethe}}--{{Salpeter}} equation
  within density matrix perturbation theory},}\ }\href
  {https://doi.org/10.1063/1.3494540} {\bibfield  {journal} {\bibinfo
  {journal} {J. Chem. Phys.}\ }\textbf {\bibinfo {volume} {133}},\ \bibinfo
  {pages} {164109} (\bibinfo {year} {2010})}\BibitemShut {NoStop}%
\bibitem [{\citenamefont {Duchemin}, \citenamefont {Deutsch},\ and\
  \citenamefont {Blase}(2012)}]{ducheminShortRangeLongRangeChargeTransfer2012}%
  \BibitemOpen
  \bibfield  {author} {\bibinfo {author} {\bibfnamefont {I.}~\bibnamefont
  {Duchemin}}, \bibinfo {author} {\bibfnamefont {T.}~\bibnamefont {Deutsch}},\
  and\ \bibinfo {author} {\bibfnamefont {X.}~\bibnamefont {Blase}},\ }\bibfield
   {title} {\enquote {\bibinfo {title} {Short-{{Range}} to {{Long-Range
  Charge-Transfer Excitations}} in the {{Zincbacteriochlorin-Bacteriochlorin
  Complex}}: {{A Bethe-Salpeter Study}}},}\ }\href
  {https://doi.org/10.1103/PhysRevLett.109.167801} {\bibfield  {journal}
  {\bibinfo  {journal} {Phys. Rev. Lett.}\ }\textbf {\bibinfo {volume} {109}},\
  \bibinfo {pages} {167801} (\bibinfo {year} {2012})}\BibitemShut {NoStop}%
\bibitem [{\citenamefont {Faber}\ \emph {et~al.}(2013)\citenamefont {Faber},
  \citenamefont {Boulanger}, \citenamefont {Duchemin}, \citenamefont
  {Attaccalite},\ and\ \citenamefont
  {Blase}}]{faberManybodyGreensFunction2013}%
  \BibitemOpen
  \bibfield  {author} {\bibinfo {author} {\bibfnamefont {C.}~\bibnamefont
  {Faber}}, \bibinfo {author} {\bibfnamefont {P.}~\bibnamefont {Boulanger}},
  \bibinfo {author} {\bibfnamefont {I.}~\bibnamefont {Duchemin}}, \bibinfo
  {author} {\bibfnamefont {C.}~\bibnamefont {Attaccalite}},\ and\ \bibinfo
  {author} {\bibfnamefont {X.}~\bibnamefont {Blase}},\ }\bibfield  {title}
  {\enquote {\bibinfo {title} {Many-body {{Green}}'s function {{GW}} and
  {{Bethe-Salpeter}} study of the optical excitations in a paradigmatic model
  dipeptide},}\ }\href {https://doi.org/10.1063/1.4830236} {\bibfield
  {journal} {\bibinfo  {journal} {J. Chem. Phys.}\ }\textbf {\bibinfo {volume}
  {139}},\ \bibinfo {pages} {194308} (\bibinfo {year} {2013})}\BibitemShut
  {NoStop}%
\bibitem [{\citenamefont {Vecharynski}\ \emph {et~al.}(2017)\citenamefont
  {Vecharynski}, \citenamefont {Brabec}, \citenamefont {Shao}, \citenamefont
  {Govind},\ and\ \citenamefont
  {Yang}}]{vecharynskiEfficientBlockPreconditioned2017}%
  \BibitemOpen
  \bibfield  {author} {\bibinfo {author} {\bibfnamefont {E.}~\bibnamefont
  {Vecharynski}}, \bibinfo {author} {\bibfnamefont {J.}~\bibnamefont {Brabec}},
  \bibinfo {author} {\bibfnamefont {M.}~\bibnamefont {Shao}}, \bibinfo {author}
  {\bibfnamefont {N.}~\bibnamefont {Govind}},\ and\ \bibinfo {author}
  {\bibfnamefont {C.}~\bibnamefont {Yang}},\ }\bibfield  {title} {\enquote
  {\bibinfo {title} {Efficient block preconditioned eigensolvers for linear
  response time-dependent density functional theory},}\ }\href
  {https://doi.org/10.1016/j.cpc.2017.07.017} {\bibfield  {journal} {\bibinfo
  {journal} {Comput. Phys. Commun.}\ }\textbf {\bibinfo {volume} {221}},\
  \bibinfo {pages} {42--52} (\bibinfo {year} {2017})}\BibitemShut {NoStop}%
\bibitem [{Note1()}]{Note1}%
  \BibitemOpen
  \bibinfo {note} {Users of LAPACK may find it convenient to use \protect
  \texttt {dsygvd}}\BibitemShut {NoStop}%
\bibitem [{Note2()}]{Note2}%
  \BibitemOpen
  \bibinfo {note} {A rank-revealing QR decomposition is convenient since it
  allows one to discard negligible components of $\protect \mathbf {Q}$. As
  $\protect \mathbf {U}$ grows large, it may be necessary to orthogonalize
  against $\protect \mathbf {U}$ before and after QR to avoid
  restarts.}\BibitemShut {Stop}%
\bibitem [{fcd(2024)}]{fcdmft}%
  \BibitemOpen
  \href@noop {} {\enquote {\bibinfo {title} {{fcDMFT library}},}\ }\bibinfo
  {howpublished} {\url{https://github.com/ZhuGroup-Yale/fcdmft}} (\bibinfo
  {year} {2024})\BibitemShut {NoStop}%
\bibitem [{\citenamefont {Zhu}, \citenamefont {Cui},\ and\ \citenamefont
  {Chan}(2020)}]{zhuEfficientFormulationInitio2020}%
  \BibitemOpen
  \bibfield  {author} {\bibinfo {author} {\bibfnamefont {T.}~\bibnamefont
  {Zhu}}, \bibinfo {author} {\bibfnamefont {Z.-H.}\ \bibnamefont {Cui}},\ and\
  \bibinfo {author} {\bibfnamefont {G.~K.-L.}\ \bibnamefont {Chan}},\
  }\bibfield  {title} {\enquote {\bibinfo {title} {Efficient {{Formulation}} of
  {{Ab Initio Quantum Embedding}} in {{Periodic Systems}}: {{Dynamical
  Mean-Field Theory}}},}\ }\href {https://doi.org/10.1021/acs.jctc.9b00934}
  {\bibfield  {journal} {\bibinfo  {journal} {J. Chem. Theory Comput.}\
  }\textbf {\bibinfo {volume} {16}},\ \bibinfo {pages} {141--153} (\bibinfo
  {year} {2020})}\BibitemShut {NoStop}%
\bibitem [{\citenamefont {Cui}, \citenamefont {Zhu},\ and\ \citenamefont
  {Chan}(2020)}]{cuiEfficientImplementationInitio2020}%
  \BibitemOpen
  \bibfield  {author} {\bibinfo {author} {\bibfnamefont {Z.-H.}\ \bibnamefont
  {Cui}}, \bibinfo {author} {\bibfnamefont {T.}~\bibnamefont {Zhu}},\ and\
  \bibinfo {author} {\bibfnamefont {G.~K.-L.}\ \bibnamefont {Chan}},\
  }\bibfield  {title} {\enquote {\bibinfo {title} {Efficient {Implementation}
  of {{Ab Initio Quantum Embedding}} in {{Periodic Systems}}: {{Density Matrix
  Embedding Theory}}},}\ }\href {https://doi.org/10.1021/acs.jctc.9b00933}
  {\bibfield  {journal} {\bibinfo  {journal} {J. Chem. Theory Comput.}\
  }\textbf {\bibinfo {volume} {16}},\ \bibinfo {pages} {119--129} (\bibinfo
  {year} {2020})}\BibitemShut {NoStop}%
\bibitem [{\citenamefont {Zhu}\ and\ \citenamefont
  {Chan}(2021{\natexlab{a}})}]{zhuInitioFullCell2021}%
  \BibitemOpen
  \bibfield  {author} {\bibinfo {author} {\bibfnamefont {T.}~\bibnamefont
  {Zhu}}\ and\ \bibinfo {author} {\bibfnamefont {G.~K.-L.}\ \bibnamefont
  {Chan}},\ }\bibfield  {title} {\enquote {\bibinfo {title} {Ab {Initio} {Full}
  {Cell} {$GW$}+{DMFT} for {Correlated} {Materials}},}\ }\href
  {https://doi.org/10.1103/PhysRevX.11.021006} {\bibfield  {journal} {\bibinfo
  {journal} {Phys. Rev. X}\ }\textbf {\bibinfo {volume} {11}},\ \bibinfo
  {pages} {021006} (\bibinfo {year} {2021}{\natexlab{a}})}\BibitemShut
  {NoStop}%
\bibitem [{\citenamefont {Sun}\ \emph {et~al.}(2018)\citenamefont {Sun},
  \citenamefont {Berkelbach}, \citenamefont {Blunt}, \citenamefont {Booth},
  \citenamefont {Guo}, \citenamefont {Li}, \citenamefont {Liu}, \citenamefont
  {McClain}, \citenamefont {Sayfutyarova}, \citenamefont {Sharma},
  \citenamefont {Wouters},\ and\ \citenamefont
  {Chan}}]{sunPySCFPythonbasedSimulations2018}%
  \BibitemOpen
  \bibfield  {author} {\bibinfo {author} {\bibfnamefont {Q.}~\bibnamefont
  {Sun}}, \bibinfo {author} {\bibfnamefont {T.~C.}\ \bibnamefont {Berkelbach}},
  \bibinfo {author} {\bibfnamefont {N.~S.}\ \bibnamefont {Blunt}}, \bibinfo
  {author} {\bibfnamefont {G.~H.}\ \bibnamefont {Booth}}, \bibinfo {author}
  {\bibfnamefont {S.}~\bibnamefont {Guo}}, \bibinfo {author} {\bibfnamefont
  {Z.}~\bibnamefont {Li}}, \bibinfo {author} {\bibfnamefont {J.}~\bibnamefont
  {Liu}}, \bibinfo {author} {\bibfnamefont {J.~D.}\ \bibnamefont {McClain}},
  \bibinfo {author} {\bibfnamefont {E.~R.}\ \bibnamefont {Sayfutyarova}},
  \bibinfo {author} {\bibfnamefont {S.}~\bibnamefont {Sharma}}, \bibinfo
  {author} {\bibfnamefont {S.}~\bibnamefont {Wouters}},\ and\ \bibinfo {author}
  {\bibfnamefont {G.~K.-L.}\ \bibnamefont {Chan}},\ }\bibfield  {title}
  {\enquote {\bibinfo {title} {{{PySCF}}: The {{Python-based}} simulations of
  chemistry framework},}\ }\href {https://doi.org/10.1002/wcms.1340} {\bibfield
   {journal} {\bibinfo  {journal} {WIREs Comput. Mol. Sci.}\ }\textbf {\bibinfo
  {volume} {8}},\ \bibinfo {pages} {e1340} (\bibinfo {year}
  {2018})}\BibitemShut {NoStop}%
\bibitem [{\citenamefont {Sun}\ \emph {et~al.}(2020)\citenamefont {Sun},
  \citenamefont {Zhang}, \citenamefont {Banerjee}, \citenamefont {Bao},
  \citenamefont {Barbry}, \citenamefont {Blunt}, \citenamefont {Bogdanov},
  \citenamefont {Booth}, \citenamefont {Chen}, \citenamefont {Cui},
  \citenamefont {Eriksen}, \citenamefont {Gao}, \citenamefont {Guo},
  \citenamefont {Hermann}, \citenamefont {Hermes}, \citenamefont {Koh},
  \citenamefont {Koval}, \citenamefont {Lehtola}, \citenamefont {Li},
  \citenamefont {Liu}, \citenamefont {Mardirossian}, \citenamefont {McClain},
  \citenamefont {Motta}, \citenamefont {Mussard}, \citenamefont {Pham},
  \citenamefont {Pulkin}, \citenamefont {Purwanto}, \citenamefont {Robinson},
  \citenamefont {Ronca}, \citenamefont {Sayfutyarova}, \citenamefont
  {Scheurer}, \citenamefont {Schurkus}, \citenamefont {Smith}, \citenamefont
  {Sun}, \citenamefont {Sun}, \citenamefont {Upadhyay}, \citenamefont {Wagner},
  \citenamefont {Wang}, \citenamefont {White}, \citenamefont {Whitfield},
  \citenamefont {Williamson}, \citenamefont {Wouters}, \citenamefont {Yang},
  \citenamefont {Yu}, \citenamefont {Zhu}, \citenamefont {Berkelbach},
  \citenamefont {Sharma}, \citenamefont {Sokolov},\ and\ \citenamefont
  {Chan}}]{sunRecentDevelopmentsPySCF2020}%
  \BibitemOpen
  \bibfield  {author} {\bibinfo {author} {\bibfnamefont {Q.}~\bibnamefont
  {Sun}}, \bibinfo {author} {\bibfnamefont {X.}~\bibnamefont {Zhang}}, \bibinfo
  {author} {\bibfnamefont {S.}~\bibnamefont {Banerjee}}, \bibinfo {author}
  {\bibfnamefont {P.}~\bibnamefont {Bao}}, \bibinfo {author} {\bibfnamefont
  {M.}~\bibnamefont {Barbry}}, \bibinfo {author} {\bibfnamefont {N.~S.}\
  \bibnamefont {Blunt}}, \bibinfo {author} {\bibfnamefont {N.~A.}\ \bibnamefont
  {Bogdanov}}, \bibinfo {author} {\bibfnamefont {G.~H.}\ \bibnamefont {Booth}},
  \bibinfo {author} {\bibfnamefont {J.}~\bibnamefont {Chen}}, \bibinfo {author}
  {\bibfnamefont {Z.-H.}\ \bibnamefont {Cui}}, \bibinfo {author} {\bibfnamefont
  {J.~J.}\ \bibnamefont {Eriksen}}, \bibinfo {author} {\bibfnamefont
  {Y.}~\bibnamefont {Gao}}, \bibinfo {author} {\bibfnamefont {S.}~\bibnamefont
  {Guo}}, \bibinfo {author} {\bibfnamefont {J.}~\bibnamefont {Hermann}},
  \bibinfo {author} {\bibfnamefont {M.~R.}\ \bibnamefont {Hermes}}, \bibinfo
  {author} {\bibfnamefont {K.}~\bibnamefont {Koh}}, \bibinfo {author}
  {\bibfnamefont {P.}~\bibnamefont {Koval}}, \bibinfo {author} {\bibfnamefont
  {S.}~\bibnamefont {Lehtola}}, \bibinfo {author} {\bibfnamefont
  {Z.}~\bibnamefont {Li}}, \bibinfo {author} {\bibfnamefont {J.}~\bibnamefont
  {Liu}}, \bibinfo {author} {\bibfnamefont {N.}~\bibnamefont {Mardirossian}},
  \bibinfo {author} {\bibfnamefont {J.~D.}\ \bibnamefont {McClain}}, \bibinfo
  {author} {\bibfnamefont {M.}~\bibnamefont {Motta}}, \bibinfo {author}
  {\bibfnamefont {B.}~\bibnamefont {Mussard}}, \bibinfo {author} {\bibfnamefont
  {H.~Q.}\ \bibnamefont {Pham}}, \bibinfo {author} {\bibfnamefont
  {A.}~\bibnamefont {Pulkin}}, \bibinfo {author} {\bibfnamefont
  {W.}~\bibnamefont {Purwanto}}, \bibinfo {author} {\bibfnamefont {P.~J.}\
  \bibnamefont {Robinson}}, \bibinfo {author} {\bibfnamefont {E.}~\bibnamefont
  {Ronca}}, \bibinfo {author} {\bibfnamefont {E.~R.}\ \bibnamefont
  {Sayfutyarova}}, \bibinfo {author} {\bibfnamefont {M.}~\bibnamefont
  {Scheurer}}, \bibinfo {author} {\bibfnamefont {H.~F.}\ \bibnamefont
  {Schurkus}}, \bibinfo {author} {\bibfnamefont {J.~E.~T.}\ \bibnamefont
  {Smith}}, \bibinfo {author} {\bibfnamefont {C.}~\bibnamefont {Sun}}, \bibinfo
  {author} {\bibfnamefont {S.-N.}\ \bibnamefont {Sun}}, \bibinfo {author}
  {\bibfnamefont {S.}~\bibnamefont {Upadhyay}}, \bibinfo {author}
  {\bibfnamefont {L.~K.}\ \bibnamefont {Wagner}}, \bibinfo {author}
  {\bibfnamefont {X.}~\bibnamefont {Wang}}, \bibinfo {author} {\bibfnamefont
  {A.}~\bibnamefont {White}}, \bibinfo {author} {\bibfnamefont {J.~D.}\
  \bibnamefont {Whitfield}}, \bibinfo {author} {\bibfnamefont {M.~J.}\
  \bibnamefont {Williamson}}, \bibinfo {author} {\bibfnamefont
  {S.}~\bibnamefont {Wouters}}, \bibinfo {author} {\bibfnamefont
  {J.}~\bibnamefont {Yang}}, \bibinfo {author} {\bibfnamefont {J.~M.}\
  \bibnamefont {Yu}}, \bibinfo {author} {\bibfnamefont {T.}~\bibnamefont
  {Zhu}}, \bibinfo {author} {\bibfnamefont {T.~C.}\ \bibnamefont {Berkelbach}},
  \bibinfo {author} {\bibfnamefont {S.}~\bibnamefont {Sharma}}, \bibinfo
  {author} {\bibfnamefont {A.~Y.}\ \bibnamefont {Sokolov}},\ and\ \bibinfo
  {author} {\bibfnamefont {G.~K.-L.}\ \bibnamefont {Chan}},\ }\bibfield
  {title} {\enquote {\bibinfo {title} {Recent developments in the {{PySCF}}
  program package},}\ }\href {https://doi.org/10.1063/5.0006074} {\bibfield
  {journal} {\bibinfo  {journal} {J. Chem. Phys.}\ }\textbf {\bibinfo {volume}
  {153}},\ \bibinfo {pages} {024109} (\bibinfo {year} {2020})}\BibitemShut
  {NoStop}%
\bibitem [{\citenamefont {Zhu}\ and\ \citenamefont
  {Chan}(2021{\natexlab{b}})}]{zhuAllElectronGaussianBasedG0W02021}%
  \BibitemOpen
  \bibfield  {author} {\bibinfo {author} {\bibfnamefont {T.}~\bibnamefont
  {Zhu}}\ and\ \bibinfo {author} {\bibfnamefont {G.~K.-L.}\ \bibnamefont
  {Chan}},\ }\bibfield  {title} {\enquote {\bibinfo {title} {All-{{Electron
  Gaussian-Based G0W0}} for {{Valence}} and {{Core Excitation Energies}} of
  {{Periodic Systems}}},}\ }\href {https://doi.org/10.1021/acs.jctc.0c00704}
  {\bibfield  {journal} {\bibinfo  {journal} {J. Chem. Theory Comput.}\
  }\textbf {\bibinfo {volume} {17}},\ \bibinfo {pages} {727--741} (\bibinfo
  {year} {2021}{\natexlab{b}})}\BibitemShut {NoStop}%
\bibitem [{\citenamefont {Lei}\ and\ \citenamefont
  {Zhu}(2022)}]{leiGaussianbasedQuasiparticleSelfconsistent2022}%
  \BibitemOpen
  \bibfield  {author} {\bibinfo {author} {\bibfnamefont {J.}~\bibnamefont
  {Lei}}\ and\ \bibinfo {author} {\bibfnamefont {T.}~\bibnamefont {Zhu}},\
  }\bibfield  {title} {\enquote {\bibinfo {title} {{Gaussian-Based
  Quasiparticle Self-Consistent $GW$ for Periodic Systems}},}\ }\href
  {https://doi.org/10.1063/5.0125756} {\bibfield  {journal} {\bibinfo
  {journal} {J. Chem. Phys.}\ }\textbf {\bibinfo {volume} {157}},\ \bibinfo
  {pages} {214114} (\bibinfo {year} {2022})}\BibitemShut {NoStop}%
\bibitem [{\citenamefont {Stoychev}, \citenamefont {Auer},\ and\ \citenamefont
  {Neese}(2017)}]{stoychevAutomaticGenerationAuxiliary2017}%
  \BibitemOpen
  \bibfield  {author} {\bibinfo {author} {\bibfnamefont {G.~L.}\ \bibnamefont
  {Stoychev}}, \bibinfo {author} {\bibfnamefont {A.~A.}\ \bibnamefont {Auer}},\
  and\ \bibinfo {author} {\bibfnamefont {F.}~\bibnamefont {Neese}},\ }\bibfield
   {title} {\enquote {\bibinfo {title} {Automatic {{Generation}} of {{Auxiliary
  Basis Sets}}},}\ }\href {https://doi.org/10.1021/acs.jctc.6b01041} {\bibfield
   {journal} {\bibinfo  {journal} {J. Chem. Theory Comput.}\ }\textbf {\bibinfo
  {volume} {13}},\ \bibinfo {pages} {554--562} (\bibinfo {year}
  {2017})}\BibitemShut {NoStop}%
\bibitem [{\citenamefont {Keller}\ \emph {et~al.}(2020)\citenamefont {Keller},
  \citenamefont {Blum}, \citenamefont {Rinke},\ and\ \citenamefont
  {Golze}}]{kellerRelativisticCorrectionScheme2020}%
  \BibitemOpen
  \bibfield  {author} {\bibinfo {author} {\bibfnamefont {L.}~\bibnamefont
  {Keller}}, \bibinfo {author} {\bibfnamefont {V.}~\bibnamefont {Blum}},
  \bibinfo {author} {\bibfnamefont {P.}~\bibnamefont {Rinke}},\ and\ \bibinfo
  {author} {\bibfnamefont {D.}~\bibnamefont {Golze}},\ }\bibfield  {title}
  {\enquote {\bibinfo {title} {Relativistic correction scheme for core-level
  binding energies from {{GW}}},}\ }\href {https://doi.org/10.1063/5.0018231}
  {\bibfield  {journal} {\bibinfo  {journal} {J. Chem. Phys.}\ }\textbf
  {\bibinfo {volume} {153}},\ \bibinfo {pages} {114110} (\bibinfo {year}
  {2020})}\BibitemShut {NoStop}%
\bibitem [{\citenamefont {Weigend}\ and\ \citenamefont
  {Ahlrichs}(2005)}]{weigendBalancedBasisSets2005}%
  \BibitemOpen
  \bibfield  {author} {\bibinfo {author} {\bibfnamefont {F.}~\bibnamefont
  {Weigend}}\ and\ \bibinfo {author} {\bibfnamefont {R.}~\bibnamefont
  {Ahlrichs}},\ }\bibfield  {title} {\enquote {\bibinfo {title} {Balanced basis
  sets of split valence, triple zeta valence and quadruple zeta valence quality
  for {{H}} to {{Rn}}: {{Design}} and assessment of accuracy},}\ }\href
  {https://doi.org/10.1039/b508541a} {\bibfield  {journal} {\bibinfo  {journal}
  {Phys. Chem. Chem. Phys.}\ }\textbf {\bibinfo {volume} {7}},\ \bibinfo
  {pages} {3297} (\bibinfo {year} {2005})}\BibitemShut {NoStop}%
\bibitem [{\citenamefont {Frisch}\ \emph {et~al.}(2016)\citenamefont {Frisch},
  \citenamefont {Trucks}, \citenamefont {Schlegel}, \citenamefont {Scuseria},
  \citenamefont {Robb}, \citenamefont {Cheeseman}, \citenamefont {Scalmani},
  \citenamefont {Barone}, \citenamefont {Petersson}, \citenamefont {Nakatsuji},
  \citenamefont {Li}, \citenamefont {Caricato}, \citenamefont {Marenich},
  \citenamefont {Bloino}, \citenamefont {Janesko}, \citenamefont {Gomperts},
  \citenamefont {Mennucci}, \citenamefont {Hratchian}, \citenamefont {Ortiz},
  \citenamefont {Izmaylov}, \citenamefont {Sonnenberg}, \citenamefont
  {Williams-Young}, \citenamefont {Ding}, \citenamefont {Lipparini},
  \citenamefont {Egidi}, \citenamefont {Goings}, \citenamefont {Peng},
  \citenamefont {Petrone}, \citenamefont {Henderson}, \citenamefont
  {Ranasinghe}, \citenamefont {Zakrzewski}, \citenamefont {Gao}, \citenamefont
  {Rega}, \citenamefont {Zheng}, \citenamefont {Liang}, \citenamefont {Hada},
  \citenamefont {Ehara}, \citenamefont {Toyota}, \citenamefont {Fukuda},
  \citenamefont {Hasegawa}, \citenamefont {Ishida}, \citenamefont {Nakajima},
  \citenamefont {Honda}, \citenamefont {Kitao}, \citenamefont {Nakai},
  \citenamefont {Vreven}, \citenamefont {Throssell}, \citenamefont
  {Montgomery}, \citenamefont {Peralta}, \citenamefont {Ogliaro}, \citenamefont
  {Bearpark}, \citenamefont {Heyd}, \citenamefont {Brothers}, \citenamefont
  {Kudin}, \citenamefont {Staroverov}, \citenamefont {Keith}, \citenamefont
  {Kobayashi}, \citenamefont {Normand}, \citenamefont {Raghavachari},
  \citenamefont {Rendell}, \citenamefont {Burant}, \citenamefont {Iyengar},
  \citenamefont {Tomasi}, \citenamefont {Cossi}, \citenamefont {Millam},
  \citenamefont {Klene}, \citenamefont {Adamo}, \citenamefont {Cammi},
  \citenamefont {Ochterski}, \citenamefont {Martin}, \citenamefont {Morokuma},
  \citenamefont {Farkas}, \citenamefont {Foresman},\ and\ \citenamefont
  {Fox}}]{g16}%
  \BibitemOpen
  \bibfield  {author} {\bibinfo {author} {\bibfnamefont {M.~J.}\ \bibnamefont
  {Frisch}}, \bibinfo {author} {\bibfnamefont {G.~W.}\ \bibnamefont {Trucks}},
  \bibinfo {author} {\bibfnamefont {H.~B.}\ \bibnamefont {Schlegel}}, \bibinfo
  {author} {\bibfnamefont {G.~E.}\ \bibnamefont {Scuseria}}, \bibinfo {author}
  {\bibfnamefont {M.~A.}\ \bibnamefont {Robb}}, \bibinfo {author}
  {\bibfnamefont {J.~R.}\ \bibnamefont {Cheeseman}}, \bibinfo {author}
  {\bibfnamefont {G.}~\bibnamefont {Scalmani}}, \bibinfo {author}
  {\bibfnamefont {V.}~\bibnamefont {Barone}}, \bibinfo {author} {\bibfnamefont
  {G.~A.}\ \bibnamefont {Petersson}}, \bibinfo {author} {\bibfnamefont
  {H.}~\bibnamefont {Nakatsuji}}, \bibinfo {author} {\bibfnamefont
  {X.}~\bibnamefont {Li}}, \bibinfo {author} {\bibfnamefont {M.}~\bibnamefont
  {Caricato}}, \bibinfo {author} {\bibfnamefont {A.~V.}\ \bibnamefont
  {Marenich}}, \bibinfo {author} {\bibfnamefont {J.}~\bibnamefont {Bloino}},
  \bibinfo {author} {\bibfnamefont {B.~G.}\ \bibnamefont {Janesko}}, \bibinfo
  {author} {\bibfnamefont {R.}~\bibnamefont {Gomperts}}, \bibinfo {author}
  {\bibfnamefont {B.}~\bibnamefont {Mennucci}}, \bibinfo {author}
  {\bibfnamefont {H.~P.}\ \bibnamefont {Hratchian}}, \bibinfo {author}
  {\bibfnamefont {J.~V.}\ \bibnamefont {Ortiz}}, \bibinfo {author}
  {\bibfnamefont {A.~F.}\ \bibnamefont {Izmaylov}}, \bibinfo {author}
  {\bibfnamefont {J.~L.}\ \bibnamefont {Sonnenberg}}, \bibinfo {author}
  {\bibfnamefont {D.}~\bibnamefont {Williams-Young}}, \bibinfo {author}
  {\bibfnamefont {F.}~\bibnamefont {Ding}}, \bibinfo {author} {\bibfnamefont
  {F.}~\bibnamefont {Lipparini}}, \bibinfo {author} {\bibfnamefont
  {F.}~\bibnamefont {Egidi}}, \bibinfo {author} {\bibfnamefont
  {J.}~\bibnamefont {Goings}}, \bibinfo {author} {\bibfnamefont
  {B.}~\bibnamefont {Peng}}, \bibinfo {author} {\bibfnamefont {A.}~\bibnamefont
  {Petrone}}, \bibinfo {author} {\bibfnamefont {T.}~\bibnamefont {Henderson}},
  \bibinfo {author} {\bibfnamefont {D.}~\bibnamefont {Ranasinghe}}, \bibinfo
  {author} {\bibfnamefont {V.~G.}\ \bibnamefont {Zakrzewski}}, \bibinfo
  {author} {\bibfnamefont {J.}~\bibnamefont {Gao}}, \bibinfo {author}
  {\bibfnamefont {N.}~\bibnamefont {Rega}}, \bibinfo {author} {\bibfnamefont
  {G.}~\bibnamefont {Zheng}}, \bibinfo {author} {\bibfnamefont
  {W.}~\bibnamefont {Liang}}, \bibinfo {author} {\bibfnamefont
  {M.}~\bibnamefont {Hada}}, \bibinfo {author} {\bibfnamefont {M.}~\bibnamefont
  {Ehara}}, \bibinfo {author} {\bibfnamefont {K.}~\bibnamefont {Toyota}},
  \bibinfo {author} {\bibfnamefont {R.}~\bibnamefont {Fukuda}}, \bibinfo
  {author} {\bibfnamefont {J.}~\bibnamefont {Hasegawa}}, \bibinfo {author}
  {\bibfnamefont {M.}~\bibnamefont {Ishida}}, \bibinfo {author} {\bibfnamefont
  {T.}~\bibnamefont {Nakajima}}, \bibinfo {author} {\bibfnamefont
  {Y.}~\bibnamefont {Honda}}, \bibinfo {author} {\bibfnamefont
  {O.}~\bibnamefont {Kitao}}, \bibinfo {author} {\bibfnamefont
  {H.}~\bibnamefont {Nakai}}, \bibinfo {author} {\bibfnamefont
  {T.}~\bibnamefont {Vreven}}, \bibinfo {author} {\bibfnamefont
  {K.}~\bibnamefont {Throssell}}, \bibinfo {author} {\bibfnamefont {J.~A.}\
  \bibnamefont {Montgomery}, \bibfnamefont {{Jr.}}}, \bibinfo {author}
  {\bibfnamefont {J.~E.}\ \bibnamefont {Peralta}}, \bibinfo {author}
  {\bibfnamefont {F.}~\bibnamefont {Ogliaro}}, \bibinfo {author} {\bibfnamefont
  {M.~J.}\ \bibnamefont {Bearpark}}, \bibinfo {author} {\bibfnamefont {J.~J.}\
  \bibnamefont {Heyd}}, \bibinfo {author} {\bibfnamefont {E.~N.}\ \bibnamefont
  {Brothers}}, \bibinfo {author} {\bibfnamefont {K.~N.}\ \bibnamefont {Kudin}},
  \bibinfo {author} {\bibfnamefont {V.~N.}\ \bibnamefont {Staroverov}},
  \bibinfo {author} {\bibfnamefont {T.~A.}\ \bibnamefont {Keith}}, \bibinfo
  {author} {\bibfnamefont {R.}~\bibnamefont {Kobayashi}}, \bibinfo {author}
  {\bibfnamefont {J.}~\bibnamefont {Normand}}, \bibinfo {author} {\bibfnamefont
  {K.}~\bibnamefont {Raghavachari}}, \bibinfo {author} {\bibfnamefont {A.~P.}\
  \bibnamefont {Rendell}}, \bibinfo {author} {\bibfnamefont {J.~C.}\
  \bibnamefont {Burant}}, \bibinfo {author} {\bibfnamefont {S.~S.}\
  \bibnamefont {Iyengar}}, \bibinfo {author} {\bibfnamefont {J.}~\bibnamefont
  {Tomasi}}, \bibinfo {author} {\bibfnamefont {M.}~\bibnamefont {Cossi}},
  \bibinfo {author} {\bibfnamefont {J.~M.}\ \bibnamefont {Millam}}, \bibinfo
  {author} {\bibfnamefont {M.}~\bibnamefont {Klene}}, \bibinfo {author}
  {\bibfnamefont {C.}~\bibnamefont {Adamo}}, \bibinfo {author} {\bibfnamefont
  {R.}~\bibnamefont {Cammi}}, \bibinfo {author} {\bibfnamefont {J.~W.}\
  \bibnamefont {Ochterski}}, \bibinfo {author} {\bibfnamefont {R.~L.}\
  \bibnamefont {Martin}}, \bibinfo {author} {\bibfnamefont {K.}~\bibnamefont
  {Morokuma}}, \bibinfo {author} {\bibfnamefont {O.}~\bibnamefont {Farkas}},
  \bibinfo {author} {\bibfnamefont {J.~B.}\ \bibnamefont {Foresman}},\ and\
  \bibinfo {author} {\bibfnamefont {D.~J.}\ \bibnamefont {Fox}},\ }\href@noop
  {} {\enquote {\bibinfo {title} {Gaussian16 {R}evision a.03},}\ } (\bibinfo
  {year} {2016}),\ \bibinfo {note} {gaussian Inc. Wallingford CT}\BibitemShut
  {NoStop}%
\bibitem [{\citenamefont {Dunning}(1989)}]{dunningGaussianBasisSets1989}%
  \BibitemOpen
  \bibfield  {author} {\bibinfo {author} {\bibfnamefont {T.~H.}\ \bibnamefont
  {Dunning}},\ }\bibfield  {title} {\enquote {\bibinfo {title} {Gaussian basis
  sets for use in correlated molecular calculations. {{I}}. {{The}} atoms boron
  through neon and hydrogen},}\ }\href {https://doi.org/10.1063/1.456153}
  {\bibfield  {journal} {\bibinfo  {journal} {J. Chem. Phys.}\ }\textbf
  {\bibinfo {volume} {90}},\ \bibinfo {pages} {1007--1023} (\bibinfo {year}
  {1989})}\BibitemShut {NoStop}%
\bibitem [{\citenamefont {Woon}\ and\ \citenamefont
  {Dunning}(1995)}]{woonGaussianBasisSets1995}%
  \BibitemOpen
  \bibfield  {author} {\bibinfo {author} {\bibfnamefont {D.~E.}\ \bibnamefont
  {Woon}}\ and\ \bibinfo {author} {\bibfnamefont {T.~H.}\ \bibnamefont
  {Dunning}, \bibfnamefont {Jr.}},\ }\bibfield  {title} {\enquote {\bibinfo
  {title} {Gaussian basis sets for use in correlated molecular calculations.
  {{V}}. {{Core}}-valence basis sets for boron through neon},}\ }\href
  {https://doi.org/10.1063/1.470645} {\bibfield  {journal} {\bibinfo  {journal}
  {J. Chem. Phys.}\ }\textbf {\bibinfo {volume} {103}},\ \bibinfo {pages}
  {4572--4585} (\bibinfo {year} {1995})}\BibitemShut {NoStop}%
\bibitem [{\citenamefont {{van Setten}}, \citenamefont {Weigend},\ and\
  \citenamefont {Evers}(2013)}]{vansettenGWMethodQuantumChemistry2013}%
  \BibitemOpen
  \bibfield  {author} {\bibinfo {author} {\bibfnamefont {M.~J.}\ \bibnamefont
  {{van Setten}}}, \bibinfo {author} {\bibfnamefont {F.}~\bibnamefont
  {Weigend}},\ and\ \bibinfo {author} {\bibfnamefont {F.}~\bibnamefont
  {Evers}},\ }\bibfield  {title} {\enquote {\bibinfo {title} {The {{GW-Method}}
  for {{Quantum Chemistry Applications}}: {{Theory}} and {{Implementation}}},}\
  }\href {https://doi.org/10.1021/ct300648t} {\bibfield  {journal} {\bibinfo
  {journal} {J. Chem. Theory Comput.}\ }\textbf {\bibinfo {volume} {9}},\
  \bibinfo {pages} {232--246} (\bibinfo {year} {2013})}\BibitemShut {NoStop}%
\bibitem [{\citenamefont {Golze}\ \emph {et~al.}(2018)\citenamefont {Golze},
  \citenamefont {Wilhelm}, \citenamefont {{van Setten}},\ and\ \citenamefont
  {Rinke}}]{golzeCoreLevelBindingEnergies2018}%
  \BibitemOpen
  \bibfield  {author} {\bibinfo {author} {\bibfnamefont {D.}~\bibnamefont
  {Golze}}, \bibinfo {author} {\bibfnamefont {J.}~\bibnamefont {Wilhelm}},
  \bibinfo {author} {\bibfnamefont {M.~J.}\ \bibnamefont {{van Setten}}},\ and\
  \bibinfo {author} {\bibfnamefont {P.}~\bibnamefont {Rinke}},\ }\bibfield
  {title} {\enquote {\bibinfo {title} {Core-{{Level Binding Energies}} from
  {{GW}}: {{An Efficient Full-Frequency Approach}} within a {{Localized
  Basis}}},}\ }\href {https://doi.org/10.1021/acs.jctc.8b00458} {\bibfield
  {journal} {\bibinfo  {journal} {J. Chem. Theory Comput.}\ }\textbf {\bibinfo
  {volume} {14}},\ \bibinfo {pages} {4856--4869} (\bibinfo {year}
  {2018})}\BibitemShut {NoStop}%
\bibitem [{\citenamefont {Golze}, \citenamefont {Keller},\ and\ \citenamefont
  {Rinke}(2020)}]{golzeAccurateAbsoluteRelative2020}%
  \BibitemOpen
  \bibfield  {author} {\bibinfo {author} {\bibfnamefont {D.}~\bibnamefont
  {Golze}}, \bibinfo {author} {\bibfnamefont {L.}~\bibnamefont {Keller}},\ and\
  \bibinfo {author} {\bibfnamefont {P.}~\bibnamefont {Rinke}},\ }\bibfield
  {title} {\enquote {\bibinfo {title} {Accurate {{Absolute}} and {{Relative
  Core-Level Binding Energies}} from {{GW}}},}\ }\href
  {https://doi.org/10.1021/acs.jpclett.9b03423} {\bibfield  {journal} {\bibinfo
   {journal} {J. Phys. Chem. Lett.}\ }\textbf {\bibinfo {volume} {11}},\
  \bibinfo {pages} {1840--1847} (\bibinfo {year} {2020})}\BibitemShut {NoStop}%
\bibitem [{\citenamefont {Li}\ \emph {et~al.}(2022{\natexlab{b}})\citenamefont
  {Li}, \citenamefont {Jin}, \citenamefont {Rinke}, \citenamefont {Yang},\ and\
  \citenamefont {Golze}}]{liBenchmarkGWMethods2022}%
  \BibitemOpen
  \bibfield  {author} {\bibinfo {author} {\bibfnamefont {J.}~\bibnamefont
  {Li}}, \bibinfo {author} {\bibfnamefont {Y.}~\bibnamefont {Jin}}, \bibinfo
  {author} {\bibfnamefont {P.}~\bibnamefont {Rinke}}, \bibinfo {author}
  {\bibfnamefont {W.}~\bibnamefont {Yang}},\ and\ \bibinfo {author}
  {\bibfnamefont {D.}~\bibnamefont {Golze}},\ }\bibfield  {title} {\enquote
  {\bibinfo {title} {Benchmark of {{GW Methods}} for {{Core-Level Binding
  Energies}}},}\ }\href {https://doi.org/10.1021/acs.jctc.2c00617} {\bibfield
  {journal} {\bibinfo  {journal} {J. Chem. Theory Comput.}\ }\textbf {\bibinfo
  {volume} {18}},\ \bibinfo {pages} {7570--7585} (\bibinfo {year}
  {2022}{\natexlab{b}})}\BibitemShut {NoStop}%
\bibitem [{\citenamefont
  {Becke}(1993)}]{beckeDensityfunctionalThermochemistryIII1993}%
  \BibitemOpen
  \bibfield  {author} {\bibinfo {author} {\bibfnamefont {A.~D.}\ \bibnamefont
  {Becke}},\ }\bibfield  {title} {\enquote {\bibinfo {title}
  {Density-functional thermochemistry. {{III}}. {{The}} role of exact
  exchange},}\ }\href {https://doi.org/10.1063/1.464913} {\bibfield  {journal}
  {\bibinfo  {journal} {J. Chem. Phys.}\ }\textbf {\bibinfo {volume} {98}},\
  \bibinfo {pages} {5648--5652} (\bibinfo {year} {1993})}\BibitemShut {NoStop}%
\bibitem [{\citenamefont {Lee}, \citenamefont {Yang},\ and\ \citenamefont
  {Parr}(1988)}]{leeDevelopmentColleSalvettiCorrelationenergy1988}%
  \BibitemOpen
  \bibfield  {author} {\bibinfo {author} {\bibfnamefont {C.}~\bibnamefont
  {Lee}}, \bibinfo {author} {\bibfnamefont {W.}~\bibnamefont {Yang}},\ and\
  \bibinfo {author} {\bibfnamefont {R.~G.}\ \bibnamefont {Parr}},\ }\bibfield
  {title} {\enquote {\bibinfo {title} {Development of the {{Colle-Salvetti}}
  correlation-energy formula into a functional of the electron density},}\
  }\href {https://doi.org/10.1103/PhysRevB.37.785} {\bibfield  {journal}
  {\bibinfo  {journal} {Phys. Rev. B}\ }\textbf {\bibinfo {volume} {37}},\
  \bibinfo {pages} {785--789} (\bibinfo {year} {1988})}\BibitemShut {NoStop}%
\bibitem [{\citenamefont {Grimme}\ \emph {et~al.}(2010)\citenamefont {Grimme},
  \citenamefont {Antony}, \citenamefont {Ehrlich},\ and\ \citenamefont
  {Krieg}}]{grimmeConsistentAccurateInitio2010}%
  \BibitemOpen
  \bibfield  {author} {\bibinfo {author} {\bibfnamefont {S.}~\bibnamefont
  {Grimme}}, \bibinfo {author} {\bibfnamefont {J.}~\bibnamefont {Antony}},
  \bibinfo {author} {\bibfnamefont {S.}~\bibnamefont {Ehrlich}},\ and\ \bibinfo
  {author} {\bibfnamefont {H.}~\bibnamefont {Krieg}},\ }\bibfield  {title}
  {\enquote {\bibinfo {title} {A consistent and accurate ab initio
  parametrization of density functional dispersion correction ({{DFT-D}}) for
  the 94 elements {{H-Pu}}},}\ }\href {https://doi.org/10.1063/1.3382344}
  {\bibfield  {journal} {\bibinfo  {journal} {J. Chem. Phys.}\ }\textbf
  {\bibinfo {volume} {132}},\ \bibinfo {pages} {154104} (\bibinfo {year}
  {2010})}\BibitemShut {NoStop}%
\bibitem [{\citenamefont {Hariharan}\ and\ \citenamefont
  {Pople}(1973)}]{hariharanInfluencePolarizationFunctions1973}%
  \BibitemOpen
  \bibfield  {author} {\bibinfo {author} {\bibfnamefont {P.~C.}\ \bibnamefont
  {Hariharan}}\ and\ \bibinfo {author} {\bibfnamefont {J.~A.}\ \bibnamefont
  {Pople}},\ }\bibfield  {title} {\enquote {\bibinfo {title} {The influence of
  polarization functions on molecular orbital hydrogenation energies},}\ }\href
  {https://doi.org/10.1007/BF00533485} {\bibfield  {journal} {\bibinfo
  {journal} {Theoret. Chim. Acta}\ }\textbf {\bibinfo {volume} {28}},\ \bibinfo
  {pages} {213--222} (\bibinfo {year} {1973})}\BibitemShut {NoStop}%
\bibitem [{\citenamefont {Diller}\ \emph {et~al.}(2013)\citenamefont {Diller},
  \citenamefont {Klappenberger}, \citenamefont {Allegretti}, \citenamefont
  {Papageorgiou}, \citenamefont {Fischer}, \citenamefont {Wiengarten},
  \citenamefont {Joshi}, \citenamefont {Seufert}, \citenamefont {{\'E}cija},
  \citenamefont {Auw{\"a}rter},\ and\ \citenamefont
  {Barth}}]{dillerInvestigatingMoleculesubstrateInteraction2013}%
  \BibitemOpen
  \bibfield  {author} {\bibinfo {author} {\bibfnamefont {K.}~\bibnamefont
  {Diller}}, \bibinfo {author} {\bibfnamefont {F.}~\bibnamefont
  {Klappenberger}}, \bibinfo {author} {\bibfnamefont {F.}~\bibnamefont
  {Allegretti}}, \bibinfo {author} {\bibfnamefont {A.~C.}\ \bibnamefont
  {Papageorgiou}}, \bibinfo {author} {\bibfnamefont {S.}~\bibnamefont
  {Fischer}}, \bibinfo {author} {\bibfnamefont {A.}~\bibnamefont {Wiengarten}},
  \bibinfo {author} {\bibfnamefont {S.}~\bibnamefont {Joshi}}, \bibinfo
  {author} {\bibfnamefont {K.}~\bibnamefont {Seufert}}, \bibinfo {author}
  {\bibfnamefont {D.}~\bibnamefont {{\'E}cija}}, \bibinfo {author}
  {\bibfnamefont {W.}~\bibnamefont {Auw{\"a}rter}},\ and\ \bibinfo {author}
  {\bibfnamefont {J.~V.}\ \bibnamefont {Barth}},\ }\bibfield  {title} {\enquote
  {\bibinfo {title} {Investigating the molecule-substrate interaction of
  prototypic tetrapyrrole compounds: {{Adsorption}} and self-metalation of
  porphine on {{Cu}}(111)},}\ }\href {https://doi.org/10.1063/1.4800771}
  {\bibfield  {journal} {\bibinfo  {journal} {J. Chem. Phys.}\ }\textbf
  {\bibinfo {volume} {138}},\ \bibinfo {pages} {154710} (\bibinfo {year}
  {2013})}\BibitemShut {NoStop}%
\bibitem [{\citenamefont {Krasnikov}\ \emph {et~al.}(2008)\citenamefont
  {Krasnikov}, \citenamefont {Sergeeva}, \citenamefont {Brzhezinskaya},
  \citenamefont {Preobrajenski}, \citenamefont {Sergeeva}, \citenamefont
  {Vinogradov}, \citenamefont {Cafolla}, \citenamefont {Senge},\ and\
  \citenamefont {Vinogradov}}]{krasnikovXrayAbsorptionPhotoemission2008}%
  \BibitemOpen
  \bibfield  {author} {\bibinfo {author} {\bibfnamefont {S.~A.}\ \bibnamefont
  {Krasnikov}}, \bibinfo {author} {\bibfnamefont {N.~N.}\ \bibnamefont
  {Sergeeva}}, \bibinfo {author} {\bibfnamefont {M.~M.}\ \bibnamefont
  {Brzhezinskaya}}, \bibinfo {author} {\bibfnamefont {A.~B.}\ \bibnamefont
  {Preobrajenski}}, \bibinfo {author} {\bibfnamefont {Y.~N.}\ \bibnamefont
  {Sergeeva}}, \bibinfo {author} {\bibfnamefont {N.~A.}\ \bibnamefont
  {Vinogradov}}, \bibinfo {author} {\bibfnamefont {A.~A.}\ \bibnamefont
  {Cafolla}}, \bibinfo {author} {\bibfnamefont {M.~O.}\ \bibnamefont {Senge}},\
  and\ \bibinfo {author} {\bibfnamefont {A.~S.}\ \bibnamefont {Vinogradov}},\
  }\bibfield  {title} {\enquote {\bibinfo {title} {An x-ray absorption and
  photoemission study of the electronic structure of {{Ni}} porphyrins and {{Ni
  N-confused}} porphyrin},}\ }\href
  {https://doi.org/10.1088/0953-8984/20/23/235207} {\bibfield  {journal}
  {\bibinfo  {journal} {J. Phys.: Condens. Matter}\ }\textbf {\bibinfo {volume}
  {20}},\ \bibinfo {pages} {235207} (\bibinfo {year} {2008})}\BibitemShut
  {NoStop}%
\bibitem [{\citenamefont {Venturella}\ \emph
  {et~al.}(2024{\natexlab{a}})\citenamefont {Venturella}, \citenamefont
  {Hillenbrand}, \citenamefont {Li},\ and\ \citenamefont
  {Zhu}}]{venturellaMachineLearningManyBody2024}%
  \BibitemOpen
  \bibfield  {author} {\bibinfo {author} {\bibfnamefont {C.}~\bibnamefont
  {Venturella}}, \bibinfo {author} {\bibfnamefont {C.}~\bibnamefont
  {Hillenbrand}}, \bibinfo {author} {\bibfnamefont {J.}~\bibnamefont {Li}},\
  and\ \bibinfo {author} {\bibfnamefont {T.}~\bibnamefont {Zhu}},\ }\bibfield
  {title} {\enquote {\bibinfo {title} {Machine {{Learning Many-Body Green}}'s
  {{Functions}} for {{Molecular Excitation Spectra}}},}\ }\href
  {https://doi.org/10.1021/acs.jctc.3c01146} {\bibfield  {journal} {\bibinfo
  {journal} {J. Chem. Theory Comput.}\ }\textbf {\bibinfo {volume} {20}},\
  \bibinfo {pages} {143--154} (\bibinfo {year}
  {2024}{\natexlab{a}})}\BibitemShut {NoStop}%
\bibitem [{\citenamefont {Perdew}, \citenamefont {Ernzerhof},\ and\
  \citenamefont {Burke}(1996)}]{perdewExactExchange1996}%
  \BibitemOpen
  \bibfield  {author} {\bibinfo {author} {\bibfnamefont {J.~P.}\ \bibnamefont
  {Perdew}}, \bibinfo {author} {\bibfnamefont {M.}~\bibnamefont {Ernzerhof}},\
  and\ \bibinfo {author} {\bibfnamefont {K.}~\bibnamefont {Burke}},\ }\bibfield
   {title} {\enquote {\bibinfo {title} {Rationale for mixing exact exchange
  with density functional approximations},}\ }\href
  {https://doi.org/10.1063/1.472933} {\bibfield  {journal} {\bibinfo  {journal}
  {The Journal of Chemical Physics}\ }\textbf {\bibinfo {volume} {105}},\
  \bibinfo {pages} {9982--9985} (\bibinfo {year} {1996})}\BibitemShut {NoStop}%
\bibitem [{\citenamefont {Brabec}\ \emph {et~al.}(2015)\citenamefont {Brabec},
  \citenamefont {Lin}, \citenamefont {Shao}, \citenamefont {Govind},
  \citenamefont {Yang}, \citenamefont {Saad},\ and\ \citenamefont
  {Ng}}]{brabecEfficientAlgorithmsEstimating2015}%
  \BibitemOpen
  \bibfield  {author} {\bibinfo {author} {\bibfnamefont {J.}~\bibnamefont
  {Brabec}}, \bibinfo {author} {\bibfnamefont {L.}~\bibnamefont {Lin}},
  \bibinfo {author} {\bibfnamefont {M.}~\bibnamefont {Shao}}, \bibinfo {author}
  {\bibfnamefont {N.}~\bibnamefont {Govind}}, \bibinfo {author} {\bibfnamefont
  {C.}~\bibnamefont {Yang}}, \bibinfo {author} {\bibfnamefont {Y.}~\bibnamefont
  {Saad}},\ and\ \bibinfo {author} {\bibfnamefont {E.~G.}\ \bibnamefont {Ng}},\
  }\bibfield  {title} {\enquote {\bibinfo {title} {Efficient {{Algorithms}} for
  {{Estimating}} the {{Absorption Spectrum}} within {{Linear Response
  TDDFT}}},}\ }\href {https://doi.org/10.1021/acs.jctc.5b00887} {\bibfield
  {journal} {\bibinfo  {journal} {J. Chem. Theory Comput.}\ }\textbf {\bibinfo
  {volume} {11}},\ \bibinfo {pages} {5197--5208} (\bibinfo {year}
  {2015})}\BibitemShut {NoStop}%
\bibitem [{\citenamefont {Yu}\ \emph {et~al.}(2024)\citenamefont {Yu},
  \citenamefont {Mei}, \citenamefont {Chen},\ and\ \citenamefont
  {Yang}}]{yu2024accurate}%
  \BibitemOpen
  \bibfield  {author} {\bibinfo {author} {\bibfnamefont {J.}~\bibnamefont
  {Yu}}, \bibinfo {author} {\bibfnamefont {Y.}~\bibnamefont {Mei}}, \bibinfo
  {author} {\bibfnamefont {Z.}~\bibnamefont {Chen}},\ and\ \bibinfo {author}
  {\bibfnamefont {W.}~\bibnamefont {Yang}},\ }\bibfield  {title} {\enquote
  {\bibinfo {title} {Accurate prediction of core level binding energies from
  ground-state density functional calculations: The importance of localization
  and screening},}\ }\href@noop {} {\bibfield  {journal} {\bibinfo  {journal}
  {arXiv preprint arXiv:2406.06345}\ } (\bibinfo {year} {2024})}\BibitemShut
  {NoStop}%
\bibitem [{\citenamefont {Polzonetti}\ \emph {et~al.}(2004)\citenamefont
  {Polzonetti}, \citenamefont {Carravetta}, \citenamefont {Iucci},
  \citenamefont {Ferri}, \citenamefont {Paolucci}, \citenamefont {Goldoni},
  \citenamefont {Parent}, \citenamefont {Laffon},\ and\ \citenamefont
  {Russo}}]{polzonettiElectronicStructurePlatinum2004}%
  \BibitemOpen
  \bibfield  {author} {\bibinfo {author} {\bibfnamefont {G.}~\bibnamefont
  {Polzonetti}}, \bibinfo {author} {\bibfnamefont {V.}~\bibnamefont
  {Carravetta}}, \bibinfo {author} {\bibfnamefont {G.}~\bibnamefont {Iucci}},
  \bibinfo {author} {\bibfnamefont {A.}~\bibnamefont {Ferri}}, \bibinfo
  {author} {\bibfnamefont {G.}~\bibnamefont {Paolucci}}, \bibinfo {author}
  {\bibfnamefont {A.}~\bibnamefont {Goldoni}}, \bibinfo {author} {\bibfnamefont
  {P.}~\bibnamefont {Parent}}, \bibinfo {author} {\bibfnamefont
  {C.}~\bibnamefont {Laffon}},\ and\ \bibinfo {author} {\bibfnamefont {M.~V.}\
  \bibnamefont {Russo}},\ }\bibfield  {title} {\enquote {\bibinfo {title}
  {Electronic structure of platinum complex/{{Zn-porphyrinato}} assembled
  macrosystems, related precursors and model molecules, as probed by {{X-ray}}
  absorption spectroscopy ({{NEXAFS}}): Theory and experiment},}\ }\href
  {https://doi.org/10.1016/j.chemphys.2003.09.036} {\bibfield  {journal}
  {\bibinfo  {journal} {Chem. Phys.}\ }\textbf {\bibinfo {volume} {296}},\
  \bibinfo {pages} {87--100} (\bibinfo {year} {2004})}\BibitemShut {NoStop}%
\bibitem [{Note3()}]{Note3}%
  \BibitemOpen
  \bibinfo {note} {2 x Intel\protect \textsuperscript {\textregistered }
  Xeon\protect \textsuperscript {\textregistered } Platinum 8268 CPU @
  2.90GHz}\BibitemShut {NoStop}%
\bibitem [{\citenamefont {Li}\ and\ \citenamefont
  {Zhu}(2024{\natexlab{a}})}]{li2024restoring}%
  \BibitemOpen
  \bibfield  {author} {\bibinfo {author} {\bibfnamefont {J.}~\bibnamefont
  {Li}}\ and\ \bibinfo {author} {\bibfnamefont {T.}~\bibnamefont {Zhu}},\
  }\bibfield  {title} {\enquote {\bibinfo {title} {Restoring translational
  symmetry in periodic all-orbital dynamical mean-field theory simulations},}\
  }\href@noop {} {\bibfield  {journal} {\bibinfo  {journal} {Faraday Discuss.}\
  } (\bibinfo {year} {2024}{\natexlab{a}})}\BibitemShut {NoStop}%
\bibitem [{\citenamefont {Li}\ and\ \citenamefont
  {Zhu}(2024{\natexlab{b}})}]{li2024interacting}%
  \BibitemOpen
  \bibfield  {author} {\bibinfo {author} {\bibfnamefont {J.}~\bibnamefont
  {Li}}\ and\ \bibinfo {author} {\bibfnamefont {T.}~\bibnamefont {Zhu}},\
  }\bibfield  {title} {\enquote {\bibinfo {title} {Interacting-bath dynamical
  embedding for capturing non-local electron correlation in solids},}\
  }\href@noop {} {\bibfield  {journal} {\bibinfo  {journal} {arXiv preprint
  arXiv:2406.07531}\ } (\bibinfo {year} {2024}{\natexlab{b}})}\BibitemShut
  {NoStop}%
\bibitem [{\citenamefont {{\c C}aylak}\ and\ \citenamefont
  {Baumeier}(2021)}]{caylakMachineLearningQuasiparticle2021}%
  \BibitemOpen
  \bibfield  {author} {\bibinfo {author} {\bibfnamefont {O.}~\bibnamefont {{\c
  C}aylak}}\ and\ \bibinfo {author} {\bibfnamefont {B.}~\bibnamefont
  {Baumeier}},\ }\bibfield  {title} {\enquote {\bibinfo {title} {Machine
  {{Learning}} of {{Quasiparticle Energies}} in {{Molecules}} and
  {{Clusters}}},}\ }\href {https://doi.org/10.1021/acs.jctc.1c00520} {\bibfield
   {journal} {\bibinfo  {journal} {J. Chem. Theory Comput.}\ }\textbf {\bibinfo
  {volume} {17}},\ \bibinfo {pages} {4891--4900} (\bibinfo {year}
  {2021})}\BibitemShut {NoStop}%
\bibitem [{\citenamefont {Venturella}\ \emph
  {et~al.}(2024{\natexlab{b}})\citenamefont {Venturella}, \citenamefont {Li},
  \citenamefont {Hillenbrand}, \citenamefont {Peralta}, \citenamefont {Liu},\
  and\ \citenamefont {Zhu}}]{venturella2024unified}%
  \BibitemOpen
  \bibfield  {author} {\bibinfo {author} {\bibfnamefont {C.}~\bibnamefont
  {Venturella}}, \bibinfo {author} {\bibfnamefont {J.}~\bibnamefont {Li}},
  \bibinfo {author} {\bibfnamefont {C.}~\bibnamefont {Hillenbrand}}, \bibinfo
  {author} {\bibfnamefont {X.~L.}\ \bibnamefont {Peralta}}, \bibinfo {author}
  {\bibfnamefont {J.}~\bibnamefont {Liu}},\ and\ \bibinfo {author}
  {\bibfnamefont {T.}~\bibnamefont {Zhu}},\ }\bibfield  {title} {\enquote
  {\bibinfo {title} {Unified deep learning framework for many-body quantum
  chemistry via green's functions},}\ }\href@noop {} {\bibfield  {journal}
  {\bibinfo  {journal} {arXiv preprint arXiv:2407.20384}\ } (\bibinfo {year}
  {2024}{\natexlab{b}})}\BibitemShut {NoStop}%
\bibitem [{\citenamefont {Biswas}\ and\ \citenamefont
  {Singh}(2024)}]{biswas2024incorporating}%
  \BibitemOpen
  \bibfield  {author} {\bibinfo {author} {\bibfnamefont {T.}~\bibnamefont
  {Biswas}}\ and\ \bibinfo {author} {\bibfnamefont {A.~K.}\ \bibnamefont
  {Singh}},\ }\bibfield  {title} {\enquote {\bibinfo {title} {Incorporating
  quasiparticle and excitonic properties into material discovery},}\
  }\href@noop {} {\bibfield  {journal} {\bibinfo  {journal} {arXiv preprint
  arXiv:2401.17831}\ } (\bibinfo {year} {2024})}\BibitemShut {NoStop}%
\end{thebibliography}%

\end{document}